\documentclass[amsmath,amssymb,11pt]{article}
\usepackage{jheppub2}
\pdfoutput=1
\usepackage{graphicx,epsfig}
\usepackage{float}
\usepackage{subfig}
\usepackage{subfloat}
\usepackage[utf8]{inputenc}
\usepackage{hyperref}
\usepackage{caption}

\usepackage{color}

\newcommand*{\affmark}[1][*]{\textsuperscript{#1}}

\newcommand{\beq}{\begin{equation}}

\newcommand{\eeq}{\end{equation}}

\title{Extending Charged Holographic R\'enyi Entropy}

\author{Andrew Svesko\affmark[a]}
\affiliation{\affmark[a]Department of Physics and Astronomy, University College London,
 Gower Street, London, WC1E 6BT, United Kingdom}
\emailAdd{a.svesko@ucl.ac.uk}

\abstract{Motivated by extended black hole thermodynamics, we generalize the R\'enyi entropy of charged holographic conformal field theories (CFTs) in $d$-dimensions. Specifically, following \cite{Johnson:2018bma}, we extend the quench description of the R\'enyi entropy of globally charged holographic CFTs by including pressure variations of charged hyperbolically sliced anti de Sitter black holes. We provide an exhaustive analysis of the new type of charged R\'enyi entropy, where we find an interesting interplay between a parameter controlling the pressure of the black hole and its charge. A field theoretic interpretation of this extended charged R\'enyi entropy is given. In particular, in $d=2$, where the bulk geometry becomes the charged Ba\~nados, Teitelboim, Zanelli black hole, we write down the extended charged R\'enyi entropy in terms of the twist operators of the charged field theory. An area law prescription for the extended R\'enyi entropy is formulated. We comment on several avenues for future work, including how global charge conservation relates to black hole super-entropicity.}

\begin{document}

\maketitle

\section{Introduction}
\label{intro}


With the advent of AdS/CFT, it has become clear there is a deep interplay between gravity and information theory. This realization is encapsulated by an entropy-area relation known as the Ryu-Takayanagi (RT) prescription \cite{Ryu:2006bv}:
\beq S_{\text{EE}}^{A}=\frac{\mathcal{A}(\gamma_{A})}{4G_{N}}\;.\label{RTent}\eeq
Here  $S^{A}_{\text{EE}}$ is the entanglement entropy of a conformal field theory (CFT) reduced to a region $A$ on the $d$-dimensional boundary of an asymptotically $(d+1)$-dimensional AdS (bulk) spacetime, and $\mathcal{A}(\gamma_{A})$ is the area of a static bulk minimal surface $\gamma_{A}$ that is homologous to boundary region $A$.  The RT formula (\ref{RTent}) and its covariant version \cite{Hubeny:2007xt} provide an information theoretic interpretation of the Bekenstein-Hawking entropy-area relation \cite{Bekenstein:1972tm,Hawking:1974rv,Hawking:1974sw}. Indeed, in an early attempt to provide a derivation of holographic entanglement entropy, Casini, Huerta, and Myers (CHM), showed that the entanglement entropy of a holographic CFT reduced to a ball in Minkowski space is equal to the thermodynamic entropy of a massless, hyperbolically sliced AdS-Schwarzschild black hole \cite{Casini:2011kv}. 
In this way, black holes provide a throughline connecting gravity, thermodynamics, and information. 

For field theories\footnote{Though we emphasize the R\'enyi entropy is defined for generic quantum systems, quantifying entanglement between two quantum subsystems $A$ and $B$.}, rather than computing the entanglement entropy directly
 one instead computes the R\'enyi entropy $S_{q}$,
\beq S_{q}=\frac{1}{1-q}\log[\text{tr}(\rho_{A}^{q})]\;,\label{renent1}\eeq
where $\rho_{A}$ is the reduced density matrix of the CFT, and $q$ is the R\'enyi index. The entropy $S_{q}$ can be used to calculate the entanglement entropy: In the limit $q\to 1$, $S_{1}=-\text{tr}(\rho\log\rho)$, the von Neumann entropy. The Reny\'i entropy, in this way, is more fundamental than the entanglement entropy. Moreover, other limits of $S_{q}$ provide additional insight into the nature of the entanglement spectrum. In the large $q$ limit (\ref{renent1}) yields the largest eigenvalue $\lambda$ of $\rho_{A}$, $S_{\infty}=-\log(\lambda)$ (the min-entropy), and as $q\to0$ we have $S_{0}=\log(d)$, where $d$ is the number of nonvanishing eigenvalues of $\rho_{A}$ (the Hartley entropy) \cite{Headrick:2010zt}. Thus, $S_{q}$ is an important diagnostic probe in information theory and condensed matter physics.

When the field theory in question carries a conserved (global) charge, the entanglement R\'enyi entropy $S_{q}$ is replaced with a grand canonical version of (\ref{renent1}),
\beq S_{q}(\mu)=\frac{1}{1-q}\log \text{tr}\left(\rho_{A}\frac{e^{\mu Q_{A}}}{n_{A}(\mu)}\right)^{q}\;,\label{charren1}\eeq
where $\mu$ is a `chemical potential' conjugate to the charge $Q_{A}$ confined to $A$, and $n_{A}(\mu)\equiv\text{tr}[\rho_{A}e^{\mu Q_{A}}]$ properly normalizes the `charged' reduced state $\rho_{A}e^{\mu Q_{A}}$. The charged R\'enyi entropy $S_{q}(\mu)$ therefore encodes how the entanglement between $A$ and its complement depends on the charge.

For CFTs with holographic duals, both the charged and uncharged entropies $S_{q}$ for general $q$ can be related to the thermodynamics of black holes and explicitly evaluated. This is accomplished in the following way \cite{Casini:2011kv}. Take the CFT ground state $\rho_{R}=e^{-K_{R}}$ reduced to a ball of radius $R$ in $d$-dimensional Minkowski space with $K_{R}$ being the modular Hamiltonian generating a \emph{local} modular flow in the causal domain of the ball. A judicious coordinate transformation takes the Minkowski background to the geometry of $\mathbb{R}\times \mathbb{H}^{d-1}$ with $\mathbb{H}^{d-1}$ a hyperbolic plane of size $R\equiv L_{0}$, up to a conformal factor, which may be eliminated with a conformal transformation. The reduced state $\rho_{R}$ is not invariant under the conformal transformation. Letting $U$ be the unitary operator acting on the CFT Hilbert space that implements the conformal transformation, we write the CHM map
\beq \rho_{R}=e^{-K_{R}}=U^{\dagger}\left(\frac{e^{-\beta H_{\tau}}}{Z}\right)U\;,\label{CHMmap}\eeq
where $H_{\tau}$ is the Hamiltonian generating time $\tau$ translations in the hyperbolic space, and $\beta$ is the inverse temperature $T_{0}=1/(2\pi L_{0})$. Since the von Neumann entropy is invariant under unitary transformations, this means the entanglement entropy across the ball $S_{\text{EE}}=-\text{tr}(\rho_{R}\log \rho_{R})$ is mapped to a thermal entropy on the hyperbolic background. With $\rho_{R}$ in Gibbs form, it is then straightforward to rewrite (\ref{renent1})  as a ``quench" of the Helmholtz free energy $F(T)$ \cite{baez:2011}
\beq S_{q}=-\frac{1}{\Delta T}[F(T_{0})-F(T_{0}/q)]=\frac{1}{\Delta T}\int^{T_{0}}_{T_{0}/q}S(T)dT\;,\label{ren2}\eeq
where $S=-\partial F/\partial T|_{V}$ and with $\Delta T=T_{0}-T_{0}/q$. In this sense, the R\'enyi entropy is an appropriately normalized measure of the free energy difference $\Delta F$ between the system at temperature $T_{0}$ and the system at temperature $T_{0}/q$, assuming the pressure is held \emph{fixed}. The connection to black hole physics is made with the aid of the CHM map (\ref{CHMmap}), in which the entropy $S(T)$ is replaced by the thermal entropy of a black hole with hyperbolic horizon \cite{Hung:2011nu}. In the charged R\'enyi case, the charged CFT is dual to a charged black hole \cite{Belin:2013uta}.

Due to the importance of both charged and uncharged R\'enyi entropies, any potential generalization is of interest. For holographic field theories a natural generalization presents itself, motivated by the program of extended black hole thermodynamics \cite{Kastor:2009wy,Kastor:2010gq,Dolan:2010ha,Dolan:2012jh,Kubiznak:2016qmn}, where the cosmological constant $\Lambda$ present in $\text{AdS}_{d+1}$ black hole backgrounds is interpreted as a pressure, $p=-\frac{\Lambda}{8\pi G_{N}}=\frac{d(d-1)}{16\pi G_{N}L^{2}}$, with $L$ being the AdS radius. Consequently, an AdS black hole has its mass $M$ identified with the enthalpy $H$ of a thermodynamic system and the first law of black hole thermodynamics is extended to $dH=TdS+Vdp$. The quantity $V=\left(\frac{\partial H}{\partial p}\right)_{S}$ is the conjugate variable to the pressure and is known as the `thermodynamic volume', since in static spacetimes of four dimensions\footnote{As we will discuss in Section \ref{sec:d2case}, not all static AdS black holes in three spacetime dimensions have a thermodynamic volume that is  equal to their naive geometric volume.} and higher it is the naive geometric volume of the black hole, while for, \emph{e.g.}, rotating systems it includes terms dependent on their spin\footnote{The thermodynamic volume, while appearing naturally as a thermodynamic variable, has its microscopic explanation shrouded in mystery. Recently, however, conditions on the thermodynamic volume  have been found to translate into restricting the number of accessible dual CFT states \cite{Johnson:2019wcq}, and, moreover, $V$ controls the complexity of formation of large black holes in either complexity equals volume/action proposals \cite{Balushi:2020wkt,Balushi:2020wjt}.}. 

According to the extended framework of black hole thermodynamics, it is then natural to extend the quench form of $S_{q}$ given in (\ref{ren2}), by relaxing the condition of fixed pressure. Using the first law $\Delta G = V\Delta p-S\Delta T$, we simply replace $\Delta F$ with a difference in Gibbs free energies $\Delta G$ evaluated at different temperatures and pressures \cite{Johnson:2018bma}
\beq S_{q,b}\equiv-\frac{G(p_{0},T_{0})-G(b^{2}p_{0},T_{0}/q)}{(\Delta T-V_{0}\Delta p/S_{0})}=\frac{1}{(\Delta T-V_{0}\Delta p/S_{0})}\int^{T_{0}}_{T_{0}/q}S(T)\left(1-\frac{V}{S}\frac{dp}{dT}\right)dT\label{Sqb1}\;,\eeq
where $\Delta p=p_{0}-b^{2}p_{0}$, with $b$ another integer `index' that plays a similar role as $q$. In the limit $b\to1$ the usual R\'enyi entropy is recovered; the limit $q\to1$ leads to a new kind of R\'enyi entropy $S_{b}$ whose interpretation is still largely mysterious. Nonetheless, as first shown in \cite{Johnson:2018bma}, the generalized R\'enyi entropy $S_{q,b}$ can be explicitly computed, and has a number of interesting properties. In particular, in the $d=2$ case, the index $b$ can be seen to undo the $q$-sheets used in the replica trick when calculating $S_{q}$, granting $S_{q,b}$ a field theory interpretation arising from a generalization of the CHM map (\ref{CHMmap}), 
\beq \rho^{(b)}_{R}=U^{\dagger}\left(\frac{e^{-(H_{\tau}+b^{2}p_{0}V_{0})/T_{0}}}{Z(T_{0},p_{0})}\right)U\;,\eeq
such that 
\beq S_{q,b}=\frac{1}{[(1-q)-q(d-1)(b^{2}-1)/2]}\log\left[\text{tr}(\rho^{(b)}_{R})^{q}\right]\;.\eeq
 This suggests $S_{1,b}$ can be computed via the replica trick as $\rho_{w}^{b}\equiv \rho_{R}^{(b)}$ for some other density matrix $\rho_{w}$ via a Euclidean path integral over a $b$ sheeted manifold. In this way, $b$ plays a role very similar to the traditional R\'enyi index $q$.

In this article, following \cite{Johnson:2018bma}, we extend the charged holographic R\'enyi entropy (\ref{charren1}) first explored in \cite{Belin:2013uta}. Our motivation of extending (\ref{charren1}) is three fold. Firstly, generalizing $S_{q}(\mu)$ to $S_{q,b}(\mu)$  provides a new entry in the holographic dictionary, in which we deepen the connection between entanglement and black hole entropy. Secondly, as the charged R\'enyi entropy plays an important role in studying phase transitions of holographic superconductors \cite{Belin:2014mva}, it's single parameter deformation $S_{q,b}(\mu)$ may shed new light into the phases of charged black hole systems. Finally, as with the uncharged generalized R\'enyi entropy, since $S_{q,b}(\mu)$ is computed starting on the gravity side, its field theoretic interpretation may provide additional insight into the replica calculation of charged R\'enyi entropies more generally. 

The outline of this article is as follows. For completeness, in Section \ref{sec:review} we briefly review the construction and computation of charged holographic R\'enyi entropies for CFT states dual to charged black holes. 
Section \ref{sec:gencharged} is devoted to extending the charged R\'enyi entropy, where we specifically consider CFT states dual to charged AdS black holes in four spacetime dimensions and higher. We find an interesting interplay between the chemical potential $\mu$ and parameter $b$, and briefly analyze the behavior of $S_{q,b}$ for imaginary chemical potentials $i\mu_{E}$. In Section \ref{sec:fieldtheoint} we provide a partial discussion of the field theory interpretation of $S_{q,b}$, including a generalization of the holographic calculation for the conformal dimension of twist operators inserted at the entangling surface. The geometry of charged $\text{AdS}_{d+1}$ black hole systems changes dramatically when $d=2$, in which case the dual R\'enyi entropy requires a more careful treatment \cite{Belin:2013uta}. In Section \ref{sec:d2case} we extend the charged R\'enyi entropy for CFTs dual to the charged Ba\~nados, Teitelboim, Zanelli (BTZ) black hole \cite{Martinez:1999qi}, where we provide a more precise field theoretic interpretation of $b$. In Section \ref{sec:gravdual} we provide an area law formulation of the extended R\'enyi entropy, akin to Dong's proposal for $S_{q}$ \cite{Dong:2016fnf}. We conclude and discuss potential avenues for future work in Section \ref{sec:conc}.


\section{Charged Holographic R\'enyi Entropy: Review}
\label{sec:review}







When we consider quantum field theories with a conserved global charge, the Reny\'i entropy is generalized to a \emph{charged} R\'enyi entropy, given in (\ref{charren1}), where $\mu$ is a `chemical potential' and $Q_{A}$ is the charge confined to the subsystem of interest \cite{Belin:2013uta}. As with ordinary Reny\'i entropies, the charged quantities can be evaluated using the replica trick, involving a Euclidean path integral on a $q$-sheeted geometry with twist operators $\sigma_{q}$ imposed at the boundary points of the cuts of the sheets where the entangling surface lives. A new ingredient is added, however, requiring a Wilson loop of the background gauge field $B$ associated with charge $Q_{A}$ to the entangling surface, thereby generalizing the twist operator to include a magnetic flux proportional to $\mu$.
The addition of a Wilson line about the Euclidean time circle is in fact a standard feature when including a chemical potential $\mu$ in the usual Euclidean path integral representation of a grand canonical thermal ensemble. 

When the quantum field theory is a CFT, the charged R\'enyi entropy can be evaluated expressed using the quench (\ref{ren2}), where the CHM map (\ref{CHMmap}) generalizes to
\beq \rho_{A}\frac{e^{\mu Q_{A}}}{n_{A}(\mu)}=U^{-1}\left(\frac{e^{-H_{\tau}/T_{0}+\mu Q_{A}}}{Z(T_{0},\mu)}\right)U\;,\quad \rho_{\text{therm}}=\frac{e^{-H_{\tau}/T_{0}+\mu Q_{A}}}{Z(T_{0},\mu)}\;.\label{CHMmapchar}\eeq
Here $Z(T_{0},\mu)$ is the thermal partition function associated with $\rho_{\text{therm}}$. The quench (\ref{ren2}) generalizes straightforwardly to 
\beq S_{q}(\mu)=\frac{1}{1-q}\log\frac{Z(T_{0}/q,\mu)}{Z(T_{0},\mu)^{q}}=\frac{q}{q-1}\frac{1}{T_{0}}\int_{T_{0}/q}^{T_{0}}S(T,\mu)dT\label{chargedren2}\;,\eeq
where we used that the free energy $F(T,\mu)$ is related to the partition function via $Z(T,\mu)=\exp(-F(T,\mu)/T_{0})$, and that the thermal entropy of the CFT is $S(T,\mu)=-\partial F/\partial T|_{V,\mu}$.


For holographic CFTs the thermal entropy $S$ is equivalent to the horizon entropy of an AdS black hole with a hypebolically sliced horizon. Via the AdS/CFT dictionary, the global symmetry due to the conserved charge $Q_{A}$ in the boundary CFT translates to a gauge field $A$ in the dual gravity theory, such that the AdS black hole is (electrically) charged. In particular, the $d+1$-dimensional bulk theory is described by an Einstein-Maxwell action
\beq I_{EM}=\frac{1}{16\pi G_{N}}\int d^{d+1}x\sqrt{-g}\left(R+\frac{d(d-1)}{L^{2}}-\frac{\ell^{2}_{\ast}}{4}F_{\mu\nu}F^{\mu\nu}\right)\;,\eeq
where $\ell_{\ast}$ is the electromagnetic coupling. This theory admits charged topological black holes with metric
\beq ds^{2}=-f(r)d\tau'^{2}+f^{-1}(r)dr^{2}+r^{2}(du^{2}+\sinh^{2}(u)d\Omega_{d-2}^{2})\;,\eeq
with
\beq f(r)=\frac{r^{2}}{L^{2}}-1-\frac{m}{r^{d-2}}+\frac{q_{e}^{2}}{r^{2d-4}}\;.\eeq
The horizon $r_{h}$ is located at the largest root of $f(r_{h})=0$. The gauge field $A$ is
\beq A=\left(\sqrt{\frac{2(d-1)}{(d-2)}}\frac{q_{e}}{\ell_{\ast}}\right)\left(\frac{1}{r_{h}^{d-2}}-\frac{1}{r^{d-2}}\right)d\tau'\;.\eeq
 The chemical potential $\mu$ is fixed by demanding $A=0$ at the horizon,
\beq \mu=2\pi\sqrt{\frac{2(d-1)}{(d-2)}}\frac{q_{e}}{\ell_{\ast}r_{h}^{d-2}}\;.\label{mure}\eeq

Writing the blackening factor $f(r)$ in terms of the horizon radius $r_{h}$, 
\beq f(r)=\frac{r^{2}}{L^{2}}-1+\frac{q^{2}_{e}}{r^{2d-4}}-\left(\frac{r_{h}}{r}\right)^{d-2}\left(\frac{r_{h}^{2}}{L^{2}}-1+\frac{q^{2}_{e}}{r_{h}^{2d-4}}\right)\;,\eeq
the temperature of the black hole is simply
\beq 
\begin{split}
T&=\frac{f'(r_{h})}{4\pi}=\frac{1}{4\pi Lx}\left(dx^{2}-(d-2)-\frac{(d-2)^{2}}{2(d-1)}\left(\frac{\mu\ell_{\ast}}{2\pi }\right)^{2}\right)
\end{split}
\label{Hawkingtemp}\;,\eeq
with $x\equiv r_{h}/L$.  In the neutral limit, when the black hole is massless $L=L_{0}$ such that $x=1$ and $T_{0}=1/(2\pi L_{0})$. The thermal entropy is the usual Bekenstein-Hawking entropy 
\beq S_{\text{BH}}=\frac{w_{d-1}L^{d-1}}{4G_{N}}x^{d-1}=\frac{L^{d-1}x^{d-1}}{4G_{N}}\Omega_{d-2}\int^{\infty}_{0}du\sinh^{d-2}(u)\;, \label{BHentunch}\eeq
where $w_{d-1}$ is the volume of $\mathbb{H}^{d-1}$ parameterized by coordinate $u\in\mathbb{R}_{+}$, and $\Omega_{d-2}$ is the volume of a unit sphere $S^{d-2}$.

The charged Reny\'i entropy is then computed via (\ref{chargedren2}), with the thermal entropy being replaced by the horizon entropy \cite{Belin:2013uta}:
\beq
\begin{split}
 S_{q}(\mu)&=\frac{q}{q-1}\frac{1}{T_{0}}\int_{x_{q}}^{x_{1}}S(x,\mu)\frac{dT}{dx}dx\\
&=\frac{q}{q-1}\frac{S_{EE}}{2}\left[x_{1}^{d-2}\left(1+\frac{(d-2)}{2(d-1)}\left(\frac{\mu\ell_{\ast}}{2\pi}\right)^{2}+x_{1}^{2}\right)-x_{q}^{d-2}\left(1+\frac{(d-2)}{2(d-1)}\left(\frac{\mu\ell_{\ast}}{2\pi }\right)^{2}+x_{q}^{2}\right)\right]\;.
\end{split}
\label{renentqcharge}\eeq 
Here $x_{q}$ is the largest solution to $T(x_{q},\mu)=T_{0}/q$, given by\footnote{Note that here the horizon entropy is \emph{not} that of an uncharged massless hyperbolic black hole. This changes the upper limit in the integral for the R\'enyi entropy from $x=1$ to $x=x_{1}$.}
\beq x_{q}=\frac{1}{qd}\left[1+\sqrt{1+q^{2}d(d-2)+\frac{q^{2}d(d-2)^{2}}{2(d-1)}\left(\frac{\mu\ell_{\ast}}{2\pi}\right)^{2}}\right]\;.\label{xqcharge}\eeq

Equation (\ref{renentqcharge}) is the charged holographic R\'enyi entropy for CFTs in dimension $d\geq3$. We emphasize by $S_{EE}$ we really mean the entanglement entropy for the neutral CFT, $S_{EE}(\mu=0)$. Holographically this is given by \cite{Casini:2011kv}
\beq S_{\text{EE}}=\left(\frac{2\Gamma(d/2)\Omega_{d-2}}{\pi^{d/2-1}}\right)a_{d}^{\ast}\int^{y_{\text{max}}}_{0}dy\frac{y^{d-2}}{\sqrt{1+y^{2}}}\;,\label{SEEint}\eeq
where $y=\sinh(u)$ and $y_{\text{max}}$ is a cutoff to regulate the UV divergence of the entanglement entropy due to correlations close to the entangling surface. The $L_{0}$ dependent factor has been replaced by the generalized central charge $a^{\ast}_{d}$ \cite{Myers:2010tj}
\beq a^{\ast}_{d}=\frac{\pi^{d/2-1}L_{0}^{d-1}}{8\Gamma(d/2)G_{N}}\;.\label{gencent}\eeq

The charged von Neumann entropy $S_{EE}(\mu)$ is simply given by the $q\to1$ limit of $S_{q}(\mu)$.
As explored in \cite{Belin:2013uta}, there are other limits of (\ref{renentqcharge}) that are of interest. In particular, the $q\to0$ limit is independent of $\mu$ and matches the same limit in the uncharged case, while the large $\mu$  behavior of $S_{q}(\mu)$ (at fixed $q$) is completely independent of the index $q$.


Charged R\'enyi entropies constructed from an imaginary chemical potential are also of interest. This can be done on the field theory side by analytically continuing $\mu\to i\mu_{E}$ for $\mu_{E}$ real. There is no issue with this analytic continuation procedure near $\mu=0$, however, one can encounter singularities along the imaginary $\mu$-axis. On the gravity side, the analytic continuation in $\mu$ corresponds to the continuation in the electric charge of the black hole $q_{e}\to iq_{e}^{E}$. There is no real problem with this except when $\mu_{E}$ exceeds the upper bound $\mu_{E}^{2}\leq\frac{8\pi^{2}(d-1)}{d-2}\left(\frac{1}{\ell_{\ast}}\right)^{2}\left(1+\frac{1}{d(d-2)q^{2}}\right)$, where $x_{q}$ goes imaginary. Consequently, the event horizon of the black hole disappears leaving a naked singularity.



\section{Physical Generalization of Charged R\'enyi Entropy}
\label{sec:gencharged}

\subsection{Extended Thermodynamics of Hyperbolic AdS-RN Black Hole}

The above holographic computations kept the AdS length scale $L$ fixed at $L_{0}$. In extended black hole thermodynamics \cite{Kastor:2009wy,Kastor:2010gq,Dolan:2010ha,Dolan:2012jh,Kubiznak:2016qmn}, a fixed length $L_{0}$ corresponds to a fixed thermodynamic pressure $p_{0}=\frac{d(d-1)}{16\pi G_{N}L_{0}^{2}}$. Equivalently, then, the quench (\ref{ren2}) is understood as a difference in \emph{Gibbs} free energies at different temperatures, but fixed pressures. It is thus natural to extend the R\'enyi entropy in (\ref{ren2}) by allowing the pressure to vary, exploiting the enlarged framework of extended black hole thermodynamics. This was first accomplished in \cite{Johnson:2018bma}. 

Similarly, we can extend the charged R\'enyi entropy (\ref{renentqcharge}) by exploiting the extended thermodynamics of hyperbolic AdS-RN black holes. The extended thermodynamics for this system has not yet appeared in the literature, however, it is straightforward to work out. We already have the temperature and entropy, (\ref{Hawkingtemp}) and (\ref{BHentunch}), respectively. The mass $M$ of the system, interpreted as the enthalpy in the enlarged framework, may be determined using the quasilocal approach of Brown and York \cite{Brown:1992br} adapted to an asymptotically AdS background
\beq M=-\frac{(d-1)w_{d-1}L^{d-2}}{16\pi G_{N}}x^{d-2}\left(1-x^{2}-\frac{(d-2)}{2(d-1)}\left(\frac{\mu\ell_{\ast}}{2\pi}\right)^{2}\right)\;.\label{massofhighd}\eeq
This expression also matches calculations using the Euclidean path integral\footnote{Since we are working with fixed chemical potential $\mu$, this corresponds to the grand canonical ensemble, \emph{i.e.}, fixed electric potential $\Phi_{e}$, such that the boundary term associated with the gauge field $A$ vanishes.} with a counterterm, first developed in \cite{Balasubramanian:1999re,Emparan:1999pm} for specific  spacetime dimensions. The electric potential $\Phi_{e}$ is defined by 
\beq \Phi_{e}=A_{t}|_{r\to\infty}-A_{t}|_{r\to r_{h}}=\frac{q_{e}}{\ell_{\ast}r_{h}^{d-2}}\sqrt{\frac{2(d-1)}{(d-2)}}=\frac{\mu}{2\pi}\;,\eeq
and the total electric charge\footnote{We use the convention $Q=\frac{\ell^{2}_{\ast}}{16\pi G_{N}}\oint\ast F$, where $\ast$ represents the Hodge dual, $\ast F=(\frac{1}{4}\sqrt{-g}\epsilon_{\alpha\beta\mu\nu})F^{\alpha\beta}dx^{\mu}\wedge dx^{\nu}$ is the totally antisymmetric Levi-Civita symbol.} $Q$ is,
\beq Q=\frac{w_{d-1}}{16\pi G_{N}}(d-2)\sqrt{\frac{2(d-1)}{(d-2)}}\ell_{\ast}q_{e}=(d-2)\frac{w_{d-1}L^{d-2}}{16\pi G_{N}}x^{d-2}\frac{\mu\ell_{\ast}^{2}}{2\pi} \;.\eeq
From the mass $M$ we find that the thermodynamic volume is
\beq V\equiv\left(\frac{\partial M}{\partial p}\right)_{S,Q}=\frac{w_{d-1}}{d}r_{h}^{d}=\frac{4G_{N}}{d}r_{h}S_{\text{BH}}\;,\label{volume1}\eeq
just as in the uncharged case. Notice that the thermodynamic volume of the charged black hole is the naive geometric volume. In fact, since the entropy only depends on the horizon radius $r_{h}$, we may conclude that paths of constant volume $V$ (isochores) are equivalent to paths of constant entropy $S_{\text{BH}}$ (adiabats) in the entire $(p,V)$ plane. This matches the uncharged hyperbolic black hole \cite{Johnson:2018amj}, as well as static AdS black holes in general \cite{Johnson:2014yja}. As we will observe later, this behavior does not hold for $d=2$ charged AdS black holes. 

The Gibbs free energy is 
\beq G=M-TS-\Phi_{e}Q\;.\label{Gbbsc}\eeq
After some minor algebra we have
\beq 
\begin{split}
G(x,L)&=-\frac{w_{d-1}L^{d-2}}{16\pi G_{N}}x^{d-2}\left[1+\frac{(d-2)}{2(d-1)}\left(\frac{\mu\ell_{\ast}}{2\pi}\right)^{2}+x^{2}\right]\;.
\end{split}
\label{freeenergycharge}\eeq

At fixed potential $\mu$, the hyperbolic AdS-RN black hole shares many of the same qualitative features as their neutral cousins. For example, the pressure $p$ for the $M=0$ curve has the same functional dependence of $V$ as shown in \cite{Johnson:2018amj}
\beq p(M=0)=\frac{\kappa}{V^{2/d}}\left[1-\frac{(d-2)}{2(d-1)}\left(\frac{\mu\ell_{\ast}}{2\pi}\right)^{2}\right]\;,\quad \kappa\equiv\frac{d(d-1)}{16\pi G_{N}}\left(\frac{w_{d-1}}{d}\right)^{2/d}\;.\eeq
Similarly, the zero temperature curve in the $(p,V)$ plane follows
\beq p(T=0)=\left(\frac{d-2}{d}\right)\left(1+\frac{d(d-2)}{2(d-1)}\left(\frac{\mu\ell_{\ast}}{2\pi}\right)^{2}\right)\frac{\kappa}{V^{2/d}}\;.\eeq
Consequently, the $p(M=0)$ and $p(T=0)$ curves never coincide with each other such that the zero temperature curve is always below the massless curve. Thus, there is an entire region within the $(p,V)$ plane below the massless curve, where the mass of the black hole is negative. 

The field theory interpretation of the curves in the $(p,V)$ plane are also largely the same as the neutral case \cite{Johnson:2018amj}, when $\mu$ is held fixed. To summarize, points on the massless curve correspond to CFT vacuum states reduced to the ball in Minkowski space, in which moving up to higher $p$ along the curve is correlated with going deeper into the IR, integrating out UV field theory degrees of freedom, equivalent to an RG flow in the dual CFT. Moreover, perturbations up and to the right of the massless curve correspond to CFT states dual to massive charged black holes. In particular, perturbations at fixed pressure correspond to perturbing the ground state of a fixed CFT (such that the extended first law of thermodynamics is dual to the extended first law of entanglement \cite{Blanco:2013joa,Wong:2013gua}), while vertical changes, keeping volume fixed, is equivalent to keeping the entanglement entropy fixed, but moving to a different CFT with a central charge $a_{d}^{\ast}$ with fewer UV degrees of freedom. Moving below the massless curve in the $(p,V)$, interestingly, describes CFT states with negative energy density such that the perturbed state is \emph{less} entangled than the unperturbed state \cite{Blanco:2013lea,Bianchi:2014qua,Rosso:2018yax,Rosso:2019lsm}.


\subsection{Extended Charged R\'enyi Entropy}

Let us now exploit the extended thermodynamics of our black hole system to generalize the charged R\'enyi entropy (\ref{renentqcharge}). First notice that as in the uncharged case \cite{Johnson:2018bma} the difference in Helmholtz free energies appearing in the quench (\ref{chargedren2}) can be replaced with a difference in Gibbs free energies, using (\ref{freeenergycharge}).  Keeping $L$ fixed, we see
\beq
 S_{q}(\mu)=-\frac{1}{\Delta T}[G(x_{1},L)-G(x_{q},L)]
\eeq
recovers the charged R\'enyi entropy (\ref{renentqcharge}).

To extend the charged R\'enyi entropy we simply allow for pressure changes\footnote{Notice the denominator (\ref{extchargedrenyi}) comes from $\Delta G=-S\Delta T+V\Delta p-Q\Delta \Phi_{e}$. Here, however, we are working in fixed $\mu$, such that $\Delta\Phi_{e}=0$. The result is the same denominator as appearing in the neutral case, however, with $V_{0}/S_{0}\sim L x_{11}$.}, such that $L_{0}$ changes to $L_{0}/b$. Then, with $\Delta T\equiv T_{0}-T_{0}/q$ and $\Delta p\equiv\frac{d(d-1)}{16\pi G_{N}}\frac{1}{L_{0}^{2}}(1-b^{2})\;,\quad \frac{V_{0}}{S_{0}}=\frac{4G_{N}L_{0}x_{11}}{d}$,
\beq 
\begin{split}
S_{q,b}(\mu)&=-\frac{1}{(\Delta T-V_{0}\Delta p/S_{0})}[G(x_{11},L_{0})-G(x_{qb},L_{0}/b)]\\
&=q\frac{S_{EE}}{2}\frac{\biggr\{x_{11}^{d-2}\left(1+\frac{(d-2)}{2(d-1)}\left(\frac{\mu\ell_{\ast}}{2\pi}\right)^{2}+x_{11}^{2}\right)-\left(\frac{x_{qb}}{b}\right)^{d-2}\left(1+\frac{(d-2)}{2(d-1)}\left(\frac{\mu\ell_{\ast}}{2\pi}\right)^{2}+x_{qb}^{2}\right)\biggr\}}{[q-1+q(d-1)(b^{2}-1)x_{11}/2]}\;,
\end{split}
\label{extchargedrenyi}\eeq
where
\beq x_{qb}=\frac{1}{qbd}\left[1+\sqrt{1+q^{2}b^{2}d(d-2)+\frac{q^{2}b^{2}d(d-2)^{2}}{2(d-1)}\left(\frac{\mu\ell_{\ast}}{2\pi}\right)^{2}}\right]\;.\label{xqbchargereal}\eeq
which is found by setting $T_{0}/q=T(\mu,x_{qb})$ and solving for $x_{qb}$. In the neutral  limit $\mu=0$, we recover the result from \cite{Johnson:2018bma}, where $x_{11}=1$. Also observe $\lim_{b\to1}S_{q,b}(\mu)$ reduces to $S_{q}(\mu)$ in (\ref{renentqcharge}). When $q\to1$, we have a genuine new kind of entropy to study $S_{1,b}(\mu)$, which we will do so in detail momentarily. Notice, moreover that when $q,b\neq1$, we have $S_{q,b}(\mu)$ diverges when the index $q$ takes the special value $q_{c}$:
\beq q_{c}=\frac{2}{(d-1)(b^{2}-1)x_{11}+2}\;,\label{qcrit}\eeq
for which $b>1$ corresponds to $q_{c}<1$, just as in the neutral case.

Let's now consider some important limits of our generalization of the charged Reny\'i entropy $S_{q,b}(\mu)$. It is useful to have 
\beq \lim_{q\;\text{or}\; b\to0} x_{qb}\to \frac{2}{qbd}\;,\label{xqbto0noL}\eeq
and 
\beq \lim_{q\,\text{or}\,b\to\infty}x_{qb}\to\sqrt{\frac{(d-2)}{d}}\sqrt{1+\frac{(d-2)}{2(d-1)}\left(\frac{\mu\ell_{\ast}}{2\pi}\right)^{2}}\equiv x_{\infty}\;,\label{xqbqinftnoL}\eeq
the former of which matches the uncharged case \cite{Johnson:2018bma}. We will also be interested in the large $\mu$ limit:
\beq \lim_{\mu\to\infty}x_{qb}\to\sqrt{\frac{(d-2)^{2}}{2d(d-1)}}\left(\frac{\mu\ell_{\ast}}{2\pi}\right)\;.\eeq

\begin{figure}[t]
  \subfloat[]{
	\begin{minipage}[1\width]{
	   0.32\textwidth}
	   \centering
	   \includegraphics[width=1.45\textwidth]{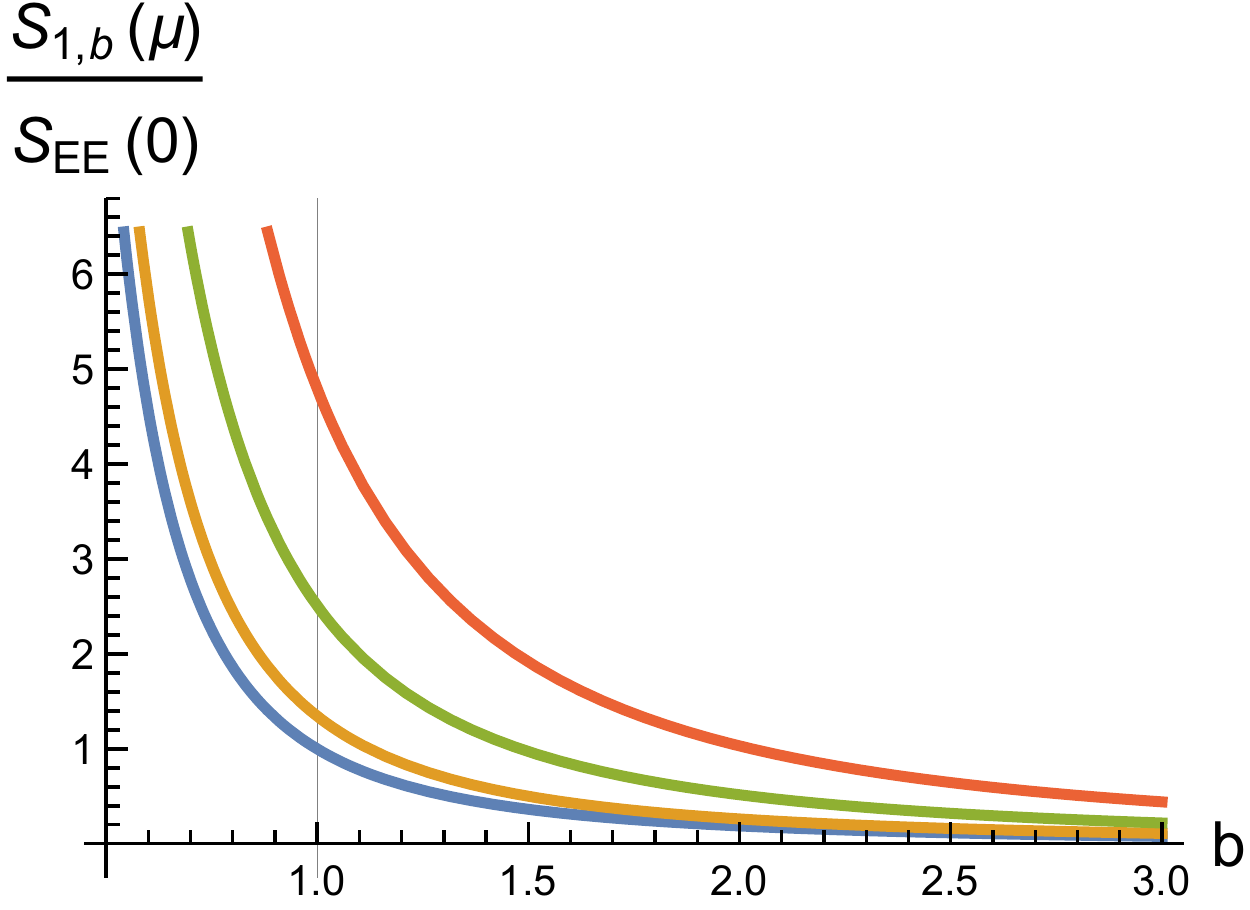}
	\end{minipage}}
 \hspace{2.5cm}
  \subfloat[]{
	\begin{minipage}[1\width]{
	   0.32\textwidth}
	   \centering
	   \includegraphics[width=1.45\textwidth]{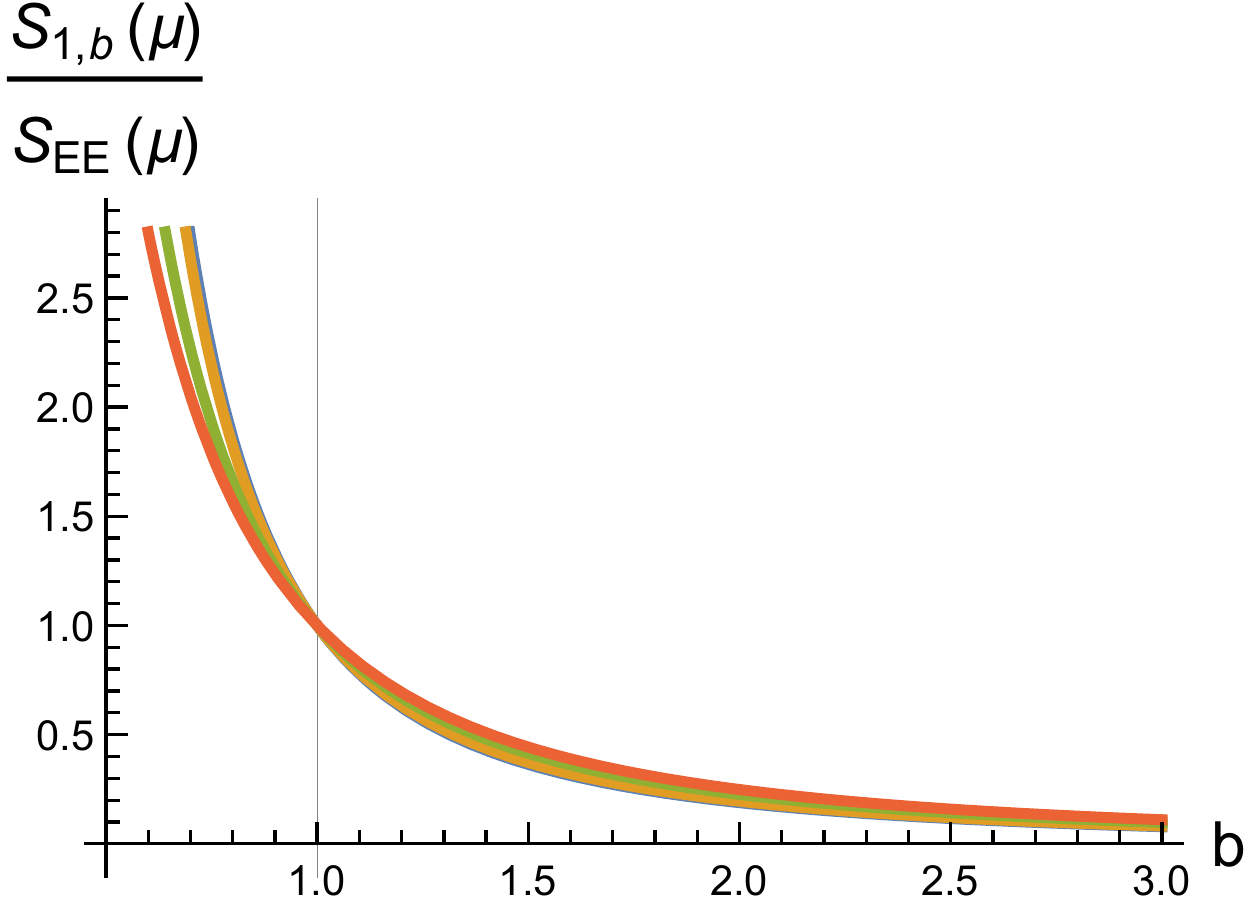}
	\end{minipage}}
\caption{Plots of $S_{1,b}(\mu)/S_{EE}(\mu)$ as a function of $b$ in $d=4$ for various $\mu$. (a) Normalized by $S_{EE}(0)$; curves have $\mu=$ 0 (blue), 1 (orange), 2 (green), 3 (red). (b) Normalized by $S_{EE}(\mu)$; same values of $\mu$ except the green and red curves correspond to $\mu=5,100$, respectively. The blue and and orange curves largely overlap. We have set $\ell_{\ast}=2\pi$.}
\label{S1bfuncb}\end{figure}

\begin{figure}[t]
  \subfloat[]{
	\begin{minipage}[1\width]{
	   0.32\textwidth}
	   \centering
	   \includegraphics[width=1.45\textwidth]{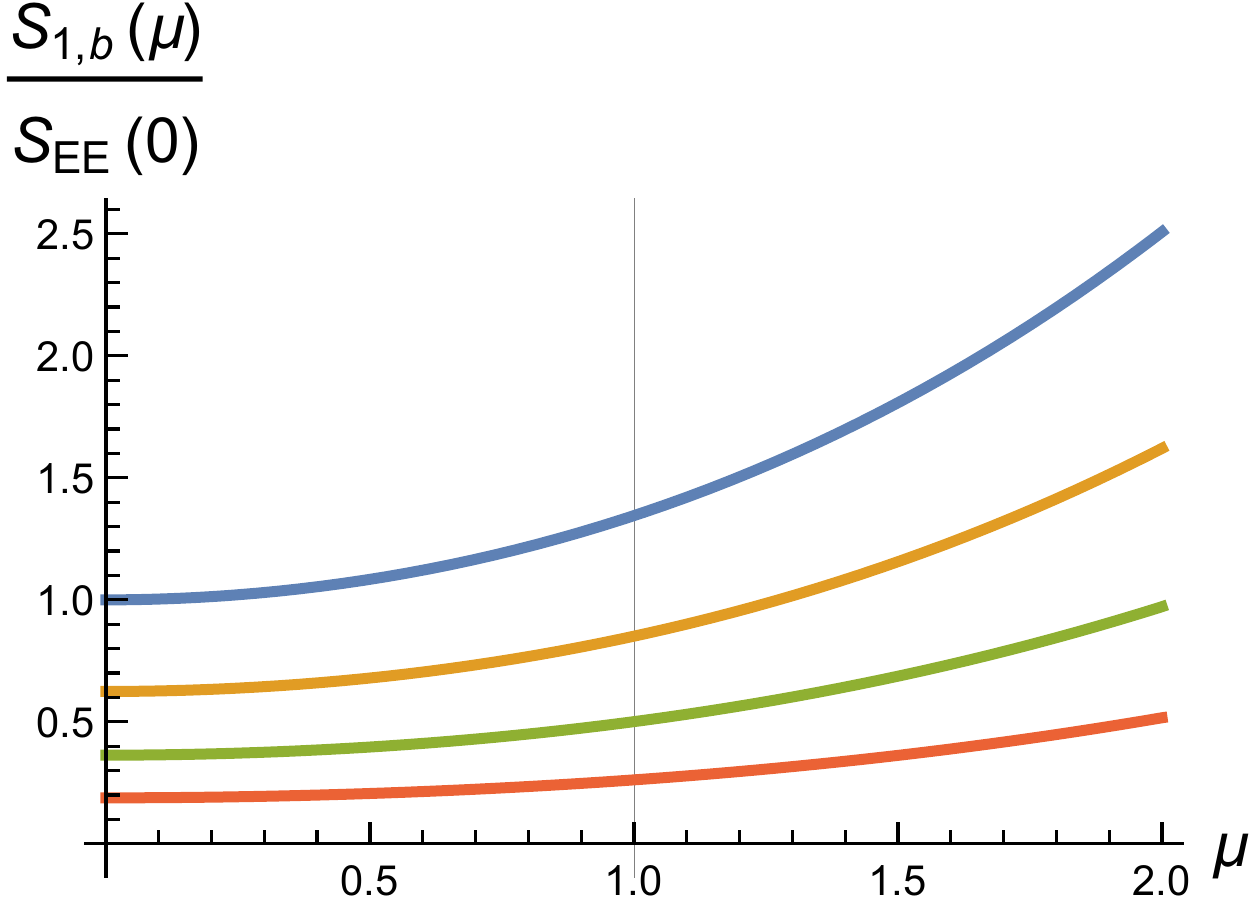}
	\end{minipage}}
 \hspace{2.5cm}
  \subfloat[]{
	\begin{minipage}[1\width]{
	   0.32\textwidth}
	   \centering
	   \includegraphics[width=1.45\textwidth]{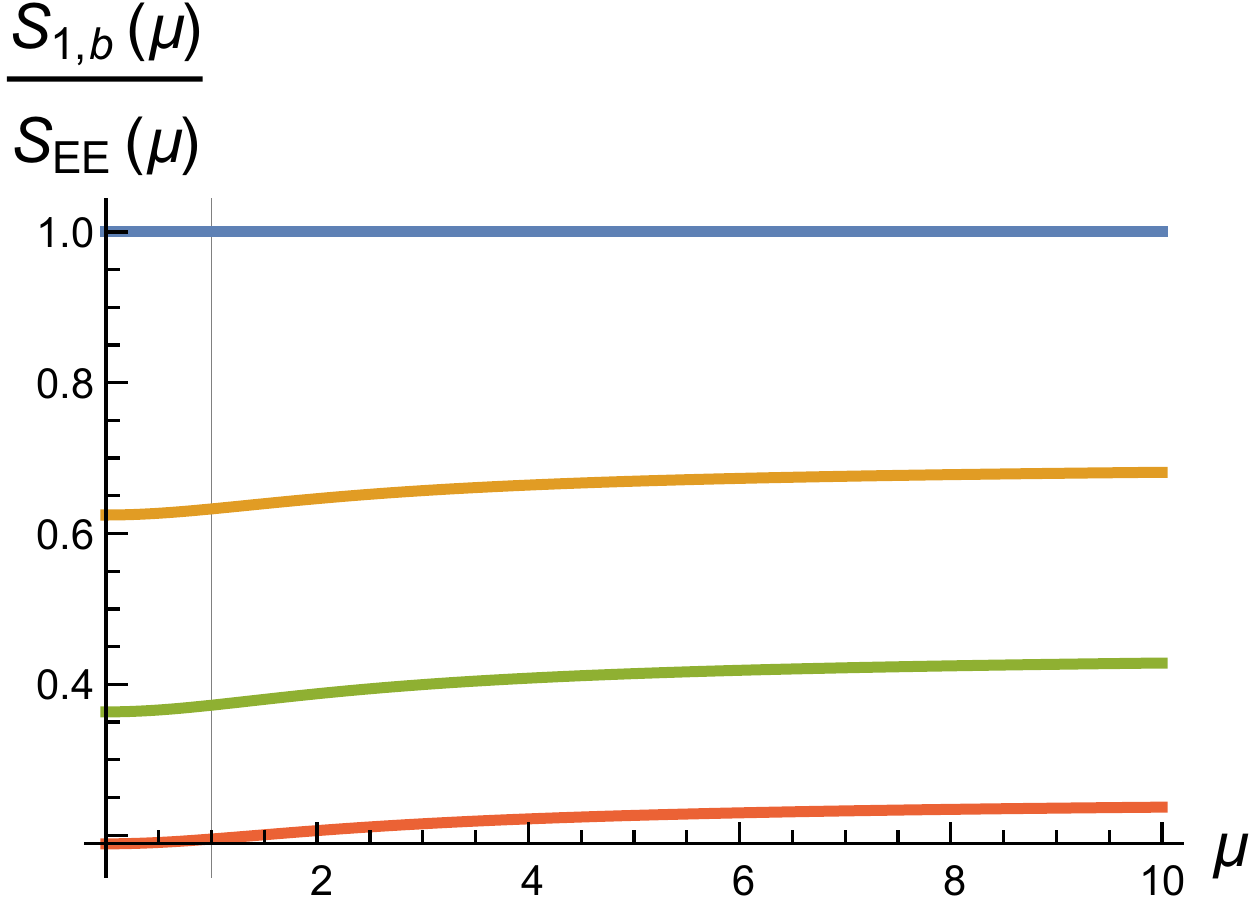}
	\end{minipage}}
\caption{Plots of $S_{1,b}(\mu)/S_{EE}(\mu)$ as a function of $\mu$ in $d=4$ for various $b$. (a) Normalized by $S_{EE}(0)$; curves have $b=$ 1 (blue), 1.2 (orange), 1.5 (green), 2 (red). (b) Normalized by $S_{EE}(\mu)$. We have set $\ell_{\ast}=2\pi$.}
\label{S1bfuncmu}\end{figure}

To take the $q,b\to1$ limits of $S_{q,b}$ the following are useful to know
\beq \frac{dx_{qb}}{dq}=-\frac{x_{qb}}{q}\frac{1}{[qbdx_{qb}-1]}\;,\quad  \frac{dx_{qb}}{db}=-\frac{x_{qb}}{b}\frac{1}{[qbdx_{qb}-1]}\;.\eeq
The limits $q\to1$ and $b\to1$ commute such that
 \beq S_{1,1}(\mu)\equiv S_{EE}(\mu)=\frac{1}{2}S_{EE}\frac{(d-2)x_{11}^{d-2}}{dx_{11}-1}\left(1+\frac{(d-2)}{2(d-1)}\left(\frac{\mu\ell_{\ast}}{2\pi}\right)^{2}+\frac{dx_{11}^{2}}{d-2}\right)\;.\label{SEEcharged}\eeq
When $\mu=0$ we recover $S_{EE}(\mu)=S_{EE}$, where $x_{11}=1$. 
 This means we will observe different behavior between $S_{q,b}(\mu)/S_{EE}(\mu)$ and $S_{q,b}(\mu)/S_{EE}(0)$, as can be seen in Figure \ref{S1bfuncb} and Figure \ref{S1bfuncmu}. Except when $\mu=0$, we consider $S_{q,b}(\mu)/S_{EE}(\mu)$ the proper normalization, in that all curves meet at $q=b=1$.

Let's consider some more interesting limits. It is straightforward to work out
\beq \lim_{q\to0}S_{q,b}(\mu)=\frac{S_{EE}}{2}\left(\frac{2}{d}\right)^{d}\left(\frac{1}{b^{2}q}\right)^{d-1}\;.\eeq
and
\beq \lim_{b\to0}S_{q,b}(\mu)=\frac{S_{EE}}{2(q(d-1)x_{11}/2 -(q-1))}\left(\frac{2}{d}\right)^{d}\left(\frac{1}{b^{2}q}\right)^{d-1}\;.\eeq
Note that the first limit is independentof $\mu$ to leading order, matching the neutral case \cite{Johnson:2018bma}. 

Next, 
\beq \lim_{q\to\infty}S_{q,b}(\mu)=\frac{S_{EE}}{2}\frac{\biggr\{x_{11}^{d-2}\left(1+\frac{(d-2)}{2(d-1)}\left(\frac{\mu\ell_{\ast}}{2\pi }\right)^{2}+x_{11}^{2}\right)-\left(\frac{x_{\infty}}{b}\right)^{d-2}\left(1+\frac{(d-2)}{2(d-1)}\left(\frac{\mu\ell_{\ast} }{2\pi }\right)^{2}+x_{\infty}^{2}\right)\biggr\}}{[1+(d-1)(b^{2}-1)x_{11}/2]}\;,\eeq
\beq \lim_{b\to\infty}S_{q,b}(\mu)=\frac{S_{EE}}{(d-1)b^{2}}x_{11}^{d-3}\left[1+\frac{(d-2)}{2(d-1)}\left(\frac{\mu\ell_{\ast}}{2\pi}\right)^{2}+x_{11}^{2}\right]\;,\label{largeblim}\eeq
\beq \lim_{\mu\to\infty}S_{q,b}(\mu)=\frac{2S_{EE}}{(b^{2}-1)}\sqrt{\frac{2d}{(d-1)}}\left(\frac{(d-2)^{2}}{2d(d-1)}\right)^{(d-2)/2}\left(\frac{\mu\ell_{\ast}}{2\pi}\right)^{d-1}\left(1-\frac{1}{b^{d-2}}\right)\;.\eeq
We observe that the $q\to\infty$ limit leads to some particular finite value, and matches  $\lim_{q\to1}S_{q}$ found in \cite{Belin:2013uta} when we set $b=1$. Moreover, we see $S_{q,b}$ will vanish for large $b$, similar to the uncharged case \cite{Johnson:2018bma}. A difference between $S_{q}(\mu)$ and $S_{q,b}(\mu)$ at large $\mu$ is $\lim_{\mu\to\infty}S_{q,b}(\mu)$ has $b$ dependence, though is independent of index $q$. 


Drastically different behavior for $S_{1,b}(\mu)$ is observed in Figure \ref{S1bfuncb} and Figure \ref{S1bfuncmu} corresponding to the way we normalize $S_{1,b}(\mu)$. Indeed, based on our expectation that $\lim_{b\to1}S_{1,b}(\mu)$ should approach unity, we find that we must normalize our $S_{q,b}(\mu)$ by (\ref{SEEcharged}), as opposed to $S_{EE}(0)$. Notice when we normalize by $S_{EE}(0)$ we see that $\mu$ has much more influence compared to when we normalize by $S_{EE}(\mu)$. In either normalization, however, the central feature evident in these figures is for $b>1$, as the index $b$ increases the entropy $S_{1,b}(\mu)$ decreases. Unlike $S_{q}(\mu)$, we see at large $b$ the entropy $S_{1,b}$ will vanish, however, the chemical potential $\mu$ can help stave off this limit, as seen by (\ref{largeblim}).




\begin{figure}[t]
  \subfloat[]{
	\begin{minipage}[1\width]{
	   0.32\textwidth}
	   \centering
	   \includegraphics[width=1.45\textwidth]{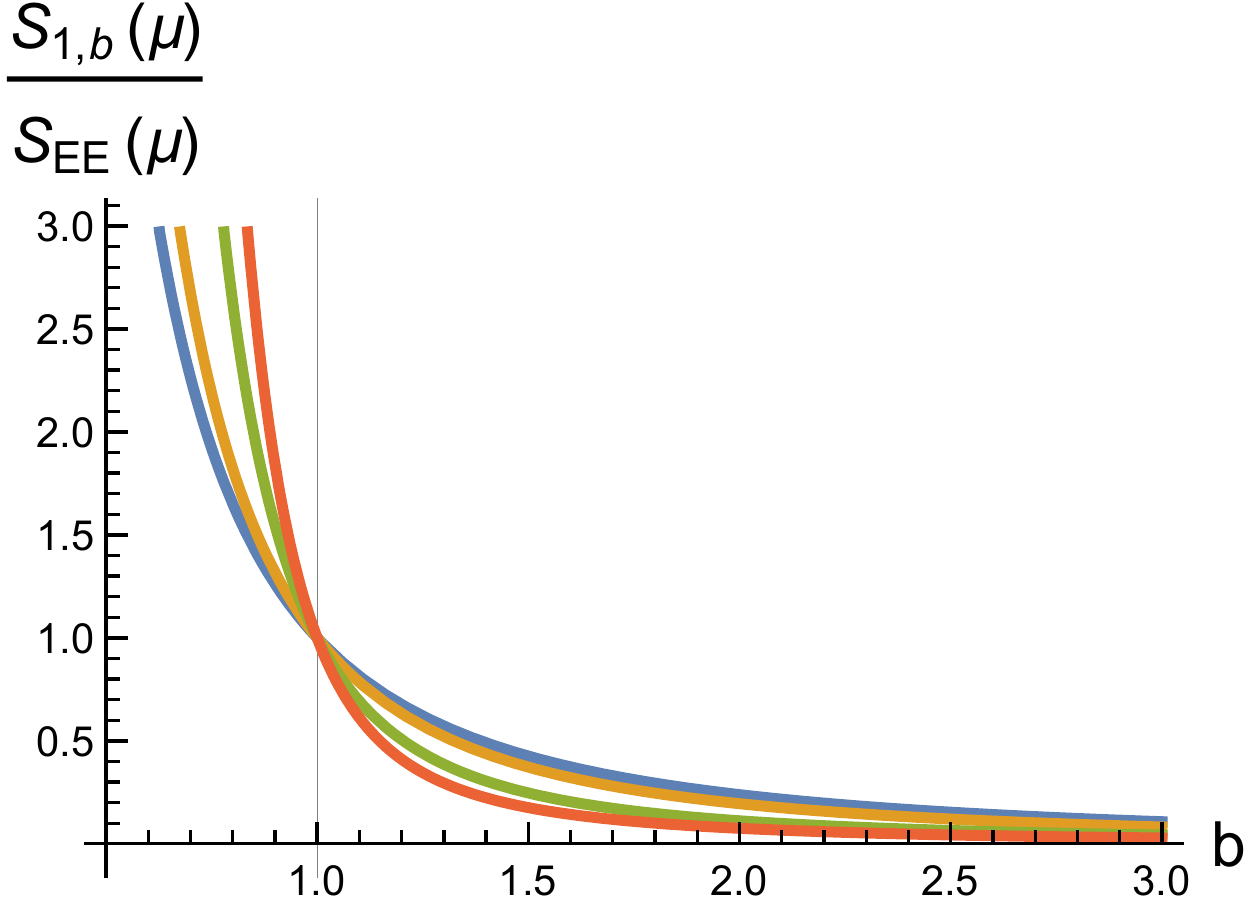}
	\end{minipage}}
 \hspace{2.5cm}
  \subfloat[]{
	\begin{minipage}[1\width]{
	   0.32\textwidth}
	   \centering
	   \includegraphics[width=1.45\textwidth]{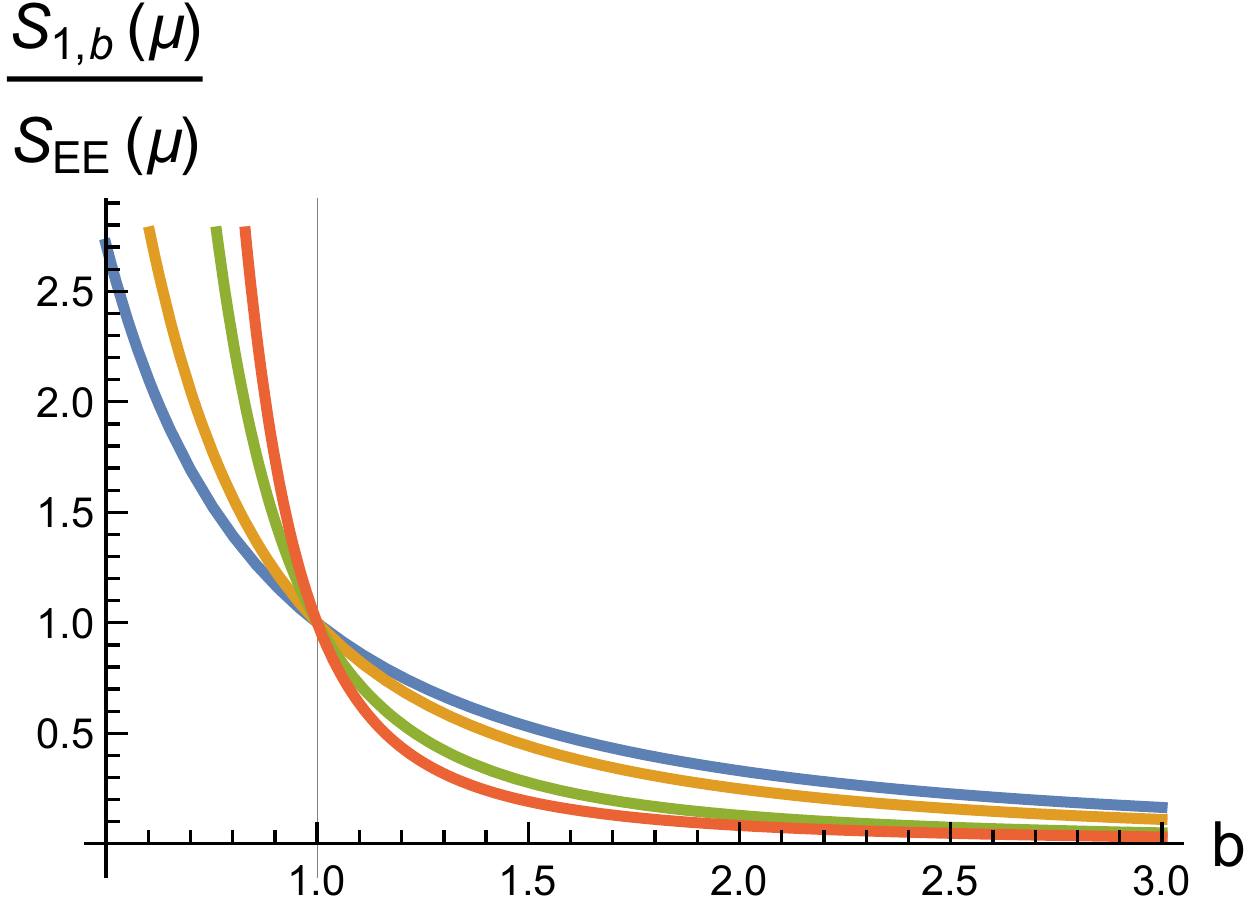}
	\end{minipage}}
\caption{A plot of $S_{1,b}(\mu)/S_{EE}$ as a function of $b$ in a variety of dimensions: $d=3$ (blue), $d=4$ (orange), $d=7$ (green), and $d=10$ (red).  Here we have normalized by $\lim_{b\to1}S_{1,b}(\mu)$ and set $\ell_{\ast}=2\pi$. (a) Chemical potential $\mu=1$, (b) $\mu=100$. }
\label{difds}\end{figure}

In Figure \ref{difds} we plot the dimension $d$ dependence of $S_{1,b}(\mu)$. For $b>1$, we see larger $d$ leads to smaller $S_{1,b}$; for $b<1$ larger $d$ corresponds to higher $S_{1,b}(\mu)$. We also observe that, starting at $b=1$ (where we have normalized by $S_{EE}(\mu)$) all curves meet, separate and then vanish as $b$ goes large. We find the expected divergences in $S_{1,b}(\mu)$ as $b$ approaches zero, where different dimensions clearly separate out $S_{1,b}(\mu)$. Notice, moreover, the spread increases between the curves above and below $b=1$ as $\mu$ increases.

The traditional R\'enyi entropy is known to satisfy a number of inequalities \cite{Hung:2011nu}, namely, 
\beq 
\begin{split}
&\frac{\partial S_{q}}{\partial q}\leq0\;,\quad \frac{\partial}{\partial q}\left(\frac{q-1}{q}S_{q}\right)\geq0\;,\\
&\frac{\partial}{\partial q}[(q-1)S_{q}]\geq0\;,\quad \frac{\partial^{2}}{\partial q^{2}}[(q-1)S_{q}]\leq0\;.
\end{split}
\eeq
These inequalities also hold for both the neutral and charged holographic R\'enyi entropies \cite{Hung:2011nu,Belin:2013uta}. The reason these inequalities hold in either case is because the CFT, via the CHM map, lives on a stable thermal ensemble; the presence of a global conserved charge in the field theory does not alter the stability of the ensemble. 

\begin{figure}[t]
  \subfloat[]{
	\begin{minipage}[1\width]{
	   0.32\textwidth}
	   \centering
	   \includegraphics[width=1.45\textwidth]{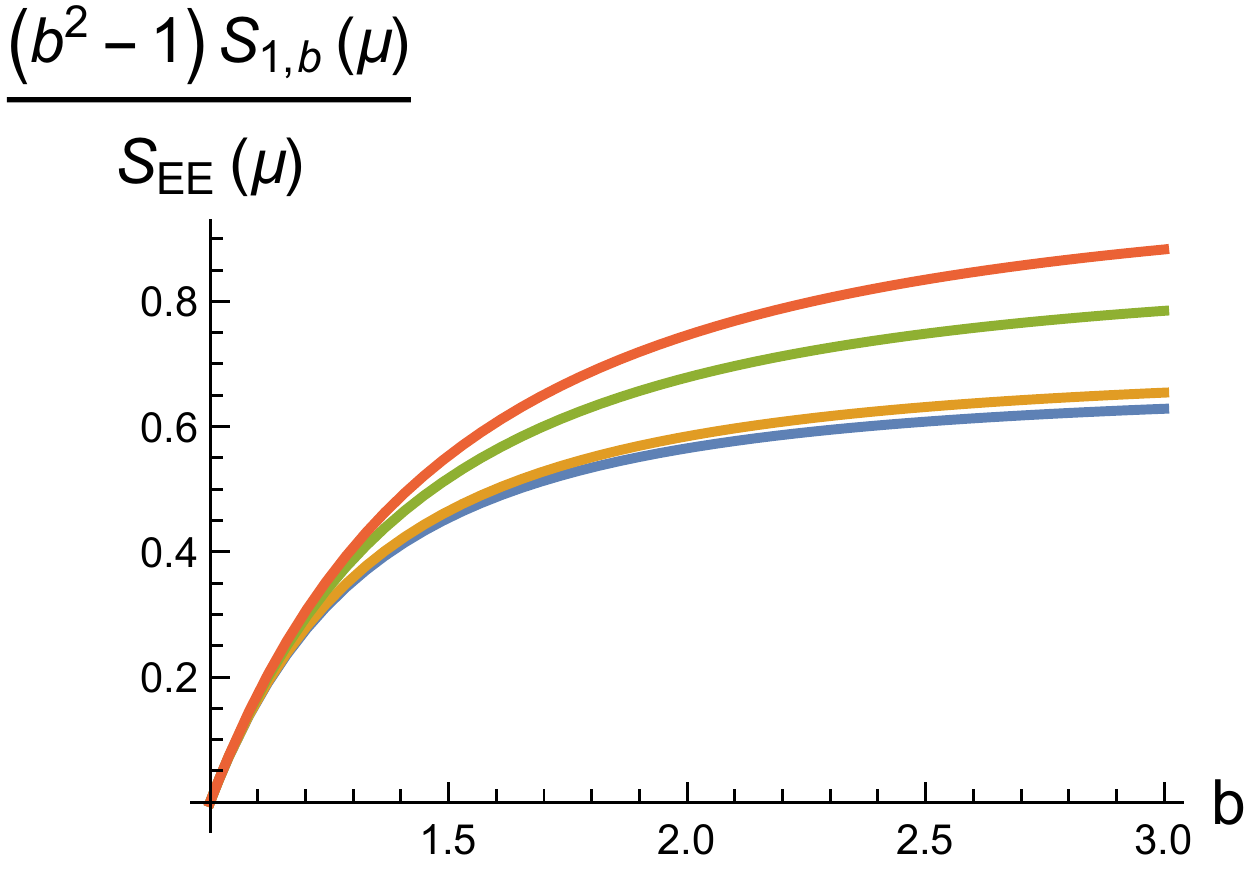}
	\end{minipage}}
 \hspace{2.5cm}
  \subfloat[]{
	\begin{minipage}[1\width]{
	   0.32\textwidth}
	   \centering
	   \includegraphics[width=1.45\textwidth]{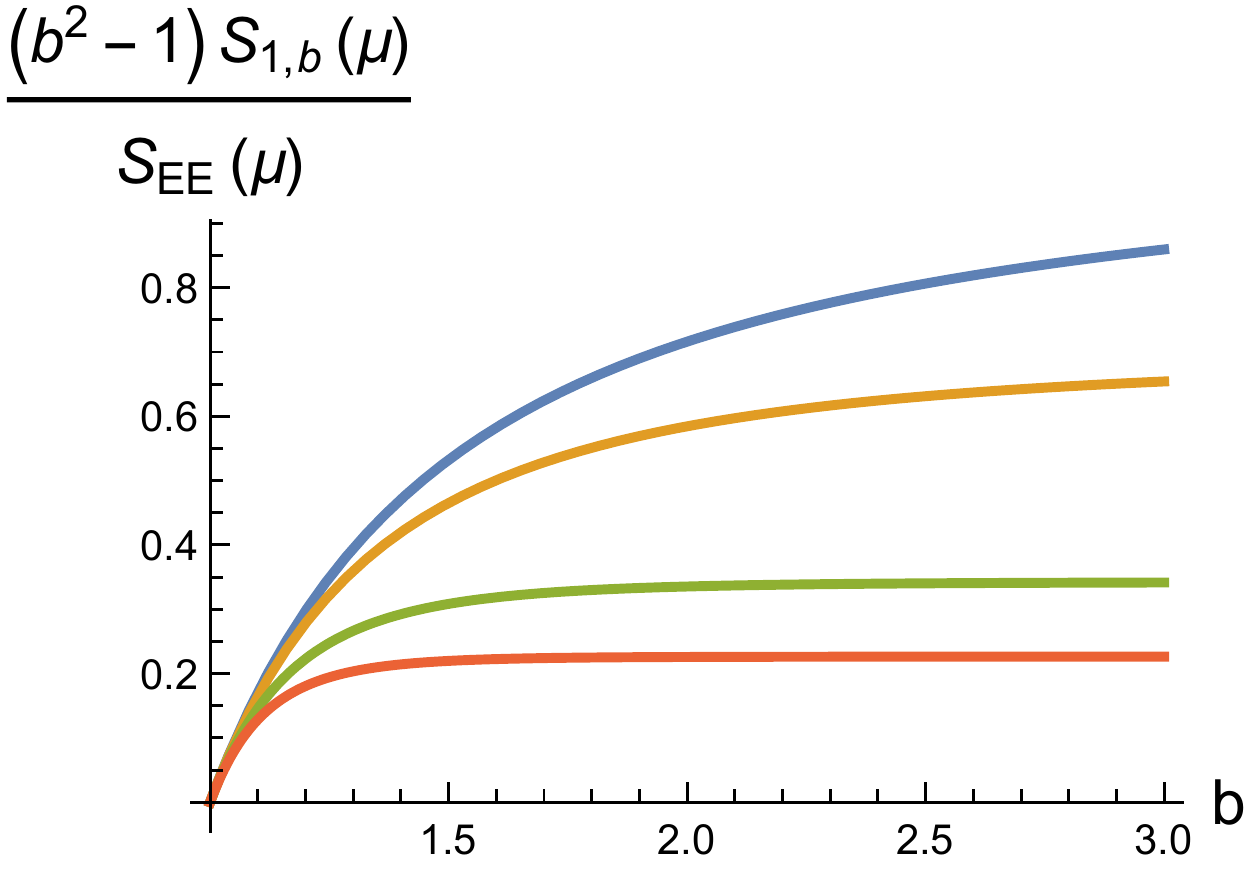}
	\end{minipage}}
\caption{A plot of $(b^{2}-1)S_{1,b}(\mu)/S_{EE}$ as a function of $b$, normalized by $S^{b\to1}_{EE}(\mu)$. (a) A variety of fixed potential $\mu$: $\mu=0$ (blue), $\mu=1$ (orange), $\mu=5$ (green), $\mu=100$ (red). (b) A variety of dimensions $d$: $d=3$ (blue), $d=4$ (orange), $d=7$ (green), and $d=10$ (red). The origin has been set at $b=1$.}
\label{bsqm1S1b}\end{figure}

It is natural to ask whether the entropy $S_{1,b}(\mu)$ will satisfy a similar set of inequalities. If so, we are more inclined to refer to the object as a R\'enyi entropy. It is evident from Figure \ref{S1bfuncb} and Figure \ref{difds} that the slope of $S_{1,b}(\mu)$ is negative, even when we are in the neutral limit $\mu=0$. In Figure \ref{bsqm1S1b} we numerically investigate the function $(b^{2}-1)S_{1,b}(\mu)$, which allows us to conclude our $S_{1,b}(\mu)$ satisfies a similar set of inequalities that $S_{q}(\mu)$ does:
\beq 
\begin{split}
&\frac{\partial S_{1,b}}{\partial b}\leq0\;,\quad \frac{\partial}{\partial b}[(b^{2}-1)S_{1,b}]\geq0\;,\quad \frac{\partial^{2}}{\partial b^{2}}[(b^{2}-1)S_{1,b}]\leq0\;.
\end{split}
\label{ineqs}\eeq

Recall from the $(p,V)$ plane that a change in $b$ corresponds to a change pressure. Even though vertical shifts in pressure below the $M=0$ curve leads to negative mass systems, there is no thermal phase transition of the hyperbolic charged black hole. Indeed, since the thermodynamic volume $V$ scales as the entropy (\ref{volume1}), the heat capacity at fixed volume $C_{V}=0$. It is unclear what the exact field theory interpretation of the inequalities presented in (\ref{ineqs}), though it is expected that it also follows from the fact the CFT living on the hyperbolic cylinder is in a stable thermal ensemble.

\subsection{Choices of $b(q)$}

\begin{figure}[t]
  \subfloat[]{
	\begin{minipage}[1\width]{
	   0.32\textwidth}
	   \centering
	   \includegraphics[width=1.45\textwidth]{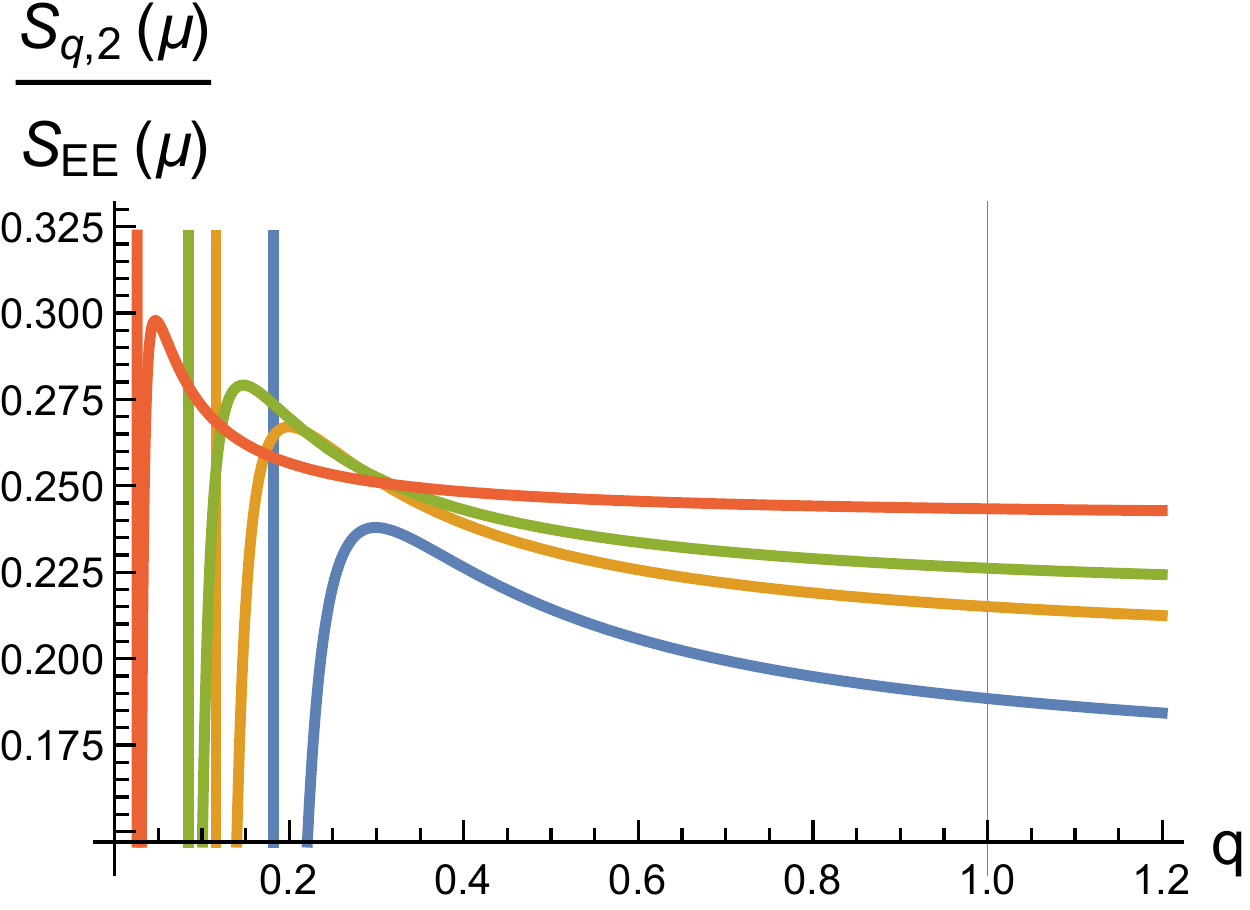}
	\end{minipage}}
 \hspace{2.5cm}
  \subfloat[]{
	\begin{minipage}[1\width]{
	   0.32\textwidth}
	   \centering
	   \includegraphics[width=1.45\textwidth]{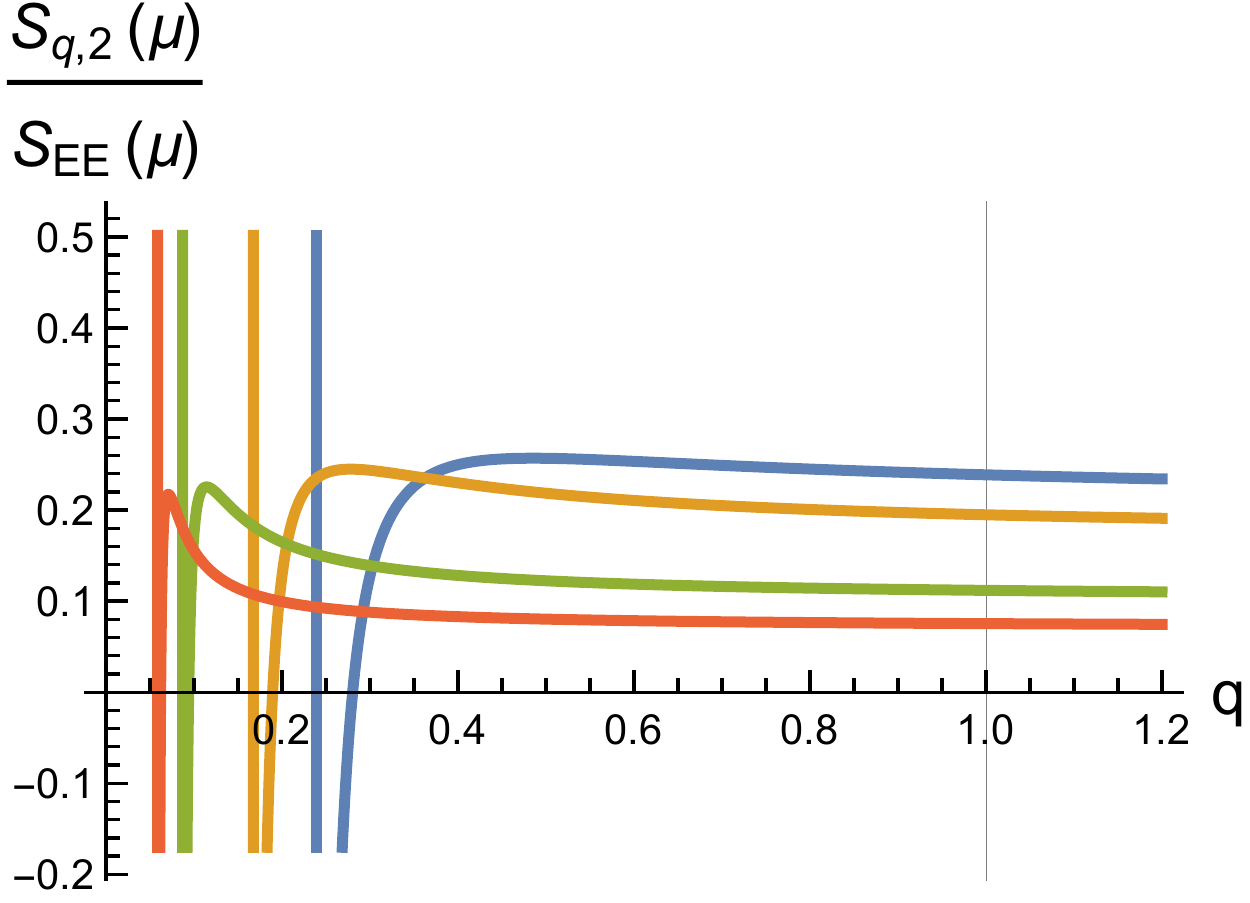}
	\end{minipage}}
\caption{A plot of $S_{q,2}(\mu)$ as a function of $q$ for different $\mu$ and $d$. (a) Fixed potential $\mu$: $\mu=0$ (blue), $\mu=1$ (orange), $\mu=1.5$ (green), $\mu=2$ (red); $d=4$. (b) Dimensions $d$: $d=3$ (blue), $d=4$ (orange), $d=7$ (green), and $d=10$ (red); $\mu=1$. Notice relatively small values of $\mu$ lead to smaller values of $q_{c}$.}
\label{Sq2mud}\end{figure}

Thus far we have really only explored the behavior of $S_{1,b}(\mu)$ as a function of the chemical potential $\mu$. We can of course leave the usual R\'enyi index $q$ turned on and consider different behaviors of the new index $b$, \emph{i.e.}, $b$-deformations of $S_{q}(\mu)$. For example, in Figure \ref{Sq2mud} we simply fix the parameter $b=2$ and analyze how $S_{q,2}(\mu)$ changes as a function of $\mu$, $q$ and $d$. As observed, selecting $b\neq1$ leads to divergences at the critical values of $q_{c}$ (\ref{qcrit}), the location of which depends on both $\mu$ and $d$. We see low values of $\mu$, compared to values of $d$,  force the the divergence in $S_{q,2}(\mu)$ to occur for smaller values of $q$. This is because at fixed $d$ as $\mu$ increases, even for relatively low $\mu$, the $\mu^{2}$ term in $x_{11}$ becomes dominant quickly since it also couples to the dimension at order $O(d^{2})$. 

\begin{figure}[t]
  \subfloat[]{
	\begin{minipage}[1\width]{
	   0.32\textwidth}
	   \centering
	   \includegraphics[width=1.45\textwidth]{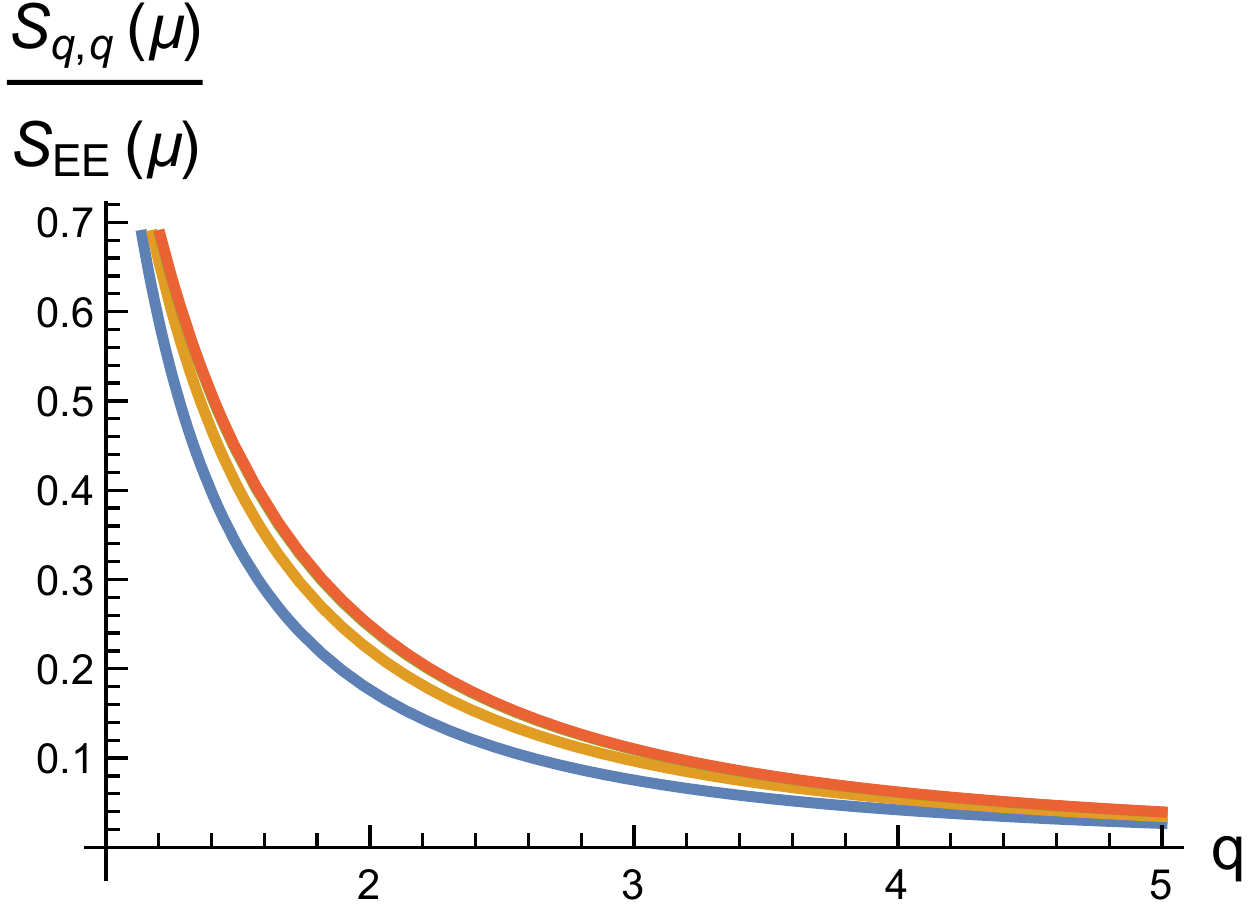}
	\end{minipage}}
 \hspace{2.5cm}
  \subfloat[]{
	\begin{minipage}[1\width]{
	   0.32\textwidth}
	   \centering
	   \includegraphics[width=1.45\textwidth]{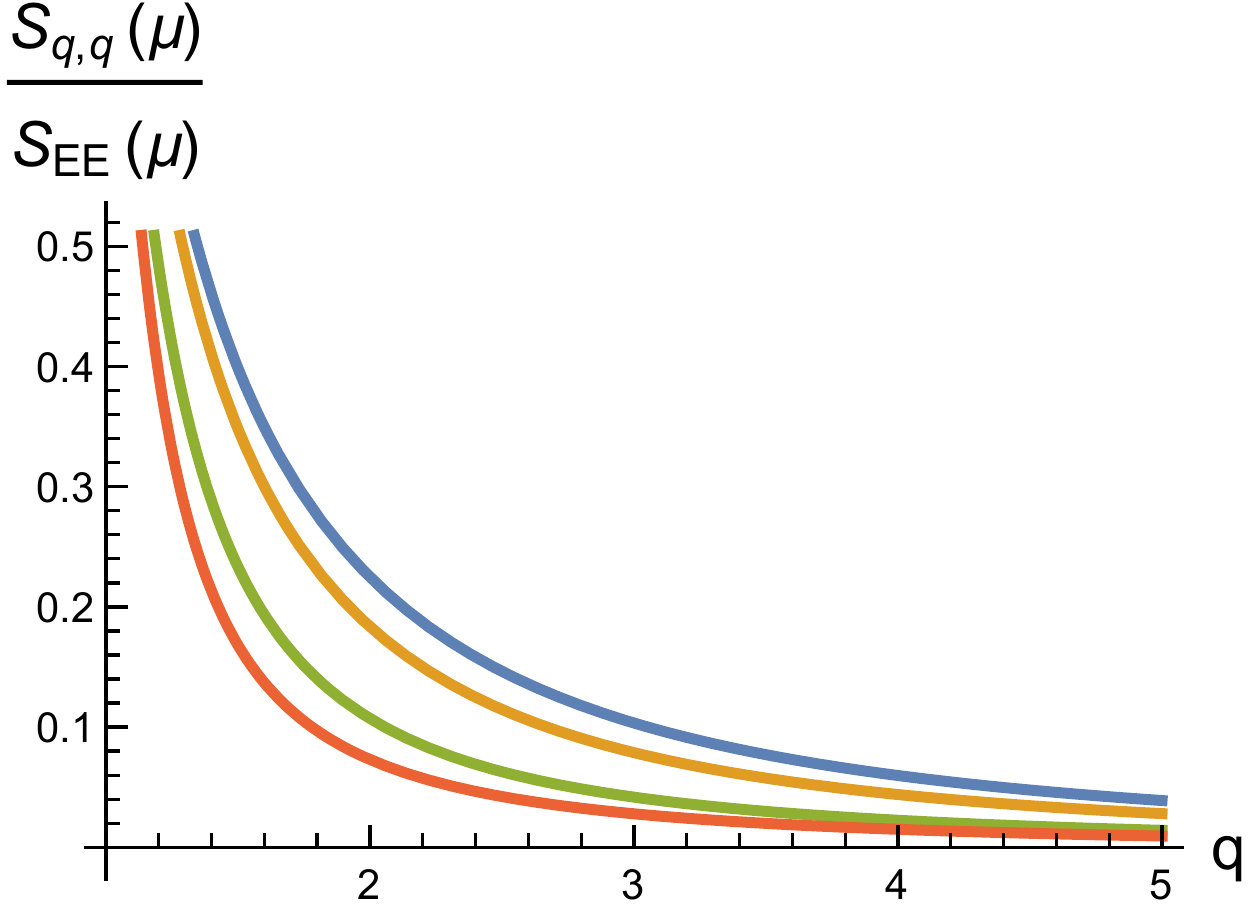}
	\end{minipage}}
\caption{A plot of $S_{q,q}(\mu)$ as a function of $q$ for different $\mu$ and $d$. (a) Fixed potential $\mu$: $\mu=0$ (blue), $\mu=5$ (orange), $\mu=50$ (green), $\mu=100$ (red); $d=4$. (b) Dimensions $d$: $d=3$ (blue), $d=4$ (orange), $d=7$ (green), and $d=10$ (red); $\mu=1$.}
\label{Sqqinvfuncmu}\end{figure}

We can also see how $S_{q,b}(\mu)$ changes when $b=b(q)$. In Figure \ref{Sqqinvfuncmu}, for example, we consider when $b=q$, for which we observe the expected concave up behavior of R\'enyi entropies. We also point out the slightly different influences between dimension $d$ and $\mu$, where higher $d$ lowers the (local) minimum of $S_{q,q}$ but higher $\mu$ raises the minimum.

 There are two specific special cases of $b(q)$ which we study in more detail below.



\subsection*{Special Cases}
\indent

As first explored in \cite{Johnson:2018bma} and reviewed in Section \ref{sec:review}, for $q,b\neq1$, there are particular choices for $b,q$ which lead to interesting insights for $S_{q,b}(\mu)$. These specific choices  pertain to types of $b$-deformations which correspond to interesting changes in the $(p,V)$ plane: (i) a completely \emph{vertical displacement} away from the $T_{0}/q$ isotherm to the $T_{0}$ line (the massless hyperbolic curve), \emph{i.e.}, no volume change. This case is interesting because it provides a special pair of values of $(q,b)$ other than $1$ such that $S_{q,b}=S_{EE}$; (ii) a change in pressure \emph{along} the $T_{0}/q$ isotherm such that we land back on the $T_{0}$ curve\footnote{Recall that for us the $T_{0}$ curve does not correspond to the $M=0$ black hole; only when $\mu=0$ do these two curves coincide.}, corresponding to the particular value $b=q^{-1}$. 

In the charged scenario we can consider these special cases as well. (i) Now a completely vertical displacement straight from the $T_{0}/q$ isotherm, where $x_{qb}=\frac{r_{h}}{L}$ and $L=L_{0}/b$, to the $T_{0}$ isotherm, where $r_{h}=L_{0}x_{11}$, corresponds to when $x_{qb}=bx_{11}$. This leads to the following relation between $b$ and $q$:
\beq (b^{2}-1)=\frac{2}{qdx_{11}}\left[1-\frac{q}{x_{11}}+\frac{dq}{2x_{11}}(1-x_{11})\right]+\frac{(d-2)^{2}}{2d(d-1)x_{11}}\left(\frac{\mu\ell_{\ast}}{2\pi}\right)^{2}\;.\eeq
Substituting this choice for $b=b(q,\mu)$ into $S_{q,b}(\mu)$ (\ref{extchargedrenyi}), we find after some algebra
 that 
\beq S_{q,b}(\mu)|_{x_{qb}=bx_{11}}=S_{EE}(\mu)\;,\eeq
where the charged von Neumann entropy $S_{EE}(\mu)$ is displayed in (\ref{SEEcharged}). Thus, as in the neutral limit, there exist special values of $b$ and $q$ not equal to 1 such that $S_{q,b}(\mu)=S_{EE}(\mu)$. From the extended black hole thermodynamics perspective this is not surprising since the constant volume paths are equivalent to constant entropy paths. 

\begin{figure}[t]
  \subfloat[]{
	\begin{minipage}[1\width]{
	   0.32\textwidth}
	   \centering
	   \includegraphics[width=1.45\textwidth]{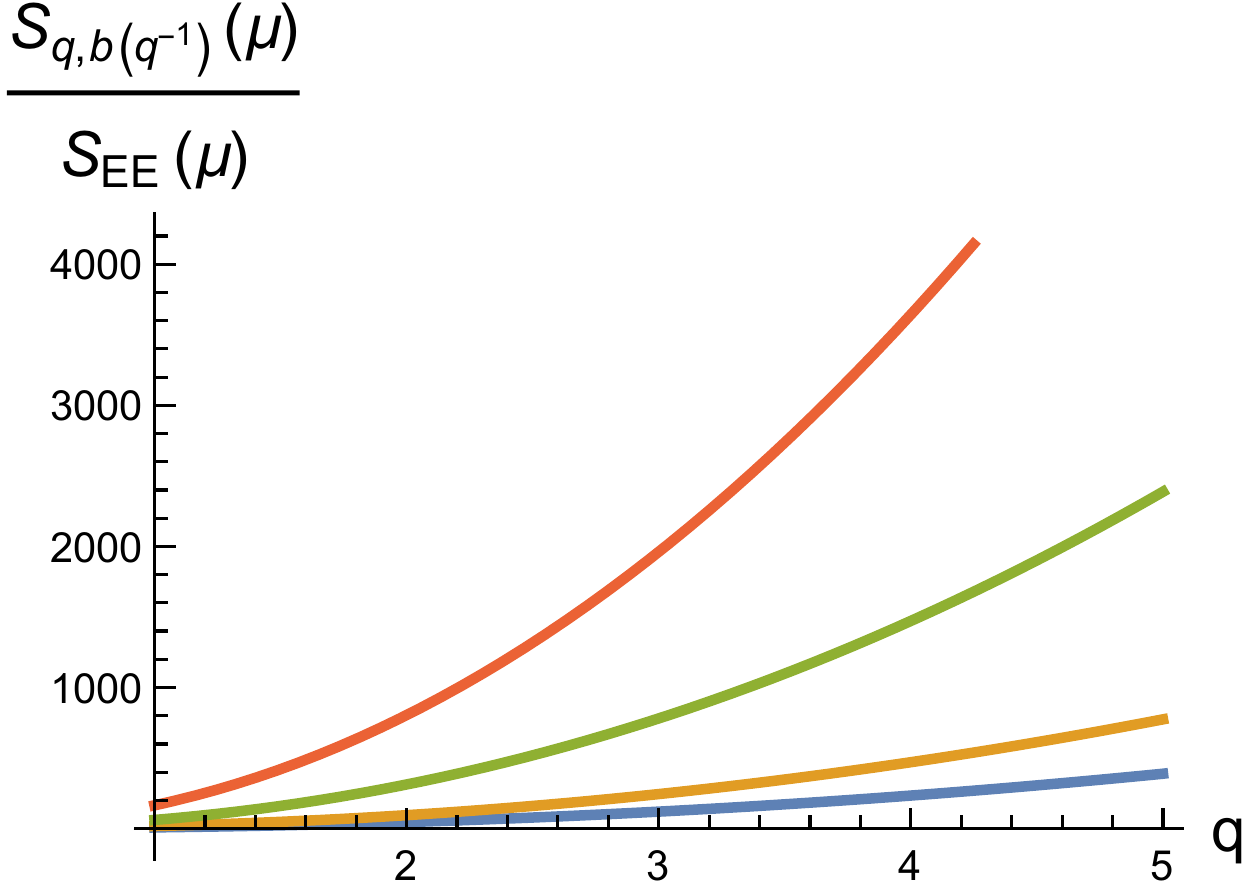}
	\end{minipage}}
 \hspace{2.5cm}
  \subfloat[]{
	\begin{minipage}[1\width]{
	   0.32\textwidth}
	   \centering
	   \includegraphics[width=1.45\textwidth]{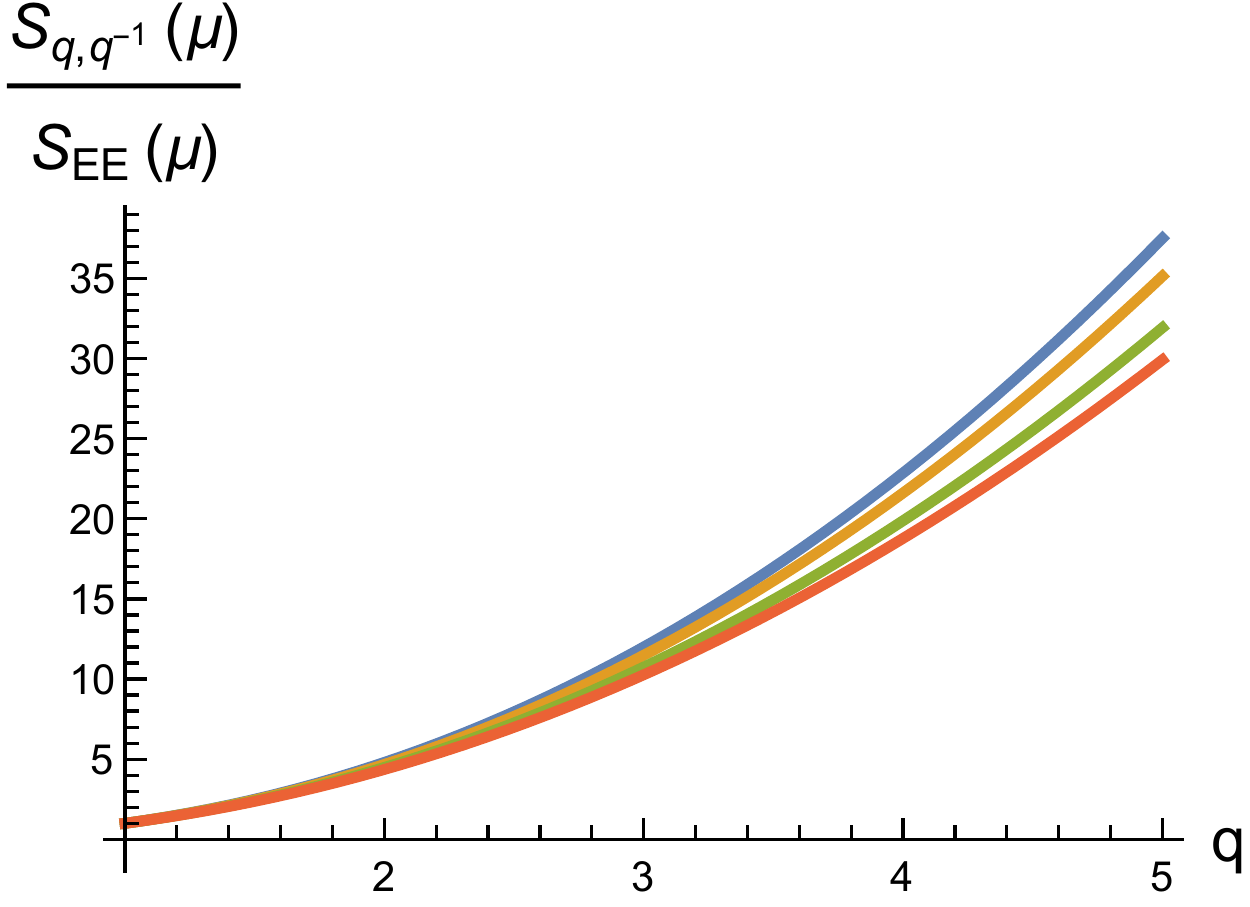}
	\end{minipage}}
\caption{A plot of $S_{q,b(q^{-1})}(\mu)$ as a function of $q$ for different $\mu$. (a) $b\propto q^{-1}$ (\ref{specb}). (b) $b=q^{-1}$. In both (a) and (b) $d=4$ and $\mu=0$ (blue), $\mu=1$ (orange), $\mu=2$ (green), $\mu=3$ (red).}
\label{Sqqinvdifmu2}\end{figure}

\begin{figure}[t]
  \subfloat[]{
	\begin{minipage}[1\width]{
	   0.32\textwidth}
	   \centering
	   \includegraphics[width=1.45\textwidth]{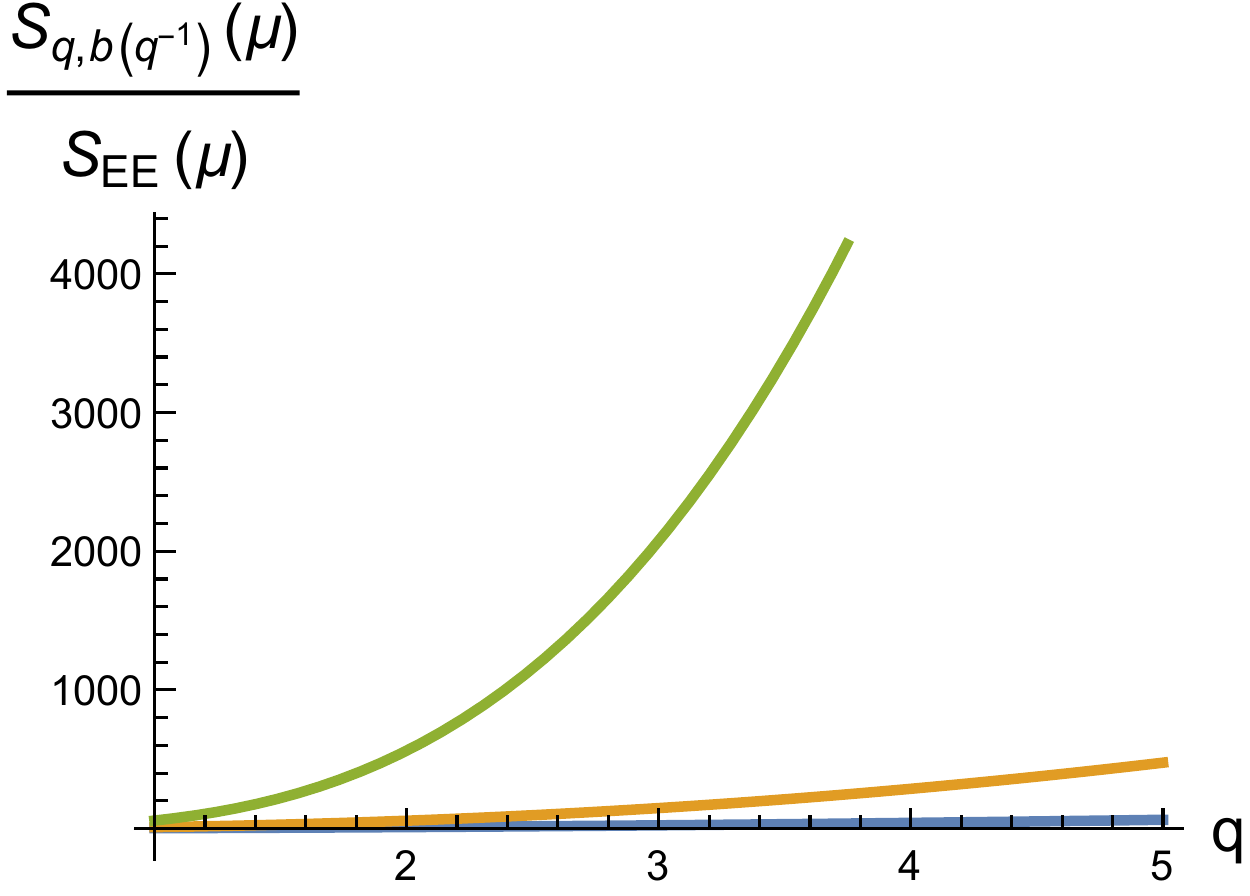}
	\end{minipage}}
 \hspace{2.5cm}
  \subfloat[]{
	\begin{minipage}[1\width]{
	   0.32\textwidth}
	   \centering
	   \includegraphics[width=1.45\textwidth]{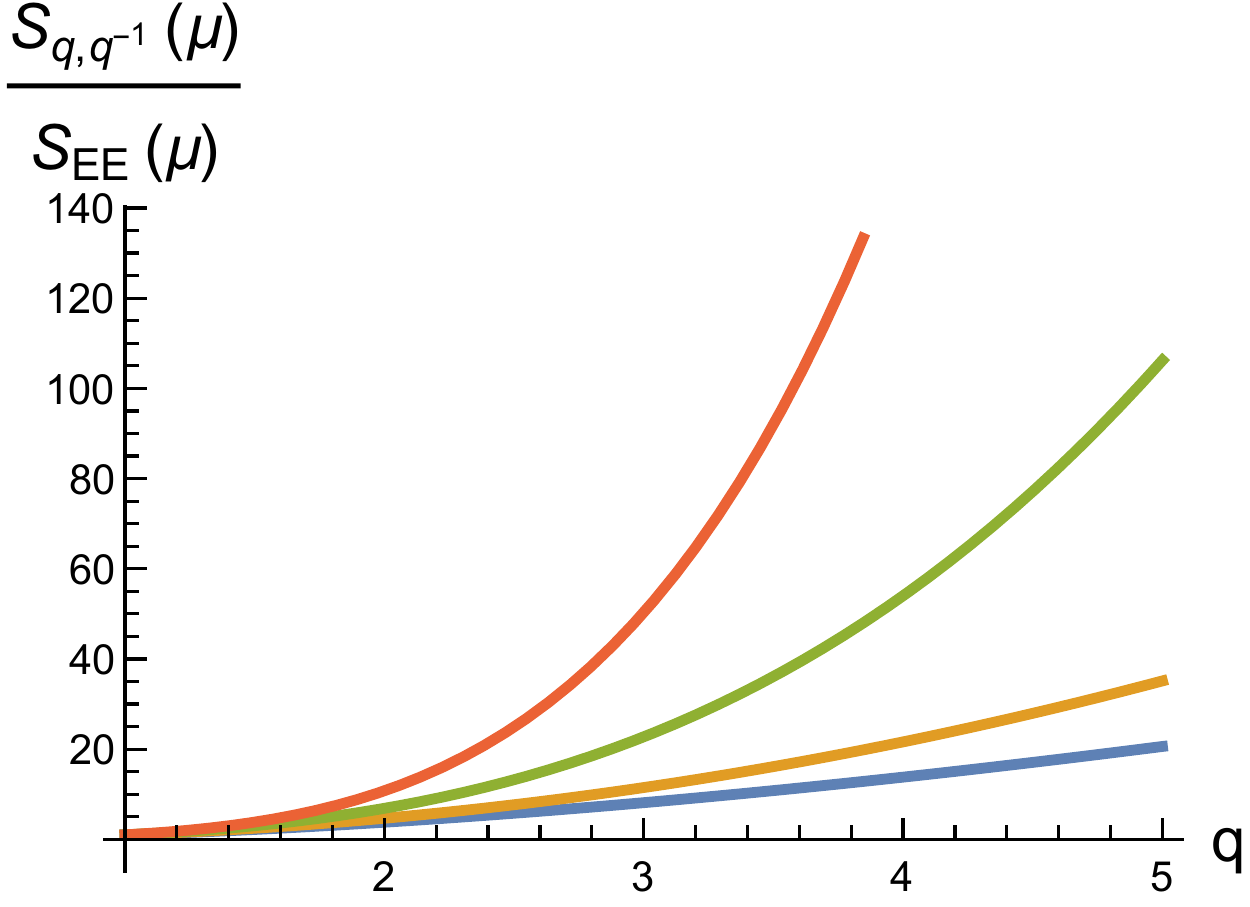}
	\end{minipage}}
\caption{A plot of $S_{q,b(q^{-1})}(\mu)$ as a function of $q$ for different $d$. (a) $b\propto q^{-1}$ (\ref{specb}) with $\mu=.5$. (b) $b=q^{-1}$ with $\mu=1$. In both cases $d=3$ (blue), $d=4$ (orange), and $d=5$ (green); in (b) we also have $d=6$ (red).}
\label{Sqqinvfuncd}\end{figure}

(ii) A change in pressure along the $T_{0}/q$ isotherm such that we return to the $T_{0}$ isotherm we have $x_{qb}=x_{11}$. This results in the following relation between $b,q$ and $\mu$:
\beq b=q^{-1}\frac{2x_{11}}{dx_{11}^{2}-(d-2)+\frac{(d-2)^{2}}{2(d-1)}\left(\frac{\mu\ell_{\ast}}{2\pi}\right)^{2}}\;.\label{specb}\eeq
Observe that in the neutral limit we recover the special case $b=q^{-1}$. Substituting (\ref{specb}) into $S_{q,b}(\mu)$ leads to a cumbersome relation which we won't express here. Importantly, from Figure \ref{Sqqinvdifmu2} and Figure \ref{Sqqinvfuncd} we see that as $q$ grows the $S_{q,b(q)}(\mu)$ grows quickly as $\mu$ or the dimension $d$ increase. This is similar to what is observed in the neutral case \cite{Johnson:2018bma}. In the neutral case, moreover, it was argued that the index $b$ can be thought of as a parameter which undoes the replica trick of creating $q$-copies of $\rho_{R}$ on flat space, hence the $b=q^{-1}$. This was explicitly verified in the $d=2$ case \cite{Johnson:2018bma}. Since the charged R\'enyi entropy can likewise be computed using the replica trick, it is also natural to interpret $b=q^{-1}$. We found, motivated by extended thermodynamics that we have $b\propto q^{-1}$, where the proportionality constant explicitly depends on the potential $\mu$. For purposes of comparison, in Figure \ref{Sqqinvdifmu2} and Figure \ref{Sqqinvfuncd} we have included plots of $S_{q,q^{-1}}(\mu)$, in which we observe similar qualitative features, however, the influence of $\mu$ and $d$ are enhanced when we choose (\ref{specb}). We of course point out a crucial difference on the dependence in $\mu$ in Figure \ref{Sqqinvdifmu2}: When $b$ is given by (\ref{specb}), larger values of $\mu$ correspond to large values of $S_{q,b}$, the \emph{opposite} of what is seen when $b=q^{-1}$.

\subsection*{Imaginary Chemical Potential}

In field theory, both real and imaginary $\mu$ are of interest. As in the non-extended case \cite{Belin:2013uta} we can simply analytically continue our above holographic calculations by setting 
\beq \mu\to i\mu_{E}\Rightarrow q_{e}\to iq^{E}_{e}\;,\eeq
where $\mu_{E}$ and $q^{E}_{e}$ are purely real. Therefore, an imaginary chemical potential is dual to an imaginary charge. The consequences of this continuation is that the root $x_{qb}$ (\ref{xqbchargereal}) will fail to exist when $\mu_{E}$ becomes too large. Specifically, 
\beq \mu_{E}\leq\frac{2(d-1)}{(d-2)}\left(\frac{2\pi}{\ell_{\ast}}\right)^{2}\left(1+\frac{1}{q^{2}b^{2}d(d-2)}\right)\;.\eeq
At fixed $q$, for $\mu_{E}$ larger than this upper bound, the event horizon will disappear leaving a naked singularity. 

\begin{figure}[t]
  \subfloat[]{
	\begin{minipage}[1\width]{
	   0.32\textwidth}
	   \centering
	   \includegraphics[width=1.45\textwidth]{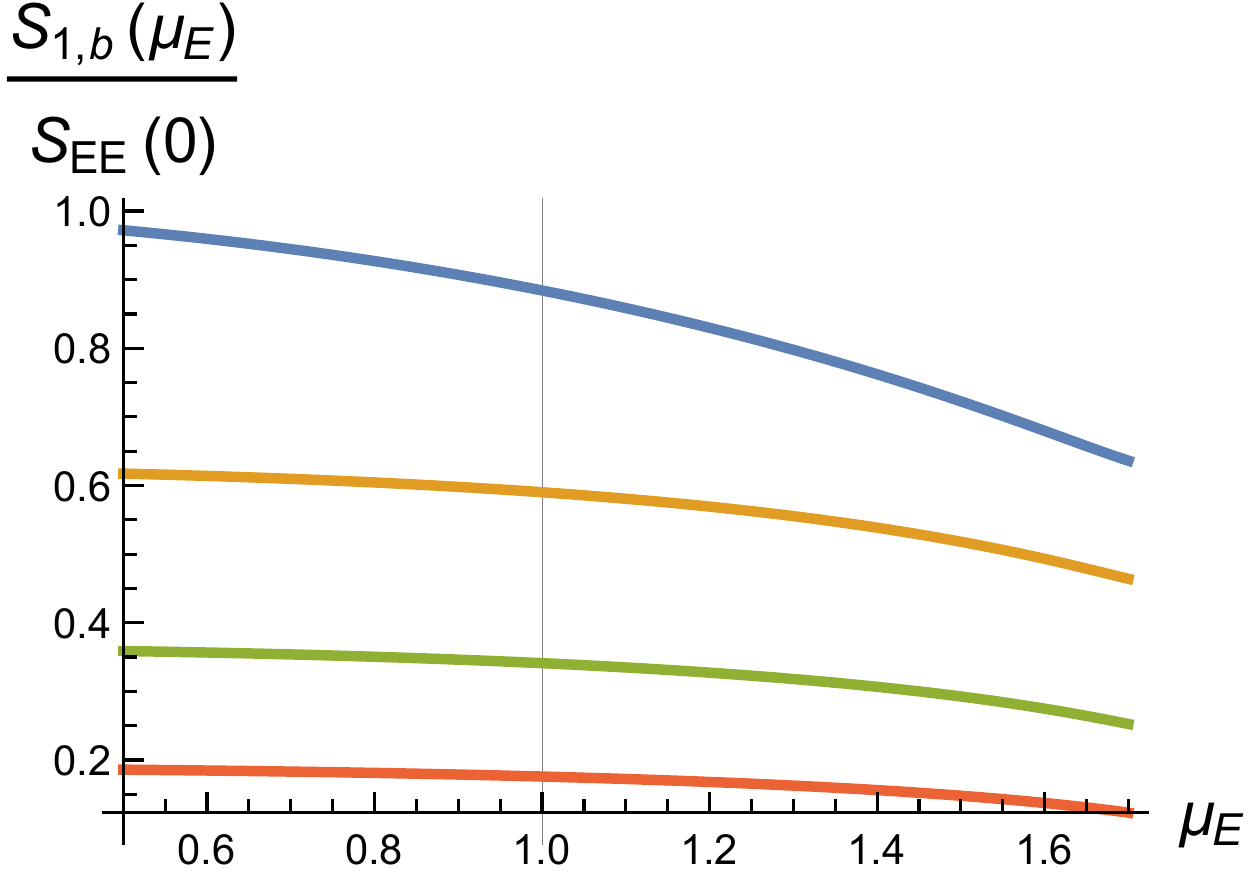}
	\end{minipage}}
 \hspace{2.5cm}
  \subfloat[]{
	\begin{minipage}[1\width]{
	   0.32\textwidth}
	   \centering
	   \includegraphics[width=1.45\textwidth]{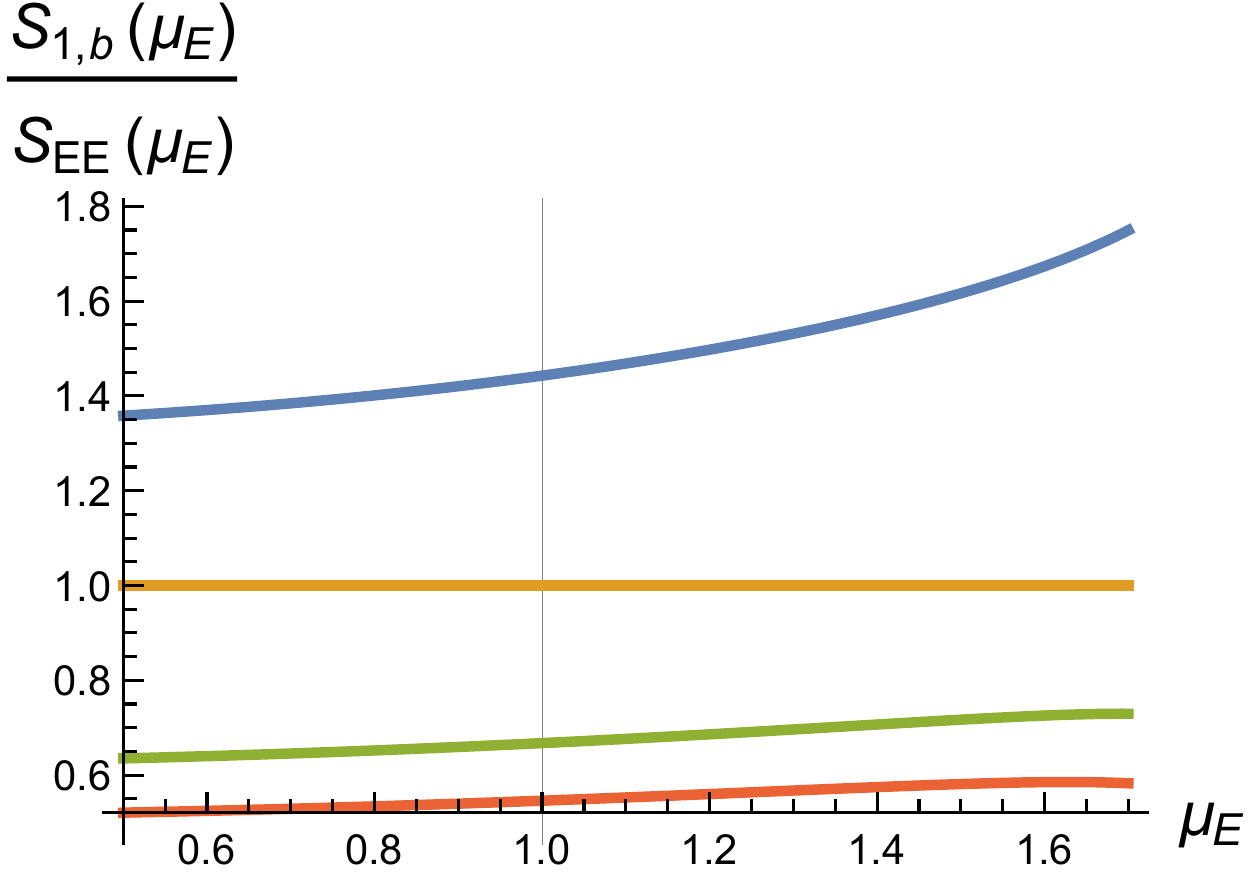}
	\end{minipage}}
\caption{A plot of $S_{1,b}(\mu_{E})$ as a function of $\mu_{E}$ for different $b$. (a) Normalized by $S_{EE}(0)$  with $b=1$ (blue), $b=1.2$ (orange), $b=1.5$ (green) and $b=2$ (red). (b) Normalized by $S_{EE}(\mu_{E})$ with $b=.9$ (blue), $b=1$ (orange), $b=1.2$ (green) and $b=1.3$ (red).  Here we have set $d=4$.}
\label{S1bIMfuncmu}\end{figure}

\begin{figure}[t]
  \subfloat[]{
	\begin{minipage}[1\width]{
	   0.32\textwidth}
	   \centering
	   \includegraphics[width=1.45\textwidth]{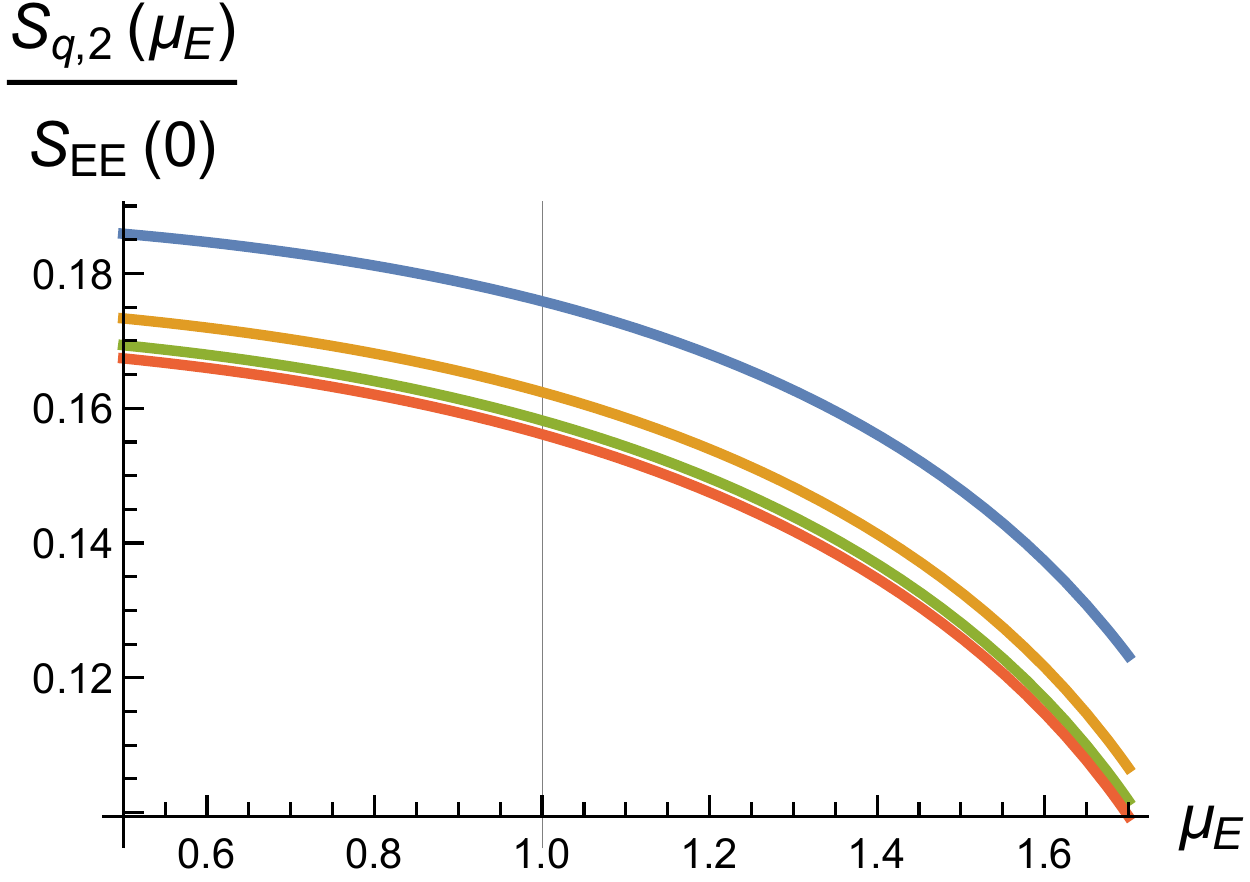}
	\end{minipage}}
 \hspace{2.5cm}
  \subfloat[]{
	\begin{minipage}[1\width]{
	   0.32\textwidth}
	   \centering
	   \includegraphics[width=1.45\textwidth]{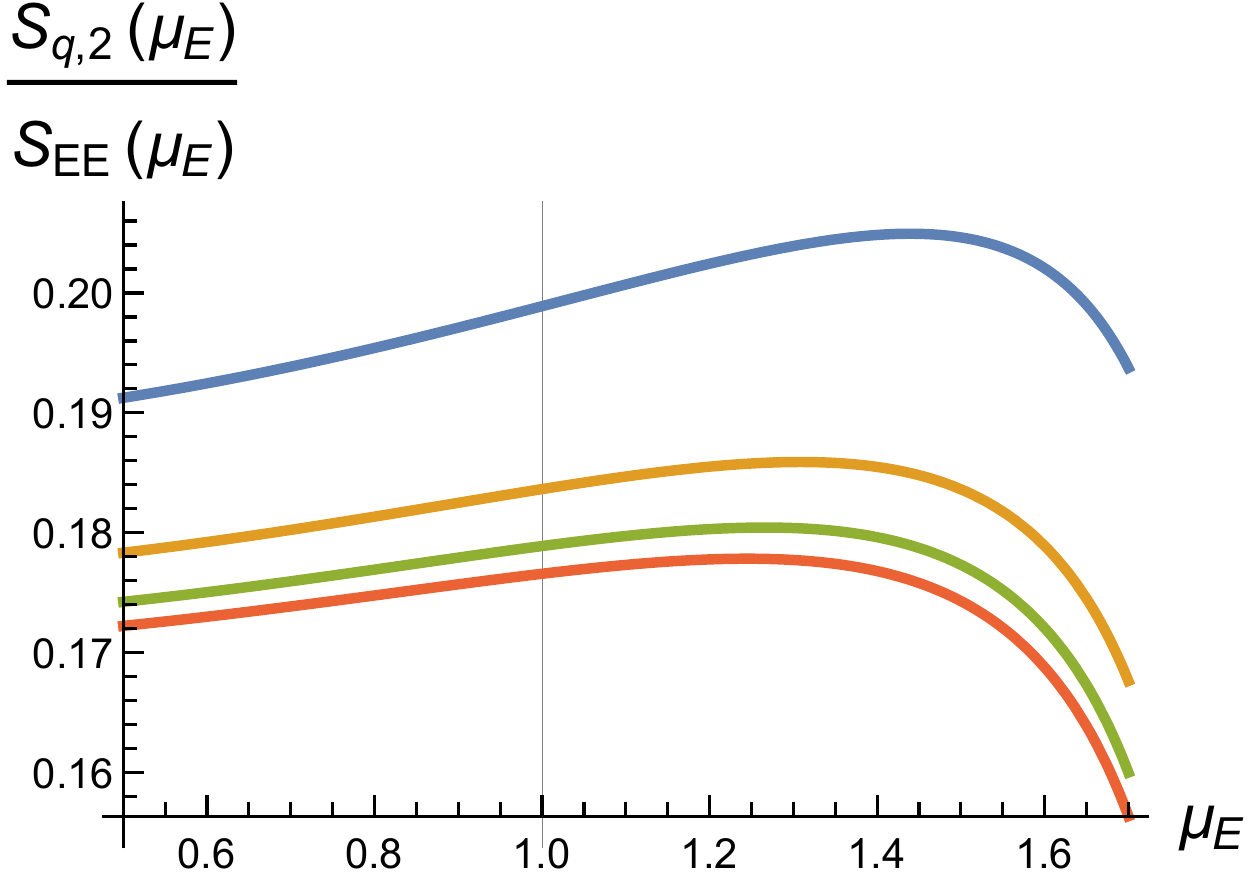}
	\end{minipage}}
\caption{A plot of $S_{q,2}(\mu_{E})$ as a function of $\mu_{E}$ for different $q$. (a) Normalized by $S_{EE}(0)$. (b) Normalized by $S_{EE}(\mu_{E})$. Here we have set $d=4$ and selected $q=1$ (blue), $q=2$ (orange), $q=3$ (green) and $q=4$ (red).}
\label{Sq2IMfuncmu}\end{figure}

In Figures \ref{S1bIMfuncmu} and \ref{Sq2IMfuncmu} we explore the behavior of $S_{q,b}(\mu_{E})$. Comparing Figure \ref{S1bIMfuncmu} to Figure \ref{S1bfuncmu}, we see that $S_{1,b}(\mu_{E})$ will increase, rather than decrease, for increasing $\mu_{E}$; similar behavior is seen in the $S_{q,1}(\mu_{E})$ case when normalized by $S_{EE}(0)$ \cite{Belin:2013uta}, however, we find no decrease in $S_{1,b}(\mu_{E})/S_{EE}(\mu_{E})$ as $\mu_{E}$ increases. Note, moreover, from Figure \ref{S1bIMfuncmu} (b) we see for $b<1$ the entropy increases as $\mu_{E}$ increases. In Figure \ref{Sq2IMfuncmu}, where we study $S_{q,2}(\mu_{E})$, we find an interesting difference between the choice of normalization. When we normalized by $S_{EE}(0)$ the entropy $S_{q,2}(\mu_{E})$ will monotonically decrease as $\mu_{E}$ increases, while normalized by $S_{EE}(\mu_{E})$ the entropy will increase before approaching a local maximum.

\section{Field Theory Interpretation}
\label{sec:fieldtheoint}

Our generalized charged R\'enyi entropy (\ref{extchargedrenyi}) was physically well motivated by the extended thermodynamics of charged AdS black holes. It is worthwhile to think about the field theory interpretation of $S_{q,b}(\mu)$, and how it would be constructed on the field theory side. This can be accomplished, at least in principle, by working backwards from our expression (\ref{extchargedrenyi}) whereby we identify the proper generalization of the CHM map (\ref{CHMmap}).

More precisely, to connect the flat space CFT state $\rho_{A}^{(b)}$ to the thermal ensemble on $\mathbb{R}\times\mathbb{H}^{d-1}$, given by $\rho_{therm}$, we would introduce the density matrix $\rho_{A}^{(b)}$ of the following form\footnote{In the case of the imaginary potential, we would write down $\rho_{A}^{(b)}=U^{\dagger}\left(\frac{e^{-H/T_{0}+b^{2}p_{0}V_{0}/T_{0}+i\mu_{E} Q_{A}}}{Z(T_{0},p_{0},\mu)}\right)U$.}:
\beq \rho_{A}^{(b)}(\mu)=U^{\dagger}\left(\frac{e^{-H/T_{0}+b^{2}p_{0}V_{0}/T_{0}+\mu Q_{A}}}{Z(T_{0},p_{0},\mu)}\right)U\;.\label{CHMmapmod}\eeq
The matrix in between the unitaries $U^{\dagger}$ and $U$ is the extended version of the thermal ensemble density matrix $\rho_{therm}$ used in the CHM map, and the thermal partition function formally given by $\text{tr}[\exp(-H/T_{0}+b^{2}p_{0} V_{0}/T_{0}+\mu Q_{A})]$. 

The charged generalization of the entropy $S_{q,b}(\mu)$ is then
\beq S_{q,b}(\mu)=\frac{1}{[(1-q)-q(d-1)x_{11}(b^{2}-1)/2]}\log\left[\text{tr}(\rho^{(b)}_{A})^{q}\right]\;,\eeq
which came from us writing $S_{q,b}(\mu)$ as a difference of logarithms of two partition functions \cite{Johnson:2018bma}
\beq S_{q,b}=\left[1+qx_{11}\left(\frac{d-1}{2}\right)\left(\frac{b^{2}-1}{q-1}\right)\right]^{-1}\frac{1}{1-q}\log\left[\frac{Z(T_{0}/q,\mu,b^{2}p_{0})}{Z(T_{0},\mu,p_{0})^{q}}\right]\;.\label{fieldtheorysqbdiff}\eeq

The interpretation of $\rho_{A}^{(b)}$ is simply the charged generalization of $\rho_{A}^{(b)}(0)$, first given in \cite{Johnson:2018bma}: It is the $b$th power of some other density matrix, denoted by $\rho_{B}$, arising from fractionating\footnote{This is analogous to what is done for the ordinary R\'enyi entropy, where to make $q$-copies of a matrix $\rho_{A}$ represented in the thermal ensemble at temperature $T_{0}$, we must fractionate the system into $q$ systems each at a temperature $T_{0}/q$, whereupon the $q$ systems are glued together in a suitable way.} the system $A$ into $b$ copies each of length $L_{0}/b$. We then presume to glue each of the $b$ copies together, creating a system of length $L_{0}$, such that 
\beq \rho_{A}^{(b)}=\rho_{B}^{b}\;\;.\label{rhob}\eeq
Notice the chemical potential $\mu$ simply goes along for the ride, supplementing the CHM map by inserting a Wilson line along the Euclidean time circle into the thermal path integral for $Z(T_{0}/q,\mu,b^{2}p_{0})$. 

Thus, the integer $b$ has a natural interpretation as another type of R\'enyi index, similar to the role of $q$, at least from the perspective of the replica trick. Moreover, while it is unclear what precisely $\rho_{B}$ is measuring, the state (\ref{rhob}) informs us of a potentially special value for $b$, namely, $b=1/q$, such that the $q$th replica of the sector at temperature $T_{0}/q$ is cancelled by the $1/q$th replica of the sector at pressure $q^{2}p_{0}$. When $\mu=0$, it was shown in  \cite{Johnson:2018bma} this interpretation holds precisely for the $d=2$ case, where an explicit CFT calculation involving twist operators reveals that $b=1/q$ undoes the $q$th replica in the temperature sector. It behooves us then to study the $d=2$ case, where it is possible to carry out a precise CFT calculation using charged twist operators \cite{Belin:2013uta}. However, we do not yet have the corresponding $d=2$ holographic computations completed since this requires us to consider the geometry of a $2+1$-dimensional charged AdS black hole, for which the geometry changes dramatically \cite{Martinez:1999qi}. We will study these holographic calculations in the next section. 

Fortunately, we can make some progress in understanding the interpretation of $b$ for $d>2$ by generalizing the conformal dimensions of the higher dimensional twist operators.



\subsection{Twist Operators in Higher Dimensions}

Just like its uncharged counterpart, the calculation of charged  Reny\'i entropies can be achieved by inserting a twist operator\footnote{Recall the twist operator $\sigma_{q}$ is a primary field in the CFT which connects the $q$-cuts in the replica trick; in the case of $d=2$, the function $\text{tr}\rho_{A}^{q}$ is then equal to the 2-point correlation function of twist operators placed at the ends of the $q$-cut.} $\sigma_{q}$ at the entangling surface. The twist operators have an associated conformal dimension $h_{q}$ which appears in correlation functions of the twist operators. In the charged case, the conformal dimension of the twist operators is generalized by considering the leading singularity in the correlator $\langle T_{\mu\nu}^{CFT}\sigma_{q}\rangle$. The leading singularity takes the form \cite{Belin:2013uta}, for example, tangentially along the entangling surface,
\beq \langle T^{CFT}_{ab}\sigma_{q}\rangle=-\frac{h_{q}}{2\pi}\frac{1}{y^{d}}\delta_{ab}\;.\eeq
Here $y$ is the perpendicular distance from $\sigma_{q}$ such that $y$ is much smaller than any scales defining the geometry of the entangling surface, $T_{ab}^{CFT}$ is the CFT energy-momentum tensor along the tangential directions $a,b$ to $\sigma_{q}$, and  $h_{q}$ is a constant representing the conformal dimension of $\sigma_{q}$. Generically $h_{q}$ is given in terms of the thermal energy density $\mathcal{E}(T,\mu)$ on the hyperbolic cylinder \cite{Hung:2011nu},
\beq h_{q}(\mu)=\frac{2\pi q L_{0}^{d}}{(d-1)}[\mathcal{E}(T_{0},\mu=0)-\mathcal{E}(T_{0}/q,\mu)]\;.\eeq

These conformal dimensions can be computed holographically using the energy densities $\mathcal{E}$ of the boundary field theory, which is proportional to the  mass $M$ of the dual hyperbolic black hole \cite{Belin:2013uta}
\beq h_{q}=\frac{2\pi q L_{0}^{d}}{(d-1)}[\mathcal{E}(T_{0},\mu=0)-\mathcal{E}(T_{0}/q,\mu)]=\frac{2\pi q}{w_{d-1}}\frac{L_{0}}{(d-1)}\left[M(T_{0},\mu=0)-M(T_{0}/q,\mu)\right]\;.\eeq

Following the spirit of \cite{Johnson:2018bma}, we can likewise obtain the conformal weights $h^{(b)}_{q}$ for the twist operators $\sigma_{q}^{(b)}$ just by using the extended thermodynamics of the charged black hole
\beq h_{q}^{(b)}=\frac{2\pi q}{w_{d-1}}\frac{L_{0}}{(d-1)}\left[M(T_{0},p_{0},\mu=0)-M(T_{0}/q,b^{2}p_{0},\mu)\right]\;.\label{confweightqb}\eeq
Since $M$ is interpreted as the enthalpy in extended black hole thermodynamics, we see $h_{q}^{(b)}$ is given as the difference of a state function. 

Evaluating our expression for the mass $M(x,L,\mu)$ (\ref{massofhighd}), we have that $M(T_{0},p_{0},\mu=0)=0$, as this corresponds to a massless hyperbolic black hole, and at $M(T_{0}/q,L_{0}/b,\mu)$, where  $x=x_{qb}$. Therefore,
\beq h_{q}^{(b)}(\mu)=\frac{qL_{0}^{d-1}}{8G_{N}}\left[\left(\frac{x_{qb}}{b}\right)^{d-2}(1-x_{qb}^{2})-\left(\frac{x_{qb}}{b}\right)^{d-2}\frac{(d-2)}{2(d-1)}\left(\frac{\mu\ell_{\ast}}{2\pi}\right)^{2}\right]\;. \label{hbqmu}\eeq
This is interpreted as the $b$-deformation of the conformal weight\footnote{Note there is a minor typo in Equation (42) in \cite{Johnson:2018bma}.} of the $d$-dimensional  twist operators $\sigma_{q}$ first explored in \cite{Hung:2014npa}.

Two remarks are in order: \textbf{(1)} Special values of $h_{q}^{(b)}(\mu)$, and \textbf{(2)} Derivatives of $h_{q}^{(b)}(\mu)$.

\subsection*{Special Values of $h_{q}^{(b)}(\mu)$}

Let's first consider the behavior of $h_{q}^{(b)}$ when $\mu=0$ for some particular special values of $q,b$. Consider when $x_{qb}=b$, such that $ b^{2}-1=2(1-q)/qd$. This choice for $b$, recall, corresponds to the special value when $S_{q,b\neq1}=S_{EE}$. The corresponding conformal weight is
\beq h^{(b)}_{q}=\frac{qL_{0}^{d-1}}{8G_{N}}(1-b^{2})=\frac{L_{0}^{d-1}}{4G_{N}}\frac{(q-1)}{d}\;.\label{confweighSEE}\eeq
Since this choice of $b(q)$ has $S_{q,b(q)}$ equal to the von Neumann entropy $S_{EE}$, we recognize there exists a replica trick to compute $\text{tr}[(\rho^{(b)}_{A})^{q}]$ involving twist operators with a conformal weight given above will lead to a direct calculation of $S_{EE}$. Indeed, we recover the $d=2$ CFT calculation presented in \cite{Johnson:2018bma}, where
\beq h^{(b)}_{q}=\frac{c}{12}q\left(1-\frac{1}{q^{2}b^{2}}\right)\;,\label{CFT2dcentcharge}\eeq
upon identifying the 2-dimensional central charge $c=3L_{0}/2G_{N}$. 

From (\ref{CFT2dcentcharge}) it is also easy to see that when $b=1/q$, \emph{i.e.}, $b$ `undoes' the $q$-replicas, we have $h^{(b)}_{q}=0$. Holographically this corresponds to $x_{qb}=1$. Note, moreover, from (\ref{hbqmu}) at $\mu=0$ we have $h^{(b=q^{-1})}_{q}=0$ for all dimensions\footnote{It is puzzling that $S_{q,q^{-1}}\neq0$ for $d>2$, given the field theory interpretation. It is thought that there is a subtlety when defining $\rho_{A}^{(b)}$, particularly when we perform the analytic continuation of $b$ from the integers to positive real numbers.}, where $S_{q,q^{-1}}|_{d>2}\neq0$, while the CFT calculation corresponds only to $d=2$, where $S_{q,q^{-1}}|_{d=2}=0$.

Let's now turn the charge $\mu$ on. There are a number of potentially interesting values of $h^{(b)}_{q}(\mu)$ given a specific $(q,b)$. First note that $h^{(b)}_{q}(\mu)=0$ when 
\beq x_{qb}=x_{qb}^{\ast}\equiv\sqrt{\frac{d}{d-2}}x_{\infty}\;,\eeq
with $x_{\infty}$ defined in (\ref{xqbqinftnoL}). Therefore, as we fractionate the system into a  large number of $q$- or $b$-copies, the conformal weight of the twist operators will vanish; this is contrary to what we see in the neutral case, where $h^{(b)}_{q}=0$ also when $x_{qb}=1$.

Next, consider when $x_{qb}=b$, such that 
\beq b^{2}-1=\frac{2}{qd}(1-q)+\frac{(d-2)}{2d(d-1)}\left(\frac{\mu\ell_{\ast}}{2\pi}\right)^{2}\;.\eeq
Then we have
\beq
\begin{split}
 h^{(b)}_{q}(\mu)&=
\frac{qL_{0}^{d-1}}{8G_{N}}\left[\frac{2}{qd}(q-1)-\frac{(d-2)(d+1)}{2d(d-1)}\left(\frac{\mu\ell_{\ast}}{2\pi}\right)^{2}\right]\;.
\end{split}
\eeq
Thus, $h^{(b)}_{q}(\mu)$ will be positive when $(q-1)>q\frac{(d-2)(d+1)}{4(d-1)}\left(\frac{\mu\ell_{\ast}}{2\pi}\right)^{2}$. Meanwhile, for $x_{qb}=1$, where
\beq b=\frac{2q^{-1}}{\left(2-\frac{(d-2)^{2}}{2(d-1)}\left(\frac{\mu\ell_{\ast}}{2\pi}\right)^{2}\right)}\;,\eeq
we have $h^{(b)}_{q}(\mu)<0$. Specifically
\beq h^{(b)}_{q}(\mu)=-\frac{qL_{0}^{d-1}}{8G_{N}b^{d-2}}\frac{(d-2)}{2(d-1)}\left(\frac{\mu\ell_{\ast}}{2\pi}\right)^{2}\;.\eeq
We see for this choice of $b(q)$, $h^{(b)}_{q}(\mu)$ will  be negative, though $S_{q,b}$ will be positive.

There are two more  particularly interesting cases to consider: (i) the special value of $b$ given in (\ref{specb}), corresponding to when $x_{qb}=x_{11}$, where $b\propto q^{-1}$, and (ii) when $x_{qb}=bx_{11}$. The first of these leads to 
\beq h^{(b)}_{q}(\mu)=\frac{qL_{0}^{d-1}}{8G_{N}}\left(\frac{x_{11}}{b}\right)^{d-2}\left[(1-x_{11}^{2})-\frac{(d-2)}{2(d-1)}\left(\frac{\mu\ell_{\ast}}{2\pi}\right)^{2}\right]\;.\eeq
Unlike the neutral case, for generic $d$ this is non-vanishing, demonstrating index $b$ in the charged framework does not purely undo the $q$-copies in the replica trick. 

Meanwhile, when $x_{qb}=bx_{11}$, in which $S_{q,b}(\mu)=S_{EE}(\mu)$, we find
\beq h^{(b)}_{q}(\mu)=\frac{qL_{0}^{d-1}}{8G_{N}}x_{11}^{d-2}\left[(1-b^{2}x_{11}^{2})-\frac{(d-2)}{2(d-1)}\left(\frac{\mu\ell_{\ast}}{2\pi}\right)^{2}\right]\;.\eeq
This is the charged analog of (\ref{confweighSEE}) which gives us the conformal dimension for the twist operator $\sigma_{q}$ necessary when computing $S_{EE}(\mu)$ using the replica trick. 


\subsection*{Derivatives of $h_{q}^{(b)}(\mu)$}

 In the neutral, non-extended case, there is a universal property for the conformal weight $h_{q}$ involving its first $q$th derivative \cite{Hung:2011nu,Hung:2014npa}
\beq \partial_{q}h_{q}|_{q=1}=\frac{2}{(d-1)}\pi^{1-\frac{d}{2}}\Gamma(d/2)a^{\ast}_{d}\;,\eeq
where $a^{\ast}_{d}$ is the central charge\footnote{Here we have chosen the normalization given in \cite{Hung:2011nu}; an alternative normalization for $a^{\ast}_{d}$ leads to $\partial_{q}h_{q}|_{q=1}=2\pi^{\frac{d}{2}+1}\frac{\Gamma(d/2)}{\Gamma(d+2)}\tilde{a}^{\ast}_{d}$ which is used in \cite{Belin:2013uta,Hung:2014npa}.} (\ref{gencent}) defined by the two point function of the CFT stress energy tensor. As noted in \cite{Belin:2013uta}, this universal behvavior does not readily extend to the conformal weight for the charged twist operators $\sigma_{q}(\mu)$. Rather, it is natural to study $h_{q}(\mu)$ expanded about $q=1$ and $\mu=0$,
\beq h_{q}(\mu)=\sum_{a,c}\frac{1}{a!c!}h_{ac}(q-1)^{a}\mu^{c}\;,\quad h_{ab}=(\partial_{q})^{a}(\partial_{\mu})^{c}h_{q}(\mu)|_{q=1,\mu=0}\;,\label{chargedimconf1}\eeq
where in particular $h_{10}|_{q=1,\mu=0}\propto a^{\ast}_{d}$. Using the holographically computed $h_{q}(\mu)$, one observes $a^{\ast}_{d}$ is precisely given by (\ref{gencent}). 

Similarly, while \emph{not} considered in \cite{Johnson:2018bma} we can study derivatives of $h_{q}^{(b)}$, where we expand $h_{q}^{(b)}$ about $q=1$ and $b=1$:
\beq h^{(b)}_{q}=\sum_{a,c}\frac{1}{a!c!}h^{(b)}_{ac}(q-1)^{a}(b-1)^{c}\;,\quad h^{(b)}_{ac}=(\partial_{q})^{a}(\partial_{b})^{c}h_{q}^{(b)}|_{q=1,b=1}\;.\label{confdim2}\eeq
Obviously, $h_{10}^{(b)}=\partial_{q}h_{q}^{(b=1)}|_{q=1}\propto a^{\ast}_{d}$, namely, 
\beq \partial_{q}h_{q}^{(1)}=\frac{L_{0}^{d-1}}{4G_{N}}\frac{1}{(d-1)}\;,\eeq
with $a^{\ast}_{d}$ as in (\ref{gencent}).  With our $b$-deformation, moreover, it is natural to compute $h_{01}^{(b)}$ holographically
\beq h_{01}^{(b)}=\partial_{b}h^{(b)}_{1}|_{b=1}=-\frac{L_{0}^{d-1}}{8G_{N}}2x_{1b}\frac{\partial x_{1b}}{\partial b}\biggr|_{b=1}=\frac{L_{0}^{d-1}}{4G_{N}}\frac{1}{(d-1)}=h_{10}^{(b)}\;.\eeq
This tells us that the first $b$-derivative of the generalized conformal weight $h_{q}^{(b)}$ likewise gives the central charge $a_{d}^{\ast}$. 

Motivated by (\ref{chargedimconf1}), it is straightforward to generalize (\ref{confdim2}) by including a charge $\mu$,
\beq h^{(b)}_{q}(\mu)=\sum_{i,j,k}\frac{1}{i!j!k!}h^{(b)}_{ijk}(\mu)(q-1)^{i}(b-1)^{j}\mu^{k}\;,\quad h_{ijk}^{(b)}=(\partial_{q})^{i}(\partial_{b})^{j}(\partial_{\mu})^{k}h_{q}^{(b)}(\mu)\biggr|_{q,b=1,\mu=0}\;.\eeq
It is simple to show 
\beq
\begin{split}
 &h_{100}^{(b)}=h_{010}^{(b)}=\frac{L_{0}^{d-1}}{4G_{N}}\frac{1}{(d-1)}\;,\quad h^{(b)}_{001}=0\;,\quad h^{(b)}_{002}=-\frac{(d-2)(2d-3)}{\pi(d-1)^{2}}\frac{\ell_{\ast}^{2}L_{0}^{d-1}}{16\pi G_{N}}\;.
\end{split}
\eeq
We don't find additional universal features for higher derivatives of the conformal weight. It may be interesting to see how these derivatives of the conformal weight match to the would be CFT calculation.


\section{Extended Charged R\'enyi Entropy in $d=2$}
\label{sec:d2case}

\subsection{Thermodynamics of Charged BTZ Black Hole}

Above we extended the charged R\'enyi entropy for CFTs dual to Einstein-Maxwell gravity in $d>2$. The restriction on the dimension $d$ was taken into account because it is well known the geometry of  charged AdS black holes in $d=2$ is markedly different from their higher dimensional counterparts. Specifically, in $d=2$ the black hole geometry is that of a charged BTZ black hole \cite{Martinez:1999qi}
\beq ds^{2}=-f(r)dt^{2}+f^{-1}(r)dr^{2}+r^{2}d\theta^{2}\;,\quad f(r)=\frac{r^{2}}{L^{2}}-2m-\frac{q_{e}^{2}}{2}\log\left(\frac{r}{L}\right)\;.\label{chargedBTZds}\eeq
The horizon radius $r_{h}$ is located at $f(r_{h})=0$. We observe that the geometry is not asymptotically AdS because of the logarithmic term appearing in $f(r)$. The gauge potential is given by 
\beq A=\frac{q_{e}}{\ell_{\ast}}\log\left(\frac{r}{L}\right)dt\;.\eeq

The extended thermodynamics\footnote{Here we follow the conventions of \cite{Johnson:2019wcq}, where we have chosen to set $G_{N}=\ell_{\ast}=1$} of the charged BTZ black hole was worked out in \cite{Frassino:2015oca}:
\beq 
\begin{split}
&M=\frac{4pS^{2}}{\pi}-\frac{q_{e}^{2}}{32}\log\left(\frac{32pS^{2}}{\pi}\right)\;,\quad S=\frac{\pi r_{h}}{2}\;,\quad p=\frac{1}{8\pi L^{2}}\;,\quad T=\frac{8pS}{\pi}-\frac{q_{e}^{2}}{16S}\;,\\
&\Phi_{e}\equiv\left(\frac{\partial M}{\partial q_{e}}\right)_{S,p}=-\frac{q_{e}}{16}\log\left(\frac{32pS^{2}}{\pi}\right)\;,\quad V\equiv\left(\frac{\partial M}{\partial p}\right)_{S,q_{e}}=\frac{4S^{2}}{\pi}-\frac{q_{e}^{2}}{32p}\;,
\end{split}
\label{extthermchargedBTZ}\eeq
where $pS^{2}=\frac{\pi x^{2}}{32}$, with $x\equiv r_{h}/L$ as usual. We emphasize that, unlike their $d>2$ counterparts, the thermodynamic volume $V$ no longer scales as the entropy $S$, \emph{i.e.}, the thermodynamic volume is not equivalent to the naive geometric volume. The implications of this observation will be discussed further momentarily.

 It is straightforward to work out the internal energy and Gibbs free energy of the charged BTZ system:
\beq U=M-pV=\frac{q_{e}^{2}}{32}\left[1-\log\left(\frac{32pS^{2}}{\pi}\right)\right]\;,\label{Ubtz}\eeq
\beq
\begin{split}
G&=M-TS-\Phi_{e}q_{e}=-\frac{4pS^{2}}{\pi}+\frac{q_{e}^{2}}{32}\left[2+\log\left(\frac{32pS^{2}}{\pi}\right)\right]\\
&=-\frac{x^{2}}{8}+\frac{q_{e}^{2}}{32}\left[2+\log(x^{2})\right]\;.
\end{split}
\label{GibbsBTZ}\eeq

Notice that demanding the temperature $T$ be positive (equivalently the $V\geq0$) results in the condition
\beq q_{e}^{2}\leq 4\eta\;,\eeq
where $\eta\equiv32pS^{2}/\pi=x^{2}$. The parameter $\eta$ also appears in $U$, which shows that for $U>0$ implies $\eta\leq1$ resulting in a bound on $q_{e}$ \cite{Johnson:2019wcq}:
\beq q_{e}^{2}\leq4\;.\label{upperboundq}\eeq

It is well known that the asymptotic symmetry group of a rotating (uncharged) BTZ black hole yields two copies of a Virasoro algebra; this is the famous result by Brown and Henneaux stating that quantum gravity in three dimensions is dual to a two dimensional CFT with left and right central charges $c_{L}=c_{R}=\frac{3L_{0}}{2}$ \cite{Brown:1986nw}. Consequently, the gravitational entropy $S$ is equal to the Cardy entropy \cite{Cardy:1986ie,Bloete:1986qm} of the dual CFT, providing a microscopic interpretation of the Bekenstein-Hawking entropy formula \cite{Strominger:1996sh,Strominger:1997eq,Carlip:1998qw,Carlip:2005zn}. Due to the presence of the logarithmic function in $f(r)$ for the charged BTZ solution (\ref{chargedBTZds}), the asymptotic symmetry group is deformed, thereby hiding the Virasoro symmetry. The Virasoro algebra can be made explicit via a renormalization procedure by enclosing the entire black hole system in a circle of radius $r_{0}$ and then take the limit $r_{0}\to\infty$ while keeping $r/r_{0}=1$ fixed \cite{Cadoni:2007ck}. The consequence is a renormalized black hole mass $M_{0}(r_{0})=M+\frac{q_{e}^{2}}{16}\log(r_{0}/L)$ such that the manifest Virasoro symmetry group is restored. As a result, the extended thermodynamics is altered so as to promote the renormalization scale $r_{0}$ to a thermodynamic variable with a corresponding thermodynamic potential $K\equiv(\frac{\partial M}{\partial r_{0}})_{S,q_{e},p}$. Including this renormalization scale, moreover, leads to a thermodynamic volume equal to the naive geometric volume \cite{Frassino:2015oca}. Thus, explicit Virasoro symmetry of the charged BTZ black hole is a byproduct of the renormalization scheme of the black hole, which affects the extended thermodynamics. 

This connection between manifest conformal symmetry and the extended thermodynamics makes an appearance in the thermodynamic stability of the charged BTZ. Recently \cite{Johnson:2019mdp} it was shown the heat capacity at constant volume $C_{V}$ may be compactly written as
\beq C_{V}=-S\left(\frac{4S^{2}-\pi V}{12S^{2}-\pi V}\right)\;,\eeq
where we see positivity of $T$ implies $4S^{2}>\pi V$, and thus, the charged BTZ black hole is thermodynamically unstable. The condition of this instability, $4S^{2}>\pi V$, is exactly the condition that the charged BTZ black hole is ``super-entropic" --  AdS black holes whose entropy exceeds the expected bound of an AdS-Schwarzschild black hole \cite{Cvetic:2010jb,Hennigar:2014cfa,Hennigar:2015cja}. In the recent work \cite{Johnson:2019wcq} it was shown super-entropicity of the charged BTZ black hole can be understood microscopically as the condition that the Bekenstein-Hawking entropy `over-counts' the number of accessible dual CFT states\footnote{This results from the fact that the naive CFT Cardy entropy formula -- which is equal to the gravitational entropy -- should be replaced with a corrected Cardy formula since the lowest eigenvalue of the zero-moded Virasoro generators is non-zero.}. We point out, moreover, using the renormalization scheme employed in \cite{Cadoni:2007ck} the charged BTZ black hole is no longer super-entropic; super-entropicity of the charged BTZ solution is intimately connected to its hidden Virasoro symmetry. 

The upper bound on the charge $q_{e}$ (\ref{upperboundq}) also has a CFT interpretation: it is a unitarity bound of the CFT associated with the (effective) central charge of the CFT \cite{Johnson:2019wcq}
\beq c_{\text{eff}}=c\left(1-\frac{q_{e}^{2}}{4}\right)\;.\eeq
Here $c$ is the familiar central charge for $\text{AdS}_{3}/\text{CFT}_{2}$ holography, $c=\frac{3L_{0}}{2}$.


\subsection{Generalized R\'enyi Entropy}

As in the higher dimensional discussion, we may apply the same quench technique and compute the R\'enyi entropy and its single parameter extension by computing the difference in Gibbs free energies
\beq S_{q,b}=-\frac{[G(L_{0},x_{11})-G(L_{0}/b,x_{qb})]}{\Delta T-\frac{V_{0}}{S_{0}}\Delta p}\;.\eeq
Using the Gibbs free energy (\ref{GibbsBTZ}) and
\beq \Delta T-\frac{V_{0}}{S_{0}}\Delta p=\frac{1}{2\pi L_{0}q}\left[(q-1)+\frac{qx_{11}}{2}(b^{2}-1)\left(1-\frac{q_{e}^{2}}{4x_{11}^{2}}\right)\right]\;,\eeq
we have the extended charged R\'enyi entropy for a globally charged two-dimensional CFT whose gravitational dual is described by the charged BTZ black hole
\beq S_{q,b}(q_{e})=\frac{q}{2}\frac{\pi L_{0}}{2}\frac{\left[(x_{11}^{2}-x_{qb}^{2})-\frac{q_{e}^{2}}{2}\log\left(\frac{x_{qb}}{x_{11}}\right)\right]}{\left[q-1+\frac{qx_{11}}{2}(b^{2}-1)\left(1-\frac{q_{e}^{2}}{4x_{11}^{2}}\right)\right]}\;.\label{SqbchargeBTZ}\eeq
Here $x_{qb}$ is found by solving $T_{0}/q=T(x_{qb})$, resulting in
\beq x_{qb}=\frac{1}{2qb}\left(1+\sqrt{1+(qbq_{e})^{2}}\right)\;.\label{xqbBTZ}\eeq

We point out that here we have chosen to keep the R\'enyi entropy $S_{q,1}$ and its extension $S_{q,b}$ as a function of the charge $q_{e}$ rather than the chemical potential $\mu$. For higher dimensions $\mu$ was chosen by requiring the gauge field $A$ vanish at the horizon, $r=r_{h}$. Due to the logarithmic running of the bulk gauge field, it is difficult to discern the chemical potential $\mu$ as it stands. We can manifestly write down $\mu$ by employing the same renormalization procedure \cite{Cadoni:2007ck} mentioned briefly above. In doing so, we would write a modified bulk gauge field as
\beq A_{0}=A-q_{e}\log\left(\frac{r_{0}}{L}\right)=q_{e}\log\left(\frac{r}{r_{0}}\right)\;.\eeq
Then, when we take the $r_{0}\to\infty$ limit, while keeping $r/r_{0}\to1$, we have $A_{0}\to0$. As further pointed out in \cite{Cadoni:2007ck} we set the renormalization scale $r_{0}=r_{h}$ such that the total energy is just the renormalized mass evaluated at the outer horizon, $M_{0}(r_{h})$. Thus, the chemical potential $\mu$, via this renormalization scheme, is simply 
\beq \mu=-q_{e}\log\left(\frac{r_{h}}{L}\right)\;.\eeq

We recognize $\mu$ then as being proportional to the electrostatic potential (\ref{extthermchargedBTZ}), as in the $d\geq3$ case. If we employ this renormalization procedure, then we must modify the extended thermodynamics of our black hole such that the thermodynamic volume becomes the geometric volume, $V\to \pi r_{h}^{2}$, and we introduce a thermodynamic potential conjugate to the renormalization scale $r_{0}$. Alternatively, we are free to not employ the aforementioned renormalization scheme \cite{Cadoni:2007ck}, thereby invoking the extended thermodynamics (\ref{extthermchargedBTZ}), but recognize a fixed chemical potential $\mu$ is equivalent to studying our system at fixed electrostatic potential $\Phi_{e}$. To summarize, explicitly introducing a $\mu$ via the renormalization scheme keeps the underlying Virasoro symmetry of the charged BTZ geometry manifest; instead we work with the system where this Virasoro symmetry is hidden, however, at fixed potential $\Phi_{e}$. 

From (\ref{xqbBTZ}), notice when $q$ or $b$ or $q_{e}$ tend to infinity we have
\beq \lim_{q,b\to\infty}x_{qb}\to \frac{q_{e}}{2}\;,\eeq
while $q$ or $b$ approaching zero leads to a $1/qb$ divergence. In the limit $q,b\to1$ we find the von Neumann limit of the R\'enyi entropy $S_{q,b}(q_{e})$ is proportional to the Bekenstein-Hawking entropy evaluated at $r_{h}=L_{0}$:
\beq \lim_{q,b\to1}S_{q,b}(q_{e})=\frac{\pi L_{0}}{2}\frac{(1+\sqrt{1+q_{e}^{2}})}{2\sqrt{1+q_{e}^{2}}}\;.\eeq
When we restore the (infinite) `volume' of the hyperbolic plane $\omega_{1}$, we recover the charged von Neumann entropy, $S_{EE}(q_{e})$.

There are a number of interesting limits to now consider, including the large $q$ and $b$ asymptotics of $S_{q,b}$. As $b\to\infty$, it is straightforward to show that $S_{q,b}(q_{e})\to\frac{1}{b^{2}}$, while large $q$ leads to 
\beq \lim_{q\to\infty}S_{q,b}(q_{e})=\frac{\pi L_{0}}{4(1+b^{2})}\left[1+\sqrt{1+q_{e}^{2}}+q_{e}^{2}\log\left(\frac{q_{e}}{1+\sqrt{1+q_{e}^{2}}}\right)\right]\;.\eeq
Thus the large $q$ limit approaches a finite asymptotic value, in stark contrast with the higher dimensional analog. As $q$ or $b$ tend to zero, moreover, the extended R\'enyi entropy diverges. 


\begin{figure}[t]
  \subfloat[]{
	\begin{minipage}[1\width]{
	   0.32\textwidth}
	   \centering
	   \includegraphics[width=1.45\textwidth]{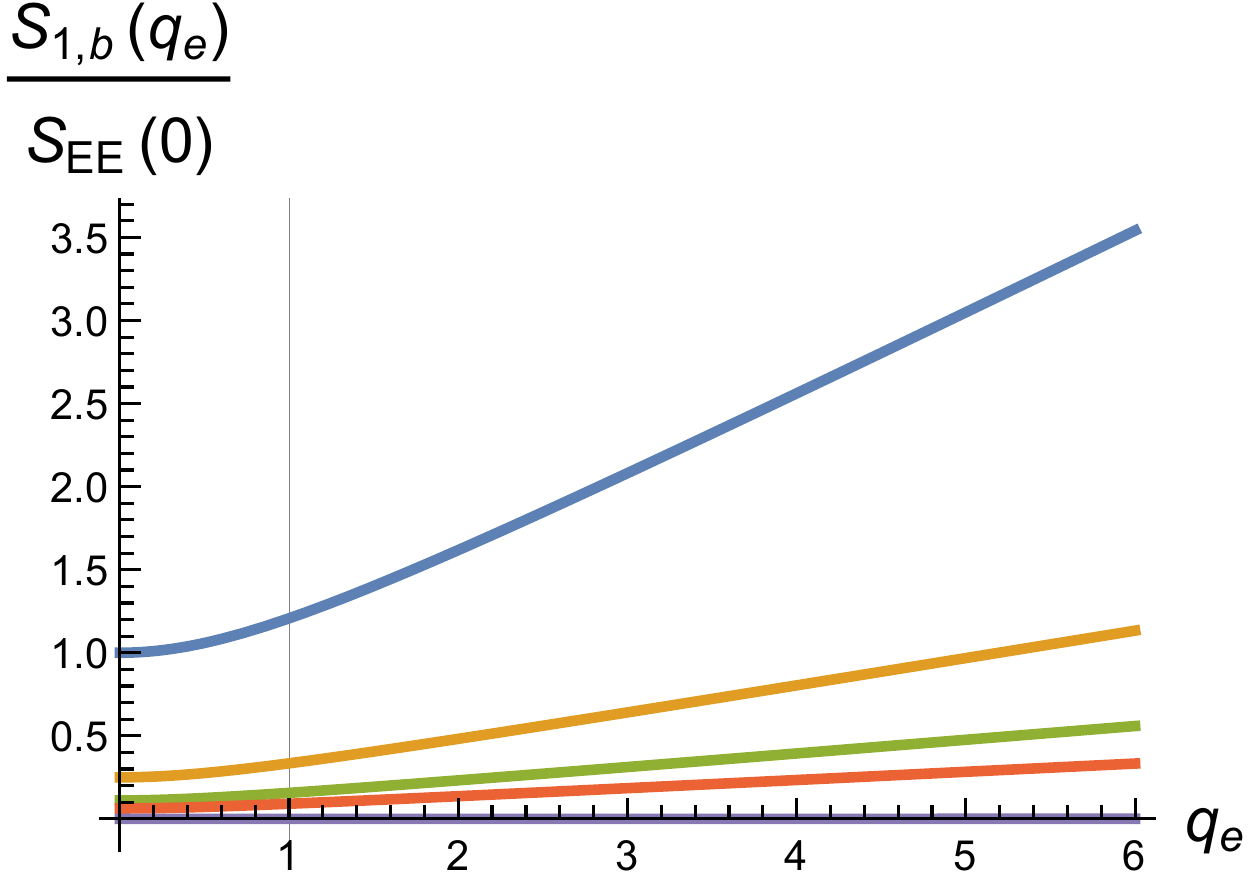}
	\end{minipage}}
 \hspace{2.5cm}
  \subfloat[]{
	\begin{minipage}[1\width]{
	   0.32\textwidth}
	   \centering
	   \includegraphics[width=1.45\textwidth]{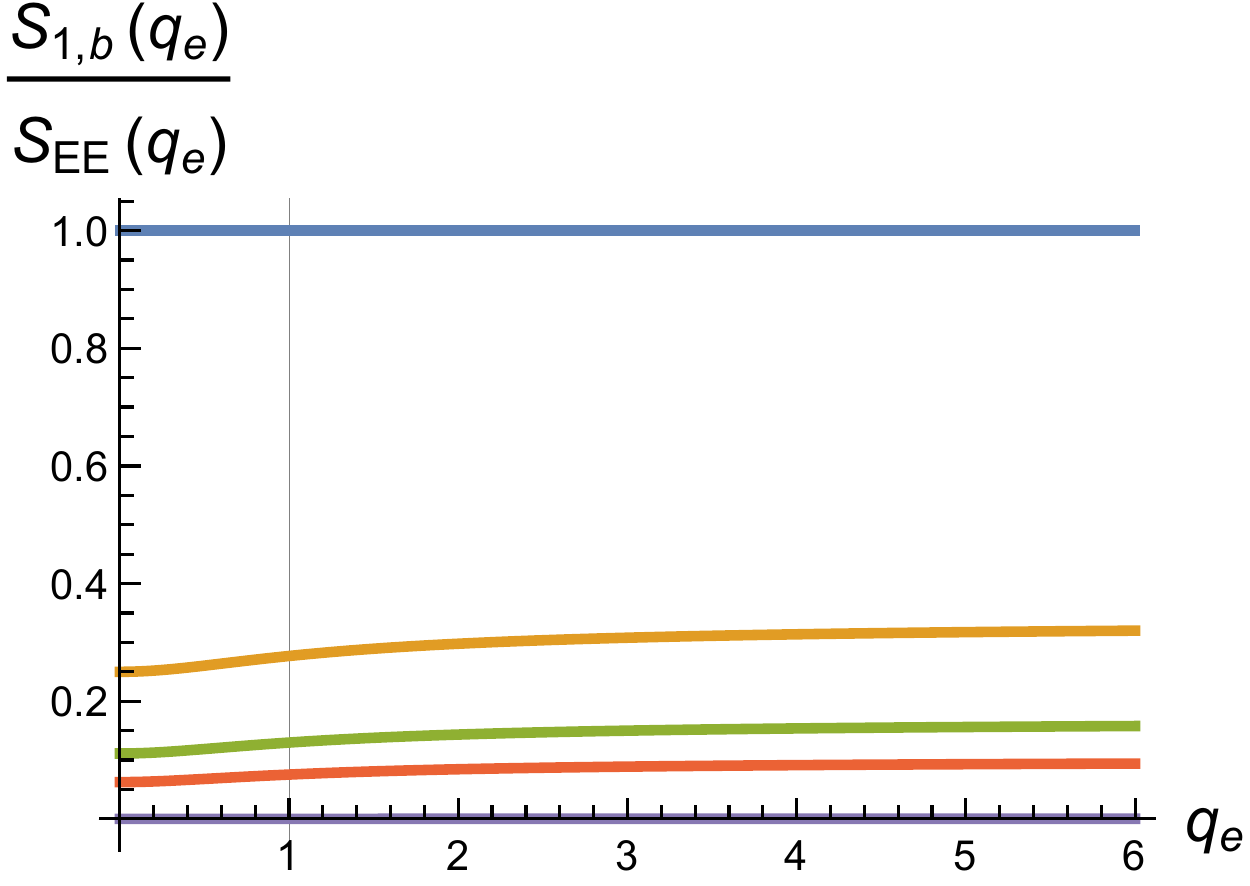}
	\end{minipage}}
\caption{A plot of $S_{1,b}(q_{e})$ as a function of $q_{e}$ for various fixed values of $b$. (a) Normalized by $S_{1,1}(0)$. (b) $S_{1,1}(q_{e})$. In both plots $b=1$ (blue), $b=2$ (orange), $b=3$ (green), $b=4$ (red), and $b=100$ (purple).}
\label{S1bBTZQ}\end{figure}

As in $d\geq3$, there exists a critical value of the R\'enyi index $q_{c}$ where $S_{q,b}$ diverges. Specifically, 
\beq q_{c}=\left[1+\frac{x_{11}}{2}(b^{2}-1)\left(1-\frac{q_{e}^{2}}{4x_{11}^{2}}\right)\right]^{-1}\;.\eeq
Notice that the sign of $q_{c}$ depends on whether $b<1$ since for any value of $q_{e}\geq0$, $1-q^{2}_{e}/4x_{11}^{2}>0$. Note, we do not find any change in this behavior as $q_{e}$ exceeds the unitarity bound $q_{e}^{2}>4$. 

\begin{figure}[t]
  \subfloat[]{
	\begin{minipage}[1\width]{
	   0.32\textwidth}
	   \centering
	   \includegraphics[width=1.45\textwidth]{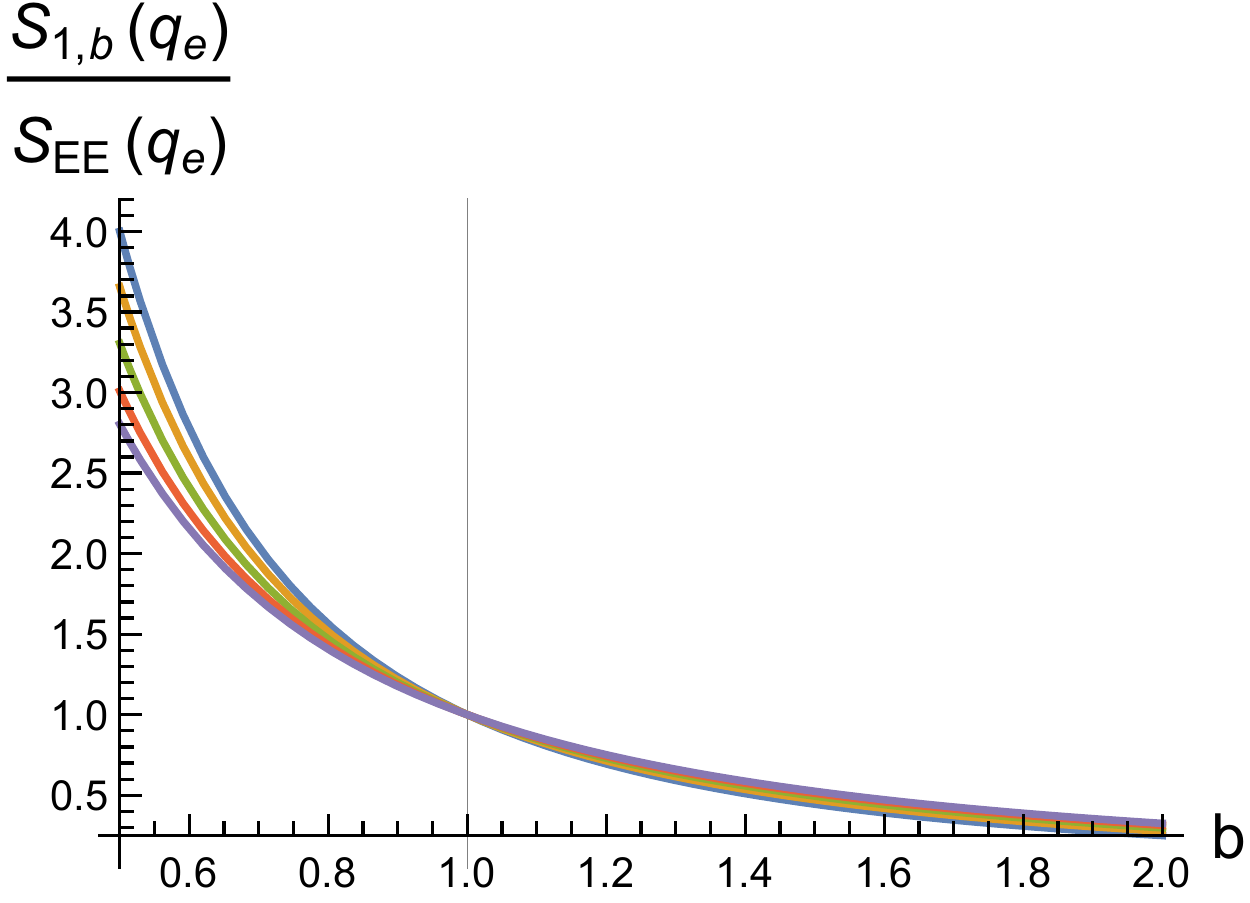}
	\end{minipage}}
 \hspace{2.5cm}
  \subfloat[]{
	\begin{minipage}[1\width]{
	   0.32\textwidth}
	   \centering
	   \includegraphics[width=1.45\textwidth]{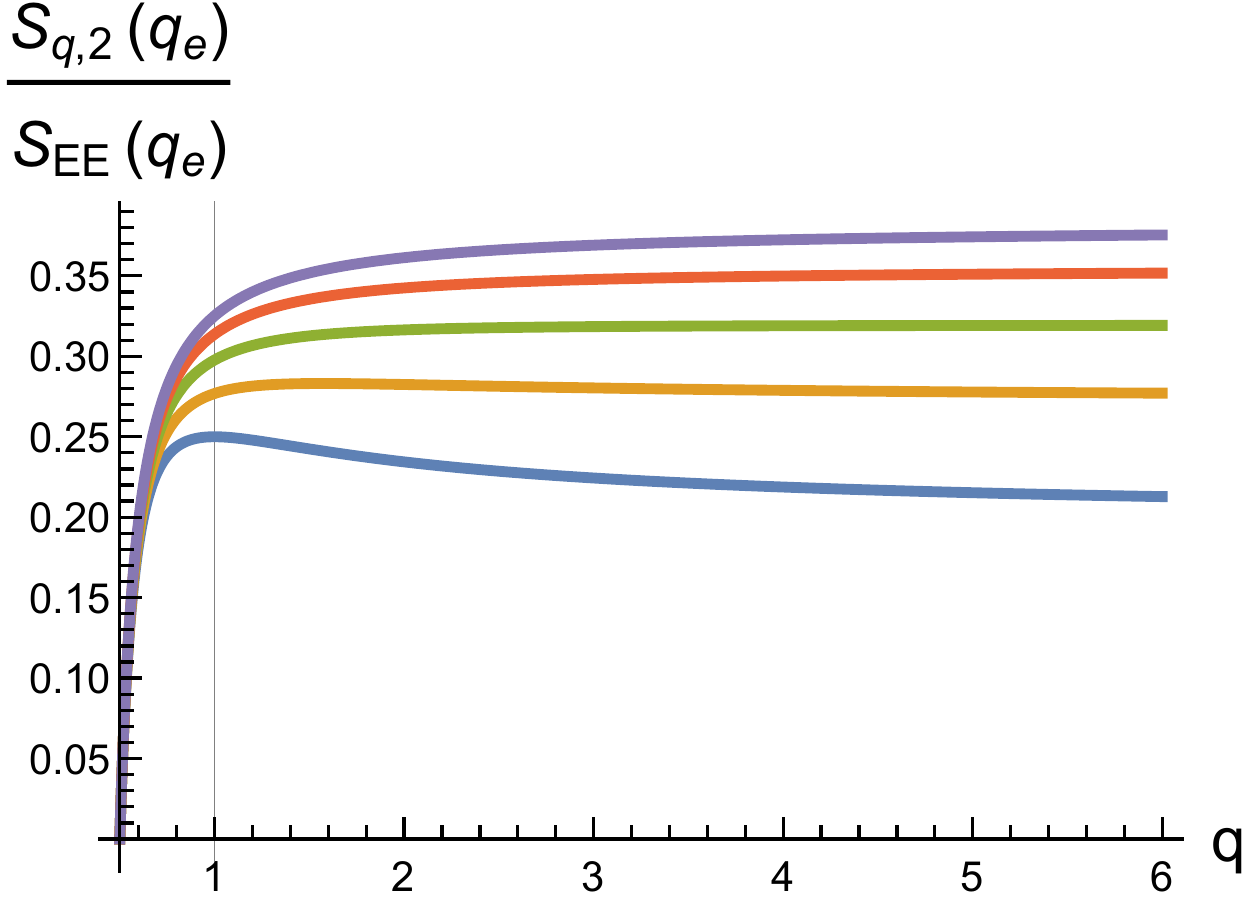}
	\end{minipage}}
\caption{(a) $S_{1,b}(q_{e})$ as a function of $b$ for various fixed values of $q_{e}$. (b) $S_{q,2}(q_{e})$ as a function of $q$ for various fixed values of $q_{e}$. In both cases $q_{e}=0$ (blue), $q_{e}=1$ (orange), $q_{e}=2$ (green), $q_{e}=4$ (red), and $q_{e}=10$ (purple).}
\label{S1bSq2BTZ}\end{figure}

We have plotted the behavior of $S_{1,b}(q_{e})$ for a range of $q_{e}$ at various fixed values of $b$ in Figure \ref{S1bBTZQ}. From Figure \ref{S1bBTZQ} we observe that, when normalized by $S_{1,1}(0)$, $S_{1,b}(q_{e})$ clearly grows linearly as a function of $q_{e}$, and as $b\to\infty$, $S_{1,b}(q_{e})$ approaches zero. The linear behavior is markedly different from the exponential growth, as seen in Figure \ref{S1bfuncmu}. A similar observation was made for $S_{q}(q_{e})$ in \cite{Belin:2013uta}. In Figure \ref{S1bSq2BTZ} we plot the behavior of $S_{1,b}(q_{e})$ as a function of $b$ for fixed values of $q_{e}$, and also $S_{q,2}(q_{e})$. We see $S_{1,b}(q_{e})$ has similar features as its higher dimensional counterparts, while $S_{q,2}(q_{e})$ displays notably different behavior; specifically we see no divergence and an initial logarithmic growth to a finite constant value.

Note we do not see any peculiar behavior for values of $q^{2}_{e}\geq4$, \emph{i.e.}, beyond the unitarity bound uncovered in \cite{Johnson:2019wcq}. As we will discuss in more detail later on, this is because we are working with a fixed electrostatic potential.


As in the $d\geq3$ case, there are potentially special values of $b=b(q)$. Perhaps the most interesting choice of $b$ is $b=q^{-1}$. In this case $x_{qq^{-1}}=x_{11}$, and we find $S_{q,q^{-1}}(q_{e})=0$, agreeing with the uncharged case in $d=2$ \cite{Johnson:2018bma}. Moreover, it was this particular finding that led to the interpretation of $b$ as a R\'enyi index, for which $b=q^{-1}$ undoes the $q$-replicas when calculating the R\'enyi entropy via the replica trick. As we will examine shortly, we will be able to arrive to the same interpretation for the charged system. 

Lastly, of interest is when we Wick rotate $q_{e}\to iq^{E}_{e}$. In doing so we have the horizon radius at which $T=T_{0}/q$ and $p=b^{2}p_{0}$,
\beq x_{qb}^{E}=\frac{1}{2qb}(1+\sqrt{1-(qbq_{e}^{E})^{2}})\;.\eeq
To maintain real solutions, we require $(q_{e}^{E})^{2}<\frac{1}{q^{2}b^{2}}$, telling us that the charge $q_{e}^{E}$ is allowed a finite range of values. Moreover, we may find from $x_{qb}^{E}$ that $(2qb x_{qb}^{E}-1)^{2}+q^{2}b^{2}(q_{e}^{E})^{2}=1$, such that we parameterize the charge $q_{e}^{E}$ via
\beq x_{qb}^{E}=\frac{1+\cos\phi}{2qb}\;,\quad q_{e}^{E}=\frac{\sin\phi}{qb}\;.\label{eucchargecondBTZ}\eeq
Thus, similar to the usual charged R\'enyi entropy $S_{q,1}(q_{e})$ \cite{Belin:2013uta}, we expect a free field calculation of $S_{q,b}(q_{e})$ will likewise have $q_{e}^{E}$ taking only a finite range of values.


\subsection{Field Theory Interpretation}

In Section \ref{sec:fieldtheoint} we considered the field theory interpretation of the extended R\'enyi entropy $S_{q,b}(\mu)$, whereby we made a generalization of the CHM map (\ref{CHMmapmod}). As in the uncharged case, a direct field theory computation of $S_{q,b}(\mu)$ is difficult and lacking for dimensions $d\geq3$. However, the calculation for $S_{q}(\mu_{E})$ has been established in $d=2$ for a free massless Dirac fermion $\psi$ on an infinite line reduced to an interval $y\in[u,v]$ \cite{Belin:2013uta}. Formally this is accomplished by evaluating the partition function for $\psi$ on a $q$-sheeted Riemann surface. The partition function may be expressed in terms of a correlation function of generalized twist operators inserted at the endpoints of the interval $y=[u,v]$. We will only summarize the details of the field theory calculation here, for more a through treatment see \cite{Belin:2013uta}.

In the uncharged case with $b=1$ the $q$-sheeted Riemann surface is described by the coordinate $w=y+i\tau$, such that the interval $[u,v]$ introduces a branch cut where $\tau$ is the periodic time coordinate with period $\beta=1/T_{0}$. Introducing a complex coordinate $\zeta=(w-u)/(w-v)$, a conformal transformation places the ends of the branch cut at $y=0$ and $y=\infty$, where each time we cross the branch cut we move from one sheet to the next. The fermion field $\tilde{\psi}_{k}$ ($k=1,...,q$) on the $k$th sheet will satisfy a set of boundary conditions coming from a phase shift in $\tilde{\psi}_{k}$ as we move from one sheet to the next, namely, 
\beq \tilde{\psi}_{m}(e^{2\pi i}(w-u))=e^{2\pi im/q}\tilde{\psi}_{m}(w-u)\;,\quad \tilde{\psi}_{m}(e^{2\pi i}(w-v))=e^{-2\pi im/q}\tilde{\psi}_{m}(w-v)\;,\label{fermbcs}\eeq
where $m=\{-(q-1)/2,-(q-1)/2+1...,(q-1)/2\}$. The phase shifts in the fermion fields are generated by the usual twist operators $\sigma_{m/q}$ and $\sigma_{-m/q}$, with conformal weights $h_{q/m}=c(m^{2}/q^{2})$. The twist operator $\sigma_{m/q}$ only acts on each $\tilde{\psi}_{m}$. The partition function $Z_{q}=\text{tr}\rho_{v}^{q}$ is proportional to the correlation function $\langle \sigma_{q}(u)\sigma_{-q}(v)\rangle\sim|(u-v)/\epsilon|^{-(h_{q}+h_{-q})}$, where $\epsilon$ is a UV regulator, and $\sigma_{q}$ is the full twist operator, $\sigma_{q}=\prod_{m}\sigma_{m/q}$. Here $h_{q}=h_{-q}$ is the conformal dimension of twist operator $\sigma_{q}$, given by
\beq h_{q}=\sum_{m=-\frac{(q-1)}{2}}^{\frac{(q-1)}{2}}h_{q/m}=\frac{cq}{12}\left(1-\frac{1}{q^{2}}\right)\;.\label{confweightd2}\eeq
In this way, the relevant calculation of the uncharged R\'enyi entropy $S_{q}=\frac{1}{q-1}(q\log Z_{1}-\log Z_{q})$ is reduced to finding the conformal dimension. 

The conformal dimension $h_{q}$ (\ref{confweightd2}) can also be found working with a single copy of the complex plane described by coordinate $z$ via the uniformization map of the $q$-sheeted Riemann surface, $z=\zeta^{1/q}$. In this scenario the conformal weight $h_{q}$ arises from the anomaly term -- proportional to a Schwarzian derivative -- of the stress energy tensor $T(w)$ associated with the CFT on the Riemann surface. 

When $b\neq1$, $S_{q,b}$ is found by computing the partition function $\text{tr}(\rho^{(b)}_{v})$ again using the correlator of twist fields. The details using the uniformization map are largely the same; all that changes is the conformal weight, which corresponds to altering the uniformization map to $z=\zeta^{1/qb}$, leading to $h^{(b)}_{q}=\frac{c}{12}q(1-\frac{1}{q^{2}b^{2}})$ \cite{Johnson:2018bma}. Importantly, $h^{(b)}_{q}\neq h_{qb}$. In light of (\ref{confweightd2}), interestingly we may reproduce $h^{(b)}_{q}$ simply by 
\beq h^{(b)}_{q}(q_{e}=0)=\frac{1}{b}h_{qb}\;.\eeq
Correspondingly, $h_{qb/m}=c(m^{2}/(qb)^{2})$, with integer $m=[-(qb-1)/2,(qb-1)/2]$, and the boundary conditions for the fermion field $\tilde{\psi}_{m}$ satisfy (\ref{fermbcs}), however, replacing $q\to qb$. 

Notice the effect of when $b=q^{-1}$. From the perspective of the uniformization map, $z=\zeta$, and $h^{(q^{-1})}_{q}=0$, such that $S_{q,q^{-1}}(q_{e}=0)=0$ in $d=2$. That is, $b$ `undoes' the $q$-replicas. This can also be seen from the modified boundary conditions for the fermion field (\ref{fermbcs}): as $b$ undoes $q$ there are no phase changes for the fermion, corresponding to $m=0$, and consequently $h_{m/qq^{-1}}=0$. We point out, moreover, when we send $q\to1$ and keep $b$, then $\tilde{\psi}_{m}$ will undergo phase changes as we move from each $k$th replica sheet, where now $k=1,...,b$. 

Let's now move on and consider charged CFTs. For the fermion system described above, it is natural to introduce a charge associated with global phase rotations of $\psi\to e^{i\theta}\psi$. The charged R\'enyi entropies are similarly computed as before, however, the introduction of a Wilson loop associated with an imaginary chemical potential $\mu_{E}$ leads to additional phase shift in the boundary conditions for fermion fields on the $m$th sheet $\tilde{\psi}_{m}$,
\beq \tilde{\psi}_{m}(e^{2\pi i}(w-u))=e^{2\pi im/q+i\mu_{E}}\tilde{\psi}_{m}(w-u)\;,\quad \tilde{\psi}_{m}(e^{2\pi i}(w-v))=e^{-2\pi im/q-i\mu_{E}}\tilde{\psi}_{m}(w-v)\;.\label{phaseshifts2}\eeq
The phase shifts arise from the introduction of twist operators $\sigma_{m/q,\mu_{E}}, \sigma_{-m/q,-\mu_{E}}$ at the endpoints of the branch cut with the same conformal weight
\beq h_{q/m}(\mu_{E})=c\left(\frac{m}{q}+\frac{\mu_{E}}{2\pi}\right)^{2}\;.\eeq
Since there is an ambiguity in defining the phase shifts (\ref{phaseshifts2}) modulo $2\pi$, we are free to add an integer $\ell_{m}$ to $h_{q}(\mu_{E})$. For small charge, \emph{i.e.}, small $\mu_{E}$, all $\ell_{m}$ may be set to zero, keeping the range of $m\in[-\frac{q-1}{2},\frac{q-1}{2}]$. Consequently, the expression for $S_{q}(\mu_{E})$ for small $\mu_{E}$ is equal to $S_{q}(\mu_{E}=0)$. As $\mu_{E}$ increases, however, not all $\ell_{m}$ can be set to zero, leading to phase transitions in $S_{q}(\mu_{E})$ occuring whenever $\mu_{E}=\frac{\pi}{q}(2k+1)$ for integer $k$. 

Now let's turn on $b$ for $q_{e}\neq0$. As in the neutral case, we can write down what $h^{(b)}_{q}(q_{e})$ in terms of $h_{qb/m}(q_{e})$, where now the range of $m$ is modified. Notice when $b=q^{-1}$, $h^{(b)}_{q}(q_{e})\neq0$ for non-zero charge. In fact,  even for small charge $q_{e}$, where we can likewise set the integer $\ell_{m}$ (such that $m\in[-\frac{qb-1}{2},\frac{qb-1}{2}]$), the conformal weight of twist operators $\sigma_{q}^{(b)}$ will be non-zero. Moreover, for  larger $q^{E}_{e}$ not all $\ell_{m}$ can be set to zero. This reveals phase transitions in $S_{q,b}(q^{E}_{e})$ whenever $q_{e}^{E}=\frac{\pi}{qb}(2k+1)$ for integer $k$, matching what we found from holographic considerations (\ref{eucchargecondBTZ}).

While we have not performed an explicit field theory calculation for the conformal weight $h^{(b)}_{q}(q_{e})$, we can compute it holographically. This is accomplished using the difference in enthalpies $M(T_{0},p_{0},q_{e}=0)-M(T_{0}/q,b^{2}p_{0},q_{e})$ (\ref{confweightqb}) using the extended thermodynamics of the charged BTZ black hole (\ref{extthermchargedBTZ})
\beq h^{(b)}_{q}(q_{e})=\frac{qc}{12}\left[(1-x_{qb}^{2})+\frac{q_{e}^{2}}{2}\log(x_{qb})\right]\;.\label{confweBTZ}\eeq
When $q_{e}=0$ we recover the uncharged result in $d=2$ \cite{Johnson:2018bma}. Notice that when $b=q^{-1}$, $h^{(b)}_{q}(q_{e})\neq0$, unlike the neutral case. Naively we might be alarmed by this since $S_{q,q^{-1}}(q_{e})=0$, however, a non-zero conformal weight when $b=q^{-1}$ is completely compatible with the result $S_{q,q^{-1}}=0$, as we now briefly demonstrate.

From the extended CHM map (\ref{CHMmapmod}) it is easy to write down $S_{q,b}(q_{e})$ in $d=2$ as
\beq S_{q,b}(q_{e})=\left[1+\frac{qx_{11}}{2}\frac{(b^{2}-1)}{q-1}\left(1-\frac{q_{e}^{2}}{4x_{11}^{2}}\right)\right]^{-1}\frac{1}{q-1}(q\log Z_{11}-\log Z_{qb})\;,\label{fieldtheSqb}\eeq
with 
\beq Z_{qb}=\left(\frac{|u-v|}{\epsilon}\right)^{-2h^{(b)}_{q}(q_{e})}\;.\eeq
In the neutral case, $b=q^{-1}$ has $\log Z_{11}=0$ since $h^{(q^{-1})}_{q}=0$. For $q_{e}\neq0$, while $\log Z_{11}\neq0$, we nonetheless find (\ref{fieldtheSqb}) with the conformal weight (\ref{confweBTZ}) reproduces the charged entropy $S_{q,b}(q_{e})$ (\ref{SqbchargeBTZ}).

It is worth pointing out that for small $q_{e}$
\beq h_{q}^{(b)}(q_{e})\approx \frac{c}{12}q\left(1-\frac{1}{q^{2}b^{2}}\right)-\frac{c}{24}q_{e}^{2}q[1+\log(qb)]\;.\eeq
Substituting this into (\ref{fieldtheSqb}) we recover, to leading order the neutral limit of $S_{q,b}$
\beq S_{q,b}(q_{e})=\frac{1}{[q-1+\frac{q}{2}(b^{2}-1)]}\left[q\left(1-\frac{1}{q^{2}b^{2}}\right)+\frac{qq_{e}^{2}}{2}\log(qb)\right]\frac{c}{6}\log\left(\frac{|u-v|}{\epsilon}\right)\;. \eeq
Meanwhile for large $q_{e}$, the conformal weight (\ref{confweBTZ}) becomes
\beq h_{q}^{(b)}(q_{e})=\frac{cq}{12}\left[\left(1-\frac{q_{e}^{2}}{4}\right)+\frac{q_{e}^{2}}{2}\log\left(\frac{q_{e}}{2}\right)\right]\;.\eeq

Intriguingly, when we saturate the unitarity bound (\ref{upperboundq}), where $q_{e}=2$, the conformal weight vanishes. However, as noted above, $\lim_{q_{e}\to\infty}S_{q,b}(q_{e})\to0$, where we see no interesting behavior for the specific value $q_{e}=2$, even when $q_{e}\gg(qb)$, with $q$ or $b$ smaller than one.


\section{Gravity Dual of Extended R\'enyi Entropy}
\label{sec:gravdual}

In the context described here, the holographic dual of the R\'enyi entropy is the difference in free energies of an appropriate AdS black hole. More generally, for holographic CFTs dual to Einstein gravity in the bulk, a quantity related to the R\'enyi entropy, which we will call the \emph{modified} R\'enyi entropy and denote as $\tilde{S}_{q}$, has been shown to satisfy an area law analogous to the Ryu-Takayanagi proposal \cite{Dong:2016fnf}
\beq \tilde{S}_{q}\equiv q^{2}\partial_{q}\left(\frac{q-1}{q}S_{q}\right)=\frac{\text{Area(Cosmic Brane)}_{q}}{4G_{N}}\;.\label{cosmicbrane1}\eeq
The `cosmic brane' is a (bulk) codimension-2 surface homologous to the boundary entangling region, with tension $T_{q}=\frac{q-1}{4qG_{N}}$, and backreacts on the ambient background geometry by generating a conical deficit \cite{Fursaev:1995ef}.  In the limit $q\to1$, where the tension $T_{q}$ vanishes, $\tilde{S}_{q}$ collapses to the von Neumann entropy, where the cosmic brane settles to the minimal Ryu-Takayanagi surface. Therefore, while in a general setting the R\'enyi entropy $S_{q}$ may not have a natural bulk interpretation, the modified R\'enyi entropy $\tilde{S}_{q}$ naturally satisfies an area law.

The modified entropy $\tilde{S}_{q}$ is derived following the method developed by \cite{Lewkowycz:2013nqa} to derive the Ryu-Takayanagi formula (see \cite{Fursaev:2013fta} for a related approach). This is accomplished by relating the partition function $Z[M_{q}]$ of the QFT on the branched cover $M_{q}$ formed from $q$-replicas of the $q=1$ boundary spacetime $M_{1}$ to the on-shell Euclidean action $I_{\text{Bulk}}[B_{q}]$ of the dominant bulk solution $B_{q}$ with boundary $M_{q}$ via
\beq Z[M_{q}]=e^{-I_{\text{Bulk}}[B_{q}]}\;.\eeq
Due to the $\mathbb{Z}_{q}$ replica symmetry, locality of the bulk action leads to $I_{\text{Bulk}}[B_{q}]=qI_{\text{Bulk}}[\hat{B}_{q}]$, where $\hat{B}_{q}$ is the orbifold $\hat{B}_{q}\equiv B_{q}/\mathbb{Z}_{q}$. Consequently, the boundary R\'enyi entropy is given in terms of the bulk action \cite{Lewkowycz:2013nqa}
\beq S_{q}=\frac{q}{q-1}\left(I_{\text{Bulk}}[\hat{B}_{q}]-I_{\text{Bulk}}[\hat{B}_{1}]\right)\;.\label{bulkintren}\eeq
Evaluating $\partial_{q}I_{\text{Bulk}}[\hat{B}_{q}]$ in the polar coordinate system $ds^{2}=dr^{2}+\frac{r^{2}}{q^{2}}d\phi^{2}+g_{ij}dy^{i}dy^{j}$, where $y^{i}$ and $g_{ij}$ are the coordinates and metric on the brane, respectively, and $\phi\in[0,2\pi]$, we arrive at (\ref{cosmicbrane1}) \cite{Dong:2016fnf}. Recently it was shown the addition of a Nambu-Goto brane action used to account for $\tilde{S}_{q}$ necessarily arises from including a boundary Hayward term in the full bulk gravity action \cite{Botta-Cantcheff:2020ywu}.

The entropy $\tilde{S}_{q}$ also has a thermodynamic interpretation. Writing the free energy as $F_{q}=-\frac{1}{q}\log\text{tr}\rho^{q}$, with temperature $\tilde{T}=1/q$, we see $\tilde{S}_{q}=-\frac{\partial F_{q}}{\partial \tilde{T}}$. For the case of spherical entangling regions, as explored here, we are really studying the integrated version of (\ref{cosmicbrane1}). Moreover, just as the quench interpretation of $S_{q}$ holds for charged CFTs, the modified entropy $\tilde{S}_{q}$ area relation is robust enough to adequately describe globally charged systems. This can be easily seen from the fact that $S_{q}(\mu)$ satisfies the inequality
\beq \tilde{S}_{q}(\mu)\equiv q^{2}\partial_{q}\left(\frac{q-1}{q}S_{q}(\mu)\right)\geq0\;.\eeq
Without going through the details, this is equal to the area of a (charged) cosmic brane.

It is natural to ask whether the extended R\'enyi entropy $S_{q,b}$ has a similar dual formulation. To study this, let's first consider when $q=1$, and let $\mu=0$. Then, defining our Gibbs free energy as $G_{1,b}=-\log[\text{tr}\rho_{A}^{(b)}]$ (where the temperature $T=1/q=1$ here), we find it is natural to define the modified entropy $\tilde{S}_{1,b}$ given in terms of a derivative of $G_{1,b}$ with respect to $pV/S=\frac{1}{2}(d-1)b^{2}$:
\beq \tilde{S}_{1,b}\equiv-\frac{\partial G_{1,b}}{\partial (pV/S)}=\frac{1}{(d-1)b}\partial_{b}[(b^{2}-1)S_{1,b}]\;.\label{modextentq1}\eeq
 From the inequalities (\ref{ineqs}) for $S_{1,b}$ we have $\tilde{S}_{1,b}$ is never negative, similar to $\tilde{S}_{q}$. Also note $\tilde{S}_{1,1}=S_{EE}$. 

Analogously, we can define the modified entropy $\tilde{S}_{q,b}$ as
\beq
\begin{split}
 \tilde{S}_{q,b}&\equiv\frac{1}{2}\left(q^{2}\partial_{q}+\frac{1}{(d-1)b}\partial_{b}\right)\left[\left(\frac{(q-1)}{q}+\frac{(d-1)x_{11}(b^{2}-1)}{2}\right)S_{q,b}\right]\\
&=-\frac{1}{2}\left(\frac{\partial}{\partial T}+\frac{\partial}{\partial (pV/S)}\right)G_{q,b}\;.
\end{split}
\label{Sqbmod}\eeq
It is easy to check that this modified entropy is positive for all $q,b>0$. The overall factor of $1/2$ normalizes $\tilde{S}_{q,b}$ such that $\lim_{q,b\to1}\tilde{S}_{q,b}\to S_{EE}$. We emphasize, however, the limits $q,b\to1$ and the $\partial_{q},\partial_{b}$ derivatives do not commute, \emph{e.g.}, taking the limit $b\to1$ does not lead to $\tilde{S}_{q}$. Thus, $\lim_{b\to1}\tilde{S}_{q,b}\neq \tilde{S}_{q}$. To include charge, all that is required is we modify $pV/S=(d-1)b^{2}x_{11}/2$.



In light of the above discussion, it is natural to wonder whether the extended modified entropy $\tilde{S}_{q,b}$ has an area-law prescription similar to $\tilde{S}_{q}$ (\ref{cosmicbrane1}). Without a more precise understanding of the field theory interpretation of $S_{q,b}$ it is unclear whether this is the case. We can make some progress, however, when we consider the $d=2$ limit, where the field theory interpretation has been more insightful than in higher dimensions, where we saw that $S_{1,b}$ can be built by evaluating the CFT partition function on a $b$-replica Riemann (boundary) surface. Indeed, this suggested $G_{q,b}\sim\log[\text{tr}(\rho_{w}^{qb})]$, for some  density matrix $\rho_{w}$ \cite{Johnson:2018bma}. 

Analogously then, we may formally follow the method used in \cite{Lewkowycz:2013nqa,Dong:2016fnf} to try and develop an area law prescription for our modified entropy $\tilde{S}_{q,b}$. For simplicity we consider the neutral limit. Let $M_{q,b}$ be the branched cover formed by $q,b$ replicas of the $q=b=1$ boundary spacetime $M_{1,1}$, and let $Z[M_{q,b}]$ be the partition function of the CFT on this replicated space. We then formally relate the CFT partition function to the on-shell bulk Euclidean action $Z[M_{q,b}]=e^{- I_{\text{Bulk}}[B_{q,b}]}$ where $B_{q,b}$ is the dominant bulk solution with boundary $M_{q,b}$. Our studies from $d=2$ suggest there is a $\mathbb{Z}_{qb}$ symmetry, such that $I_{\text{Bulk}}[B_{q,b}]=qbI_{\text{Bulk}}[\hat{B}_{q,b}]$, with $\hat{B}_{q,b}=B_{q,b}/\mathbb{Z}_{qb}$. Then, from the form of $S_{q,b}$ in (\ref{fieldtheorysqbdiff}), where $S_{q,b}\sim q\log Z[M_{1,1}]-\log Z[M_{q,b}]$, we have
\beq 
\left[\frac{(q-1)}{q}+\frac{1}{2}(b^{2}-1)\right]S_{q,b}=\left(bI_{\text{Bulk}}[\hat{B}_{q,b}]-I_{\text{Bulk}}[\hat{B}_{1,1}]\right)\;.
\label{modsqbv3}\eeq

Here we take the bulk action to be the same as given in \cite{Dong:2016fnf}, $I_{\text{Bulk}}=-\frac{1}{16\pi G_{N}}\int d^{3}X\sqrt{G}R$, $X$, $G_{\mu\nu}$, and $R$ are, respectively, denote the coordinates, metric and Ricci scalar of the bulk geometry. If we leave $b$ fixed, we find the $q$ derivative of this object is 
proportional to the area of a cosmic brane, up to a factor of $q^{-2}$ (the $b$ index is cancelled), where we have evaluated $\partial_{q}I_{\text{Bulk}}[\hat{B}_{q,b}]$ with the polar coordinate system $ds^{2}=dr^{2}+\frac{r^{2}}{q^{2}b^{2}}d\phi^{2}+g_{yy}(y)dy^{2}$.  Explicitly \cite{Dong:2016fnf},
\beq \partial_{q}I_{\text{Bulk}}[\hat{B}_{q,b}]=\frac{1}{16\pi G_{N}}\int d^{2}yd\phi\sqrt{\gamma}\hat{n}^{\mu}(\nabla^{\nu}\partial_{q}G_{\mu\nu}-G^{\nu\rho}\nabla_{\mu}\partial_{q}G_{\nu\rho})\;,\eeq
where we evaluate the the integral on a thin codimension-1 tube around the cosmic brane, with $\gamma$ being the induced metric on this tube and $\hat{n}^{\mu}=\partial^{\mu}_{r}$ is the outward pointing normal away from the brane. Then, in our polar coordinae system, where $\hat{n}^{\mu}\nabla^{\nu}\partial_{q}G_{\mu\nu}=\frac{2}{qr}$ we find
\beq \partial_{q}\left[\frac{(q-1)}{q}+\frac{1}{2}(b^{2}-1)\right]S_{q,b}=b\partial_{q}I_{\text{Bulk}}[\hat{B}_{q,b}]=\frac{1}{q^{2}}\frac{\text{Area(Cosmic Brane)}_{q,b}}{4G_{N}}\;.\label{alawsq1}\eeq
Taking the $b\to1$ limit on the left hand side simply yields $\tilde{S}_{q}$.

Alternatively, note that a $b$-derivative on (\ref{modsqbv3}) leads to an additional term
\beq \partial_{b}\left(\left[\frac{(q-1)}{q}+\frac{1}{2}(b^{2}-1)\right]S_{q,b}\right)=b(\partial_{b}I_{\text{Bulk}}[\hat{B}_{q,b}])+I_{\text{Bulk}}[\hat{B}_{q,b}]\;.\eeq
Since $g_{yy}=g_{yy}(y)$, the $I_{\text{Bulk}}[\hat{B}_{q,b}]$ vanishes on-shell\footnote{Crucially, this is a feature of the fact we are working in $d=2$; for higher dimensions, this additional term will be present in general and will lead to a non-trivial modification of the area relation to $\tilde{S}_{1,b}$.}, and we find the remaining term is proportional to the area of the cosmic brane. Specifically, 
\beq qb\partial_{b}\left(\left[\frac{(q-1)}{q}+\frac{1}{2}(b^{2}-1)\right]S_{q,b}\right)=\frac{\text{Area(Cosmic Brane)}_{q,b}}{4G_{N}}\;.\label{alaws1b}\eeq
Taking the $q\to1$ limit on the left hand side, we recognize this quantity as $\tilde{S}_{1,b}$ (\ref{modextentq1}). 

Substituting results (\ref{alawsq1}) and (\ref{alaws1b}) into (\ref{Sqbmod}) yields an area-law relation for $\tilde{S}_{q,b}$:
\beq \tilde{S}_{q,b}=\frac{1}{2}\left(1+\frac{1}{qb^{2}}\right)\frac{\text{Area(Cosmic Brane)}_{q,b}}{4G_{N}}\;.\label{cosmicbraneSqb}\eeq
As noted before, $\lim_{q\to1}\tilde{S}_{q,b}\neq \tilde{S}_{1,b}$, however, when $q,b\to1$ we recover $\tilde{S}_{1,1}=S_{EE}$, with $S_{EE}$ given by the Ryu-Takayanagi formula. 

On one hand, the result (\ref{cosmicbraneSqb}) is not particularly surprising in that our extended R\'enyi entropy follows from the thermal entropy of a hyperbolically sliced black hole, whose horizon is the minimal Ryu-Takaynagi surface. Alternately, our latter derivation did not make explicit use of the black hole geometry, but instead that of the geometry near the cosmic brane, likewise characterized by a Nambu-Goto action. We should also stress that while our above derivation was a special case, (neutral and $d=2$) we expect that we will find a similar area prescription for $\tilde{S}_{q,b}$ when the dual CFT is charged. Working in higher dimensions may be more difficult, as the additional term $I_{\text{Bulk}}[\hat{B}_{q,b}]$ may lead to a non-trivial modification to the area law contribution to $\tilde{S}_{1,b}$. It would be interesting to persue this further in the future. 


\section{Summary and Future Work}
\label{sec:conc}

Here we have extended the charged holographic R\'enyi entropy $S_{q}(\mu)$ of a globally charged holographic CFT characterized by a chemical potential $\mu$. Our generalization was motivated by the extended black hole thermodynamics of charged AdS black holes with hyperbolically sliced horizons. The resulting quantity $S_{q,b}(\mu)$ can be understood as a single parameter deformation of the usual R\'enyi entropy $S_{q}(\mu)$, where $b$ corresponds to changes in the pressure of a black hole whose charge and electrostatic potential is proportional to $\mu$. Collectively our work extends both \cite{Belin:2013uta} and \cite{Johnson:2018bma}. We exhaustively analyzed the behavior of $S_{q,b}(\mu)$ as a function of each of its parameters, $q,b$, and $\mu$, and found that $S_{q,b}(\mu)$ behaves in many ways like a R\'enyi entropy. Of particular interest is the $q\to1$ limit $S_{1,b}$, which we showed acts like the usual R\'enyi entropy $S_{q}$, satisfying a similar set of inequalities $S_{q}$ is expected to satisfy. Our analysis suggests an apt field theoretic interpretation of $b$ is that of a genuine R\'enyi index. This conclusion was confirmed by our holographic computations of the conformal weights $h^{(b)}_{q}$ of higher dimensional twist operators, as well as a field theoretic calculation of $h^{(b)}_{q}$ in the $d=2$ case, where the charged AdS black hole geometry is that of a charged BTZ black hole. Ultimately we found that many of the general features of the extended R\'enyi entropy are present even in the presence of a global charge, a result similar to what was observed in \cite{Belin:2013uta}. Finally, we introduced a modified entropy $\tilde{S}_{q,b}$ and found an area law prescription in terms of a backreacting cosmic brane when restricted to $d=2$.

Let's now outline possible avenues for future work. 

\vspace{2mm}

\textbf{CFTs Dual to Higher Curvature Theories of Gravity}

\vspace{2mm}

Here we only considered CFTs that are dual to general relativity. There are, of course, holographic CFTs whose bulk description is given by higher curvature theories of gravity, \emph{e.g.}, Gauss-Bonnet or quasi-topological gravity. In fact, the neutral holographic R\'enyi entropy $S_{q}$ has already been studied in the context of such bulk theories \cite{Hung:2011nu}, where the R\'enyi entropy is a complicated non-linear function of the generalized central charge $a_{d}^{\ast}$ \cite{Myers:2010tj}, computed using the Wald entropy of the black hole. It would be interesting to see what effects a global charge would introduce, both for $S_{q}$ and its extension $S_{q,b}$, for example, how the conformal weights of the generalized twist operators behave.

 As with the bulk theory we considered here, special attention would need to be given to theories in $d=2$ dimensions. This is because determining the generalized central charge $a_{d}^{\ast}$ requires the solution of the bulk theory to be \emph{locally} AdS, for which the charged BTZ black hole is not. It turns out when the bulk theory is Einstein-Maxwell the expression for $a_{d}^{\ast}$ does not change for the charged BTZ black hole because the form of the Wald entropy functional satisfies in effect the locally AdS symmetry condition\footnote{A similar observation was made when computing the thermodynamic volume of the charged BTZ black hole as a pressure derivative of the holographic entanglement entropy \cite{Rosso:2020zkk}.}. For higher curvature theories of gravity in $d=2$, such as new massive gravity and its generalizations \cite{Bergshoeff:2009hq,Bergshoeff:2009aq,Sinha:2010ai}, we expect understanding how the R\'enyi entropy depends on the $a_{d}^{\ast}$ will be more elusive. 

For the case of $d=2$ it might also be interesting to consider how the extended R\'enyi entropy $S_{q,b}(q_{e})$ changes when the bulk theory in question is Chern-Simons gravity. It is plausible that $S_{q,b}$ will be independent of the charge $q_{e}$, similar to $S_{q}$ \cite{Belin:2013uta}, as the gauge potential $A$ does not couple to the metric such that $A$ is constant in the bulk, and without a bulk source this constant is zero. Since $S_{q,b}$, however, depends on both the entropy and the thermodynamic volume $V$, it is conceivable the inclusion of $V$ will lead to a $S_{q,b}$ which still depends on $q_{e}$. It would be interesting to study this scenario as it may provide an example in which the behaviors of $S_{q}$ and $S_{q,b}$ differ dramatically. 

\vspace{2mm}

\textbf{Varying Potential, Virasoro Symmetry and Super-entropicity}

\vspace{2mm}

We considered holographic CFTs with a conserved global charge, described in the grand canonical ensemble with chemical potential $\mu$. For any dimension $d$, this condition translated to working with a charged black hole in the grand canonical ensemble at fixed electrostatic potential $\Phi_{e}$. On the gravity side, it is quite natural to study systems for which the potenial $\Phi_{e}$ or charge $q$ are left unfixed; in fact, the extended thermodynamics of charged AdS black holes is by now well-known to be rich and complex, behaving as a van der Waals fluid \cite{Dolan:2011xt,Kubiznak:2012wp}. Allowing the potential $\Phi_{e}$ to vary would immediately change the quench expression for the difference in Gibbs free energies, in which the denominator includes a term $-Q\Delta \Phi_{e}$. However, since $\Phi_{e}\propto \mu$, it is unclear what the field theoretic meaning of this difference in Gibbs free energies corresponds to, as this would imply the global charge of the CFT is not conserved. 


Studying the set-up for varying potential $\Phi_{e}$ would be particularly interesting in the $d=2$ case. In fact, note we did not see any peculiar behavior for values of $q^{2}_{e}\geq4$, \emph{i.e.}, beyond the unitarity bound uncovered in \cite{Johnson:2019wcq}. Naively we might expect to see a transition in behavior for $q^{2}_{e}$ above and below the unitarity bound considering this is when the BTZ black hole has a thermodynamic instability, and is correspondingly super-entropic. Since we  held $\Phi_{e}$ fixed, this corresponds to subtly using the renormalization scheme \cite{Cadoni:2007ck}. Consequently, the charged BTZ black hole is notably \emph{not} super-entropic, thermodynamically stable, and, correspondingly, $q^{2}_{e}$ does not affect the unitarity bound
 of the dual $\text{CFT}_{2}$. Moreover, employing this renormalization scheme has the charged BTZ black hole exhibiting explicit Virasoro symmetry in its asymptotic structure. Effectively, then, allowing $\Phi_{e}$ to vary means we ignore the renormalization scheme, whereby the Virasoro symmetry of the underlying CFT becomes hidden and the black hole becomes super-entropic and thermodynamically unstable. Provided there exists a suitable interpretation of $S_{q}(\mu)$ for varying $\mu$, it would be interesting to see whether the thermodynamic instability affects the R\'enyi entropy, perhaps leading to new phase transitions in $S_{q}$ in $d=2$. Moreover, we may gain further insight into the microscopic understanding of black hole super-entropicity, building off of the recent work \cite{Johnson:2019wcq}.


\vspace{2mm}

\textbf{Holographic Phase Transitions}

\vspace{2mm}

It is well known that charged hyperbolic AdS black holes do not exhibit a Hawking-Page phase transition \cite{Cai:2004pz}. This can be observed from the Gibbs free energy, which is everywhere negative except in the extremal limit when it becomes zero. Moreover, the specific heat at fixed $\mu$ is always positive. Therefore, the charged topological AdS black hole does not exhibit any phase transitions. As such, the charged R\'enyi entropy $S_{q}(\mu)$ and its extension $S_{q,b}(\mu)$ do not exhibit any phase transitions induced by the bulk geometry. 

In spite of this, when a (charged) light scalar field is present in the bulk, (charged) black holes become unstable in the near extremal limit when the mass of the scalar field is below the BF bound of the $AdS_{2}$ sector of the near horizon geometry. Consequently, the scalar field condenses leading to a hairy black hole, signalling a dual phase transition in the (charged) R\'enyi entropy \cite{Belin:2013dva, Belin:2014mva}. It would be interesting to similarly study the phase transitions of $S_{q,b}$ -- both the neutral and charged cases -- that arise from the same scalar field condensation, as it may expand our understanding of holographic superconductors. An analytic study of such holographic phase transitions is currently underway. 

\vspace{2mm}

\textbf{Information Theoretic Meaning of $S_{q,b}$}

\vspace{2mm}

Perhaps the most worthwhile future research avenue is to develop a concrete information theoretic interpretation of $S_{q,b}$. Here we exploited the quench definition of the R\'enyi entropy, and naturally generalized it by replacing the difference in Helmholtz free energies with a difference in Gibbs free energies.  As noted in \cite{baez:2011}, written as a quench, it is evident $S_{q}$ is proportional to the `$q$-derivative' of the free energy $F(T)$, $S_{q}=-\left(\frac{\partial F}{\partial T_{0}}\right)_{q^{-1}}$, where the $q$-derivative of a function $f$ is defined as 
\beq \left(\frac{\partial f}{\partial x}\right)_{q}=\frac{f(qx)-f(x)}{(q-1)x}\;.\eeq
The $q$-derivative appears often in mathematics literature whenever one `$q$-deforms' some ordinary structure, \emph{e.g.}, $q$-deformed Lie groups produce quantum groups, and is prevalent in the theory of quantum calculus \cite{cheung2002}. The $q$-derivative interpretation of information and statistical entropies has been known for some time \cite{Abe1997} and continues to be of interest \cite{Marinho:2020sbc}.

Similarly, we can interpret $S_{1,b}$ as a $q$-derivative of the Gibbs free energy, specifically, $S_{1,b}=-(1/V_{0}/S_{0})\left(\frac{\partial G(T_{0},p_{0})}{\partial p_{0}}\right)_{b^{2}}$. We point out, however, that $S_{q,b}$ is not quite a $b$-deformation of $S_{q}$ in the mathematical sense \cite{cheung2002}; rather $S_{q,b}$ is akin to a simultaneous $q$-derivative of $G(T_{0},p_{0})$ in both of its arguments. It would be interesting to see whether $S_{q,b}$ can be precisely formulated with $q$-calculus, as it applies to more generic thermodynamic systems, and may lead to a new kind of information entropy, whose holographic dual would have an immediate interpretation.

\section*{Acknowledgements}

It is a pleasure to thank Nikhil Monga and Clifford Johnson for feedback on this manuscript, and Victoria Martin and Felipe Rosso for additional discussions.  AS is supported by the Simons Foundation \emph{It from Qubit} collaboration (under Jonathan Oppenheim).


\bibliography{referencesbhphase}

\end{document}